\pdfoutput=1

\documentclass[11pt]{article}

\usepackage[english]{babel}
\usepackage{xcolor}
\usepackage{graphicx}

\usepackage[colorlinks,allcolors=blue,linktocpage=true,draft=false]{hyperref}

\usepackage{cite}
\usepackage{subcaption}
\usepackage{amsmath}

\usepackage{amssymb}

\usepackage{booktabs}   

\usepackage[top=15mm,bottom=12mm,left=30mm,right=30mm,foot=12mm,includefoot]{geo
metry}

\allowdisplaybreaks[3]

\numberwithin{equation}{section}

\newcommand{\sect}[1]{section~#1}
\newcommand{\sects}[1]{sections~#1}
\newcommand{\app}[1]{appendix~#1}

\newcommand{\fig}[1]{figure~#1}
\newcommand{\figs}[1]{figures~#1}

\newcommand{\tab}[1]{table~#1}

\newcommand{\eqn}[1]{equation~#1}
\newcommand{\eqs}[1]{equations~#1}

\newcommand{\als}{\alpha_s}

\newcommand{\gev}{\operatorname{GeV}}

\newcommand{\microsec}{\operatorname{\mu s}}

\newcommand{\ms}{\mskip 1.5mu}
\newcommand{\bs}{\mskip -1.5mu}

\newcommand{\tvec}[1]{\boldsymbol{#1}}
\newcommand{\vect}[1]{\boldsymbol{#1}}

\newcommand{\mvec}[1]{\smash{\vec{\mskip 0.5mu #1}}\mskip 1.5mu}

\newcommand{\ftil}{\tilde{f}}
\newcommand{\otil}{\widetilde{\omega}}
\newcommand{\Atil}{\widetilde{A}}
\newcommand{\htil}{\tilde{h}}

\newcommand{\ser}{\sigma}

\newcommand{\chilipdf}{\textsc{ChiliPDF}}
\newcommand{\bestlime}{\textsc{BestLime}}

\graphicspath{{plots/}}

\newcommand{\rev}[1]{#1}

\renewcommand{\arraystretch}{1.3}

\begin{document}

\begin{flushright}
DESY-24-057 \\
\href{https://arxiv.org/abs/2405.08616}{arXiv:2405.08616 [hep-ph]}
\end{flushright}

\begin{center}
\vspace{4\baselineskip}
\textbf{\Large Efficient computation of Fourier-Bessel transforms \\[0.2em]
for transverse-momentum dependent parton distributions \\[0.4em]
and other functions} \\
\vspace{3\baselineskip}
Markus~Diehl and Oskar Grocholski

\vspace{\baselineskip}

Deutsches Elektronen-Synchrotron DESY, Notkestr.~85, 22607 Hamburg, Germany

\vspace{3\baselineskip}

\parbox{0.9\textwidth}{
We present a method for the numerical computation of Fourier-Bessel transforms
on a finite or infinite interval.  The function to be transformed needs to be
evaluated on a grid of points that is independent of the argument of the Bessel
function.  We demonstrate the accuracy of the algorithm for a wide range of
functions, including those that appear in the context of transverse-momentum
dependent parton distributions in Quantum Chromodynamics.

}
\end{center}

\vfill

\newpage

\tableofcontents


\newpage


\section{Introduction}
\label{sec:intro}

The Fourier-Bessel transform --- also known as Hankel transform --- appears in
many branches of physics; in particular it arises from the multi-dimensional
Fourier transform of functions with rotational symmetry.  It has the form
\begin{align}
   \label{gen-int-f}
   I(q)
   &=
   \int_{z_a}^{z_b} dz \, J_{\nu}(q z) \, \ftil(z)
   \,,
\end{align}
where $J_\nu$ is a Bessel function of the first kind.  In general the
integration runs from $0$ to $\infty$, but it may be reduced to a finite
interval by the support properties of $\ftil(z)$.

In Quantum Chromodynamics, the Fourier-Bessel transform prominently appears in
the context of transverse-momentum dependent parton distributions (TMDs), where
cross sections are proportional to integrals of the form \eqref{gen-int-f} with
$\nu = 0, 1, 2, 3$ and $\ftil(z)$ equal to $z^{\nu+1}$ times the product of two
TMDs,\footnote{%
   We note that in the TMD literature, the spatial distance is
   usually called $b$ rather than $z$.}
see e.g.\ \sect{2.11} in \cite{Boussarie:2023izj}.
The integration limits are $0$ and $\infty$ in that case.  The method proposed
in \cite{Ebert:2022cku} involves similar integrals over TMDs from $0$ to some
finite value, and from that finite value to $\infty$.
A more involved application is given by two-parton TMDs, which appear in the
description of double parton scattering \cite{Buffing:2017mqm}.  In this case,
two nested Fourier transforms occur, the radial parts of which can be converted
into two nested Fourier-Bessel integrals, $\int d z_1 \int
d z_2\, J_{\nu_1}(q_1 z_1) \, J_{\nu_2}(q_2 \ms z_2) \, \ftil(z_1, z_2)$.

In the cases just mentioned, one typically needs to compute the transform
\eqref{gen-int-f} for many settings of external parameters, namely longitudinal
parton momentum fractions and hard scales. For uncertainty estimates on TMDs
fitted to data (using the Hessian method or replicas) one must repeat the
calculation for many TMD sets. The numerical computation of the integrands
itself can be rather costly; at small transverse distances $z$ it involves
Mellin convolutions in the longitudinal momentum fractions, with lengthy kernels
at higher perturbative orders (see \sect{2.8} in \cite{Boussarie:2023izj} and
references therein).

In the TMD context, the variable $q$ in \eqref{gen-int-f} is a measured
transverse momentum, for instance of the produced lepton pair in the Drell-Yan
process.   The process to be described is characterised by a hard scale $Q$,
which can range from a few $\gev$ for semi-inclusive deep inelastic scattering
or non-resonant Drell-Yan production to about $100 \gev$ for the production
of a $W$, $Z$, or Higgs boson.  For the production of a pair of such bosons, the
hard scale $Q$ is given by the invariant mass of the pair and can be
significantly larger.
The TMD formalism describes cross sections in the region $q \ll Q$, whilst a
description in terms of conventional collinear parton distributions is adequate
for $q \sim Q$.  A smooth transition between the two regimes typically
requires evaluating the TMD expression for $q$ up to order $Q$, see
\sect{4.7} in \cite{Boussarie:2023izj} for an overview and \cite{Cal:2023mib}
for a recent example.

In summary, the calculation of cross sections with TMDs may involve a large
number of Fourier-Bessel transforms of complicated functions for a possibly wide
range of $q$ values.  In such a scenario, integration methods with nodes that
depend on $q$ come at a rather high computational cost since the integrand needs
to be computed many times.  This is the case for the method of Ogata
\cite{Ogata:2005}, which is used for TMDs for instance in the codes
\texttt{artemide} \cite{Scimemi:2017etj} or \texttt{ResBos2}
\cite{Isaacson:2023iui}, the latter employing the algorithm of Kang
et al.~\cite{Kang:2019ctl}.\footnote{%
   The respective codes and documentation can be found at
   \url{https://github.com/VladimirovAlexey/artemide-public} for
   \cite{Scimemi:2017etj} and at
   \url{https://github.com/UCLA-TMD/Ogata} and
   \url{https://ucla-tmd.github.io/Ogata} for \cite{Kang:2019ctl}.}
A method using fixed nodes in $z$ for a whole set of $q$ values is the discrete
Hankel transform, see \cite{Baddour:2015, Baddour:2019} for a mathematical
description and \cite{Venkat:2010by} for an application in hadron physics.
Accessing high $q$ requires a large number of nodes in this method.

In the present work, we adapt a method for oscillatory integrands due to Levin
\cite{Levin:1982, Levin:1996} to the specific case of Fourier-Bessel transforms.
We show that high accuracy can be achieved by combining this method with
Chebyshev interpolation (which was already done in \cite{Li:2010, Li:2011},
albeit not for Bessel weighted integrals).
Specifically, we present a method to compute integrals of the form
\begin{align}
   \label{gen-int-g}
   I(q)
   &=
   \int_{z_a}^{z_b} dz \, J_{\nu}(q z) \,
      \biggl( \frac{1+z}{z} \biggr)^{\nu}\, f(z)
   \,,
   &
   \nu \ge 0, \ \,
   q > 0, \ \,
   z_a < z_b
\end{align}
with the following specifications:
\begin{itemize}
\item $z_a$ may be zero or positive, and $z_b$ may be finite or $\infty$.

We find that it can be useful for both accuracy and computation time to evaluate
the integral over $z \in [0,\infty]$ as the sum of integrals over two or more
subintervals.
\item $\nu$ may be integer or non-integer.  Notice that for half integer $\nu$,
the integral \eqref{gen-int-g} can be rewritten in terms of the spherical Bessel
functions $j_n(x) = \sqrt{\pi / (2 x)} \, J_{n + 1/2}(x)$.
\item $f(z)$ must be finite on the full integration interval, including the end
points $z_a$ and $z_b$.  To obtain accurate integrals, $f(z)$ should should be
sufficiently smooth and not have fast oscillations.  We will see that a
non-analytic behaviour like $f(z) \sim z^{1 - \delta}$ at $z = 0$ is amenable to
the method if $\delta$ is not too large, although the accuracy tends to degrade
with increasing~$\delta$.
\item The function $f(z)$ is evaluated on a discrete set of $n$ points from
$z_a$ to $z_b$ (specified in \sect{\ref{sec:cheb}}).  We find that with a
suitable choice of discretisation grid, good integration accuracy can be
obtained for a rather wide range of~$q$.
\end{itemize}
We have implemented this method in a C++ library named
\bestlime,\footnote{\textbf{Bes}sel \textbf{t}ransformation with
\textbf{L}evin's \textbf{i}ntegration \textbf{me}thod.}
which can be downloaded from \cite{BestLime:release} and is briefly discussed in
\sects{\ref{sec:code}} and \ref{sec:timing} of the present paper.
Although our primary physics motivation is the application to TMDs, we think
that the method may be of interest to other domains.  We show in
\sect{\ref{sec:benchmarks}} that it can handle a large variety of functions
$f(z)$.

Our paper is organised as follows.  In \sect{\ref{sec:method}} we develop our
method starting from the original one in \cite{Levin:1996}, and in
\sect{\ref{sec:method-final}} we present its final form.  We discuss general
numerical aspects in \sect{\ref{sec:numerics}} and integrals appearing in TMD
cross sections in \sect{\ref{sec:tmd-numerics}}.  \rev{In
\sect{\ref{sec:special-int}} we consider integrals from $0$ to $\infty$ that do
not converge but can be defined in the distributional sense.}  Our findings are
summarised in \sect{\ref{sec:summary}}.  Various mathematical details are given
in \app{\ref{sec:functions}} \rev{to \app{\ref{sec:distribution}}}.

\section{Developing the method}
\label{sec:method}

\subsection{The original Levin method}
\label{sec:Levin-orig}

To begin with, let us present the original method of Levin \cite{Levin:1996} for
the case of Fourier-Bessel integrals.  We use a slightly different notation than
the cited work.  The integral to be evaluated has the form
\begin{align}
   \label{gen-int-Levin}
   I(q)
   &=
   \int_{z_a}^{z_b} dz\,
   \Bigl[\,
      \otil_1(z, q) \, \ftil_1(z) + \otil_2(z, q) \, \ftil_2(z)
   \,\Bigr]
\end{align}
with weights
\begin{align}
   \label{Levin-weights}
   \otil_1(z, q)
   &=
   J_{\nu}(q z)
   \,,
   &
   \otil_2
   &=
   J_{\nu+1}(q z)
   \,,
\end{align}
where it is assumed that the functions $\ftil_1(z)$ and $\ftil_2(z)$ are
\emph{not} strongly oscillating.
The Fourier-Bessel transform \eqref{gen-int-f} is obtained by setting $\ftil_1 =
\ftil$ and $\ftil_2 = 0$.  Setting $\ftil_1 = 0$ and $\ftil_2 = \ftil$ instead,
one obtains the analogue of \eqref{gen-int-f} for $J_{\nu+1}$ instead of
$J_{\nu}$. This requires only a minor additional computational effort; a bonus
that is preserved by our adaptation of Levin's method.

To proceed, we introduce the vector notation
\begin{align}
   \tvec{\otil}
   &=
   \bigl( \otil_1 \,, \otil_2 \bigr)^T
   \,,
   &
   \tvec{\ftil}
   &=
   \bigl(  \ftil_1 \,, \ftil_2 \bigr)^T
   \,,
\end{align}
where a superscript $T$ denotes transposition.  Using the relations
\eqref{Bessel-basic}, one readily derives
\begin{align}
   \label{ODE-for_A}
   \frac{d}{d z} \,  \tvec{\otil}(z, q)
   &=
   \Atil^{\, T}(z, q) \cdot \tvec{\otil}(z, q)
\end{align}
with the matrix
\begin{align}
   \label{Atilde}
   \Atil(z, q)
   &=
   \begin{pmatrix}
      \nu / z \,, & q \\
      -q \,,      & - (\nu + 1) / z
   \end{pmatrix}
   \,.
\end{align}

If one finds a solution for the system
\begin{align}
   \label{gen-Levin-ODE}
   \tvec{\ftil}(z)
   &=
   \biggl[ \frac{d}{d z} + \Atil(z, q) \biggr] \, \tvec{\htil}(z, q)
\end{align}
of ordinary differential equations (ODEs), one can evaluate the integral as
\begin{align}
   \label{gen-Levin-int}
   I(q)
   &=
   \int_{z_a}^{z_b} dz\,
      \tvec{\otil}^T(z, q) \cdot \tvec{\ftil}(z)
   =
   \Bigl[\ms \tvec{\otil}^T(z, q) \cdot \tvec{\htil}(z, q)
   \ms\Bigr]_{z_a}^{z_b}
   \;\;,
\end{align}
where we use the common notation
\begin{align}
   \bigl[ F(z, q) \bigr]_{z_a}^{z_b}
   &=
   F(z_b, q) - F(z_a, q)
   \,.
\end{align}
This is readily seen by inserting \eqref{gen-Levin-ODE} in the first expression
of \eqref{gen-Levin-int}, integrating by parts, and then using
\eqref{ODE-for_A}.

As argued in \cite{Levin:1982, Levin:1996}, the system \eqref{gen-Levin-ODE} has
one particular solution with functions $\htil_i(z, q)$ that are not rapidly
oscillating.
One can find this solution by collocation, which consists in approximating
$\htil_1(z, q)$ and $\htil_2(z, q)$ by a linear combination of $n$ basis
functions $b_i(z)$ and by determining the coefficients of these functions by
evaluating the system \eqref{gen-Levin-ODE} at a set of $n$ suitably chosen
collocation points $z_i$.  This yields a coupled system of linear equations that
can be solved by standard linear algebra methods (see \sect{\ref{sec:linalg}}
for details).
For the method to give accurate results, it is essential that the approximation
of $\htil_1$ and $\htil_2$ is sufficiently precise, in particular at the
end points $z_a$ and $z_b$, where these functions are evaluated in the final
result \eqref{gen-Levin-int}.

As it stands, this method is not suitable for a lower integration boundary $z_a
= 0$, where the diagonal elements of $\Atil$ diverge.  We circumvent this
problem by rescaling, as described in the next subsection.


\subsection{Rescaling: general form}
\label{sec:rescale}

The representation \eqref{gen-Levin-int} of $I(q)$ remains valid if one replaces
$\tvec{\ftil}$, $\tvec{\htil}$ and $\tvec{\otil}$ by the rescaled functions
\begin{align}
   \label{rescale-fcts}
   \tvec{f}(z)
   &=
   r^{\nu}(z) \, \tvec{\ftil}(z)
   \,,
   &
   \tvec{h}(z)
   &=
   r^{\nu}(z) \, \tvec{\htil}(z)
   \,,
   \notag \\[0.2em]
   \tvec{\omega}(z, q)
   &=
   r^{-\nu}(z) \, \tvec{\otil}(z, q)
   \,,
\end{align}
where $r(z)$ will be specified below.
Using the explicit form \eqref{Atilde} of $\Atil$, one can rewrite the system of
ODEs \eqref{gen-Levin-ODE} as
\begin{align}
   \label{ODE-h2}
   f_1
   &=
   \frac{d}{dz} \, h_1
   + q h_2
   + \nu \ms \bigg[\ms \frac{1}{z} - \frac{r'}{r} \ms\biggr] \, h_1
   \,,
   \notag \\
   f_2
   &=
   \frac{d}{dz} \, h_2
   - q h_1
   - \biggl[\ms \frac{\nu+1}{z} + \nu \, \frac{r'}{r} \ms\biggr] \, h_2
   \,,
\end{align}
where we have omitted the arguments $z$ or $(z, q)$ of the functions and
abbreviated $r' = d r / dz$.
The singularity at $z=0$ in the first equation of \eqref{ODE-h2} is removed if
\begin{align}
   r(z)
   &
   \sim z
   &
   \text{ for } z \to 0
   \,,
\end{align}
and the singularity in the second equation is avoided by setting
\begin{align}
   h_2(z, q)
   &=
   r(z) \, h_3(z, q)
   \,.
\end{align}
The final form of the ODE system then reads
\begin{align}
   \label{ODE-h3}
   f_1
   &=
   \frac{d}{dz} \, h_1
   + q \ms r \ms h_3
   + \nu \ms \bigg[\ms \frac{1}{z} - \frac{r'}{r} \ms\biggr] \, h_1
   \,,
   \notag\\
   f_2
   &=
   r \, \frac{d}{dz} \, h_3
   - q h_1
   - \biggl[\ms (\nu+1) \, \frac{r}{z} + (\nu-1) \, r' \ms\biggr] \, h_3
   \,,
\end{align}
and the integral $I(q)$ is given by
\begin{align}
   \label{Levin-int-resc}
   I(q)
   &=
   \int_{z_a}^{z_b} dz \; \,
   r^{-\nu}(z) \,
   \Bigl[ J_{\nu}(q z) \, f_1(z)
      + J_{\nu+1}(q z) \, f_2(z) \Bigr]
   \nonumber \\
   &=
   \biggl[\ms J_{\nu}(q z) \, r^{-\nu}(z) \, h_1(z, q)
      + J_{\nu+1}(q z) \, r^{-\nu+1}(z)  \, h_3(z, q)
   \ms\biggr]_{z_a}^{z_b}
   \,.
\end{align}
We will show that if $f_1(z)$ and $f_2(z)$ are sufficiently well-behaved
at $z\to 0$, the solutions $h_1(z, q)$ and $h_3(z, q)$ of \eqref{ODE-h3} are
well behaved at $z\to 0$ as well.  This enables one to solve the system by
collocation with basis functions that have a Taylor expansion in $z$ around
$z=0$.  We anticipate that our method will use basis functions with this
property, postponing details to \sect{\ref{sec:cheb}}.


\subsection{Rescaling with \texorpdfstring{$r(z) = z$}{r(z) = z}}
\label{sec:rescale_lin}

The main purpose of this subsection is to explicitly solve the system
\eqref{ODE-h3} of ODEs for particular choices of $f_1$ and $f_2$.  This is an
intermediate step, and a reader mainly interested in final version of our method
may skip forward to \sect{\ref{sec:method-final}}.

In the present subsection, we set
\begin{align}
   \label{r-linear}
   r(z) &= z
   \,,
\end{align}
such that the system of ODE takes the rather simple form
\begin{align}
   \label{ODE-h3-lin-f1}
   f_1
   &=
   \frac{d}{dz} \, h_1
   + q z \ms h_3
   \,,
   \\
   \label{ODE-h3-lin-f2}
   f_2
   &=
   z \ms \frac{d}{dz} \, h_3
   - q h_1
   - 2\nu \ms h_3
   \,.
\end{align}
One can eliminate $h_1$ by taking the derivative of \eqref{ODE-h3-lin-f2} and
then inserting $d h_1/ d z$ from \eqref{ODE-h3-lin-f1}.  The result is
\begin{align}
   \label{ODE-h3-only}
   z^2 \ms \frac{d^2}{dz^2} \, (z^{-\nu} \ms h_3)
   + z \ms \frac{d}{dz} \, (z^{-\nu} \ms h_3)
   + \bigl( q^2 z^2 - \nu^2 \bigr) \, (z^{-\nu} \ms h_3)
   &=
   z^{1-\nu} \ms \bigl( q f_1^{} + f^{\ms \prime}_2 \ms\bigr)
\end{align}
with $f^{\ms \prime}_2 = d\mskip 0.1mu f_2^{} / dz$.

If one sets $f_1 = f_2 = 0$, this is the Bessel equation for the function
$z^{-\nu} h_3$.  The \emph{homogeneous} solution of \eqs{\eqref{ODE-h3-lin-f1}}
and \eqref{ODE-h3-lin-f2} is therefore given by
\begin{align}
   \label{hom-solution}
   \begin{pmatrix} h_1 \\ h_3 \end{pmatrix} \Bigg|_{\text{hom}}
   &=
   c_J \ms \begin{pmatrix} {}- (q z)^{\nu+1} \, J_{\nu+1}(q z) \\
                           q \, (q z)^{\nu} \, J_{\nu}(q z) \end{pmatrix}
   + c_Y \ms \begin{pmatrix} {}- (q z)^{\nu+1} \, Y_{\nu+1}(q z) \\
                             q \, (q z)^{\nu} \, Y_{\nu}(q z) \end{pmatrix}
\end{align}
with arbitrary coefficients $c_J$ and $c_Y$, where $Y_{\nu}$ denotes the Bessel
functions of the second kind.  We obtained the solution for $h_1$ by inserting
the one for $h_3$ into \eqref{ODE-h3-lin-f2} and using the relations
\eqref{Bessel-basic} and \eqref{Bessel-diff}.

At small $z$, the Bessel functions $J_\nu$ and $Y_\nu$ can be represented as
sums of different power series in $z^2$, some of which are multiplied by
functions of $z$.  To represent this in a compact form, we write
\begin{align}
   \label{gen-series-def}
   \ser(x)
   &=
   \sum_{k=0}^\infty  a_k \ms x^{k}
\end{align}
to denote a power series in the variable $x$.  It is understood that the
series may terminate, so that one has a polynomial in $x$.  We use the notation
\eqref{gen-series-def} in a similar way as the familiar symbol $\mathcal{O}(x)$,
i.e., different occurrences of $\ser(x)$ (even within a single formula) may have
different coefficients $a_k$, and these coefficients may depend on variables
other than $x$ (such as $q$ and $\nu$ in our context).

With this notation, the Bessel functions can be written in the form
\eqref{Bessel-J-Y-series}, which implies a small-$z$ behaviour
\begin{align}
   \label{hom-solution-small-z}
   \begin{pmatrix} h_1 \\ h_3 \end{pmatrix} \Bigg|_{\text{hom}}
   &=
   \begin{pmatrix}
      \ser(z^2) + z^{2 \nu + 2} \, \ser(z^2)
         + z^{2 \nu + 2} \ms \ln(z) \, \ser(z^2) \\
      \ser(z^2) + z^{2 \nu} \, \ser(z^2)
         + z^{2 \nu} \ms \ln(z) \, \ser(z^2)
   \end{pmatrix}
\end{align}
of the homogeneous solution \eqref{hom-solution}, where we assumed $\nu \ge 0$.
The terms with $\ln(z)$ only appear for integer values of $\nu$.
We see that the homogeneous solutions are non-analytic at $z=0$, except for
half-integer $\nu$.  Recalling that the basis functions we will use for
collocation have a Taylor expansion around $z=0$, we require
\begin{align}
   \nu & \ge 1
\end{align}
when using \eqref{Levin-int-resc}, which ensures that the functions
\eqref{hom-solution-small-z} and their first derivatives in $z$ are finite at
that point.  Integrals with $J_\nu$ for $\nu < 1$ can be evaluated using integration by parts, as specified in \eqn{\eqref{IBP}} below.


\subsubsection{Solution if \texorpdfstring{$f_1$}{f1} or
   \texorpdfstring{$f_2$}{f2} are powers of \texorpdfstring{$z$}{z}}
\label{sec:f1-or-f2-power}

Let us first consider the case where $f_2 = 0$ and
\begin{align}
   \label{pow-fct-1}
   f_1(z) &= c_1 \ms z^{\mu + \nu}
   \,,
\end{align}
which corresponds to the integral of $c_1 \ms z^{\mu} J_{\nu}(q z)$ in
\eqref{Levin-int-resc}.
Notice that a finite value of $f_1(z)$ at $z = 0$ requires $\mu + \nu \ge 0$.

The r.h.s.\ of \eqref{ODE-h3-only} now equals $c_1 q z^{1 + \mu}$, and we
recognise the Lommel differential equation \eqref{Lommel-ODE} for the function
$z^{-\nu} h_3$ (up to a constant on the r.h.s.).
The equation is solved by $c_1 \ms q^{-\mu}$ times the Lommel functions of the first
or second kind, $s_{\mu, \nu}(q z)$ or $S_{\mu, \nu}(q z)$, which differ by a
linear combination $c_J \ms J_{\nu}(q z) + c_Y \ms Y_{\nu}(q z)$ of the
homogeneous solutions, see \eqref{2nd-Lommel-def}.

\begin{figure}[t]
\centering
\subfloat[$\nu = 1$, \ $x < 2$]{
   \includegraphics[width=0.47\textwidth,trim=0 0 40 57,clip]{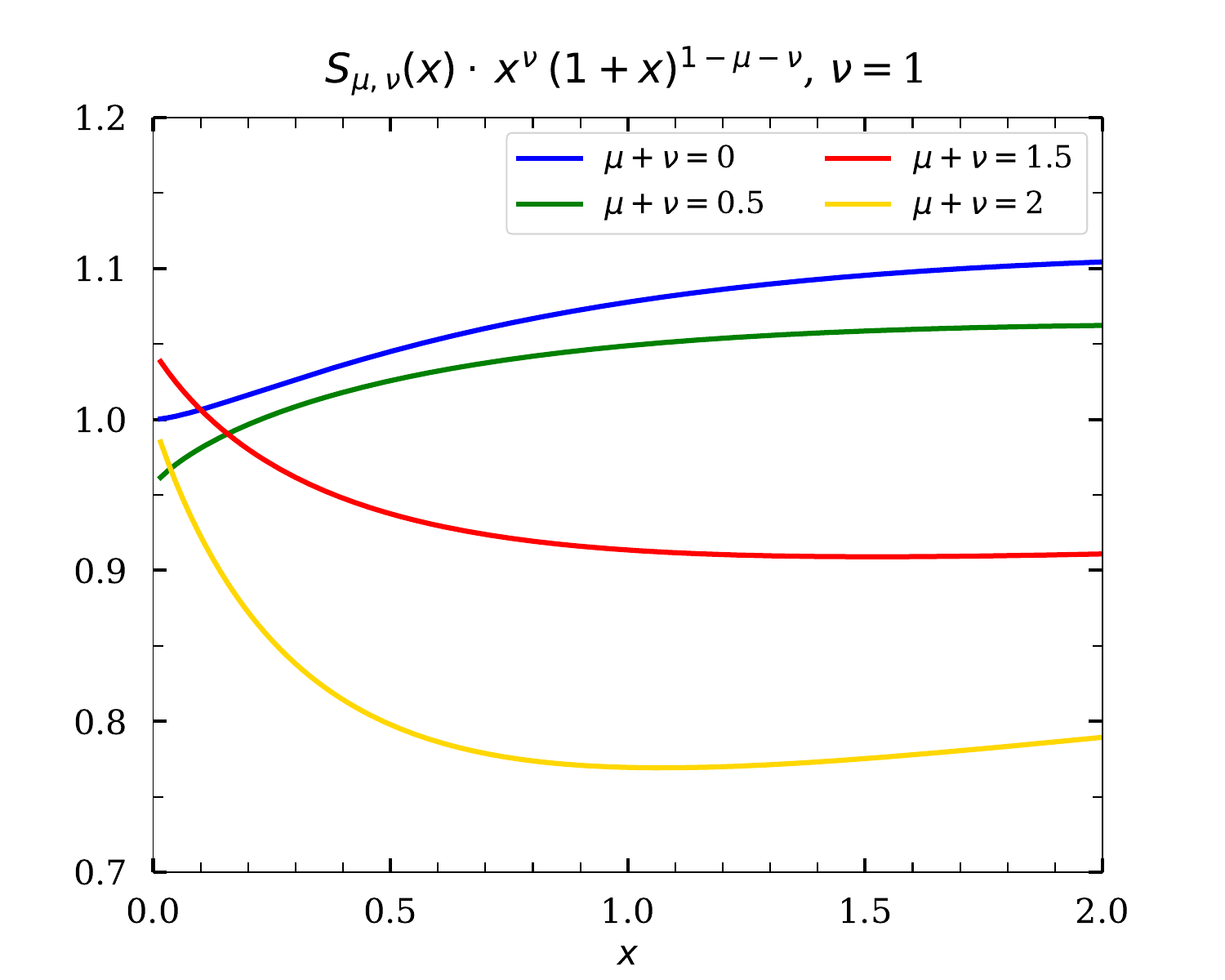}
}
\subfloat[$\nu = 1$, \ $x > 2$]{
   \includegraphics[width=0.47\textwidth,trim=0 0 40 57,clip]{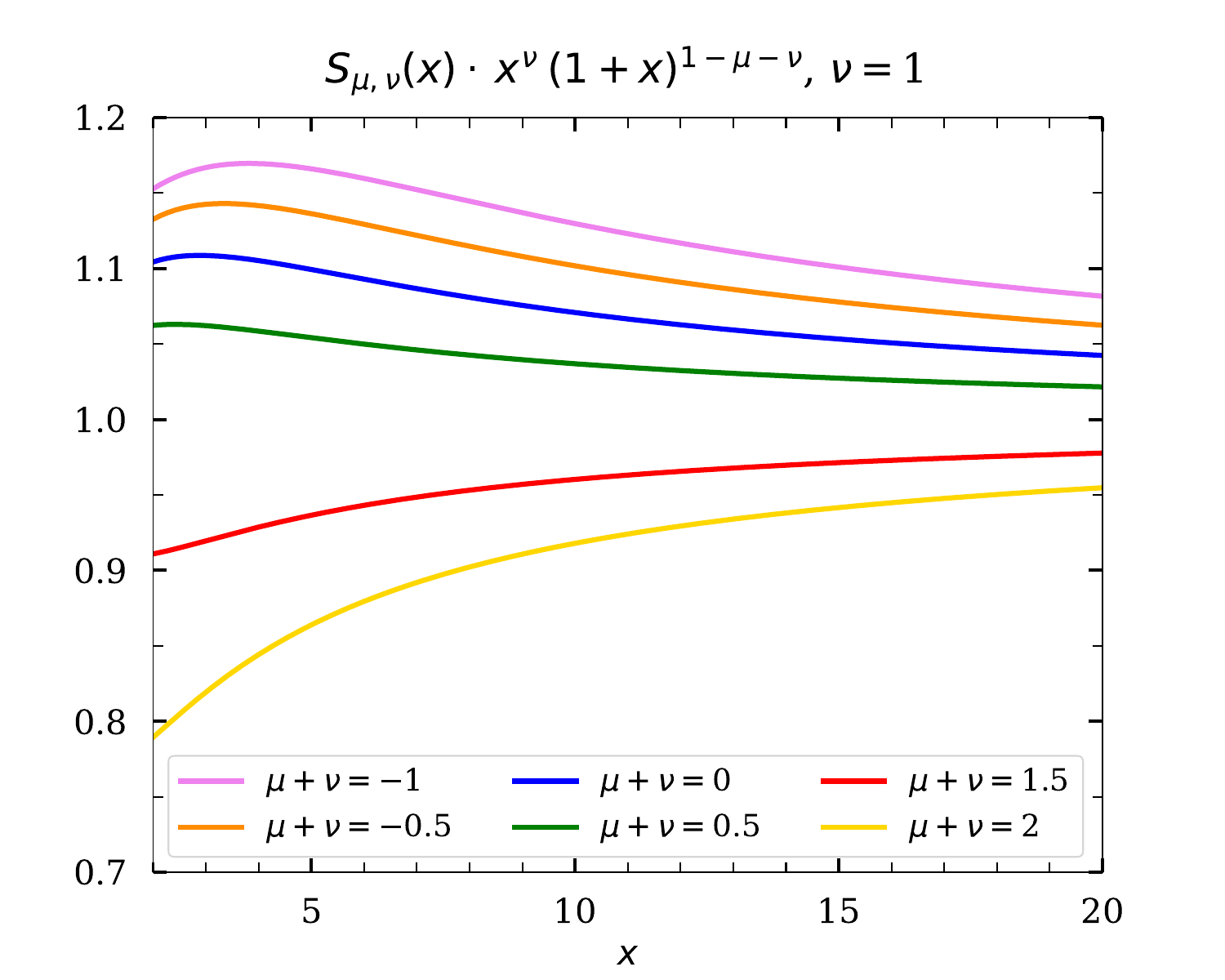}
}
\\[1.5em]
\subfloat[$\nu = 3/2$, \ $x < 2$]{
   \includegraphics[width=0.47\textwidth,trim=0 0 40 57,clip]{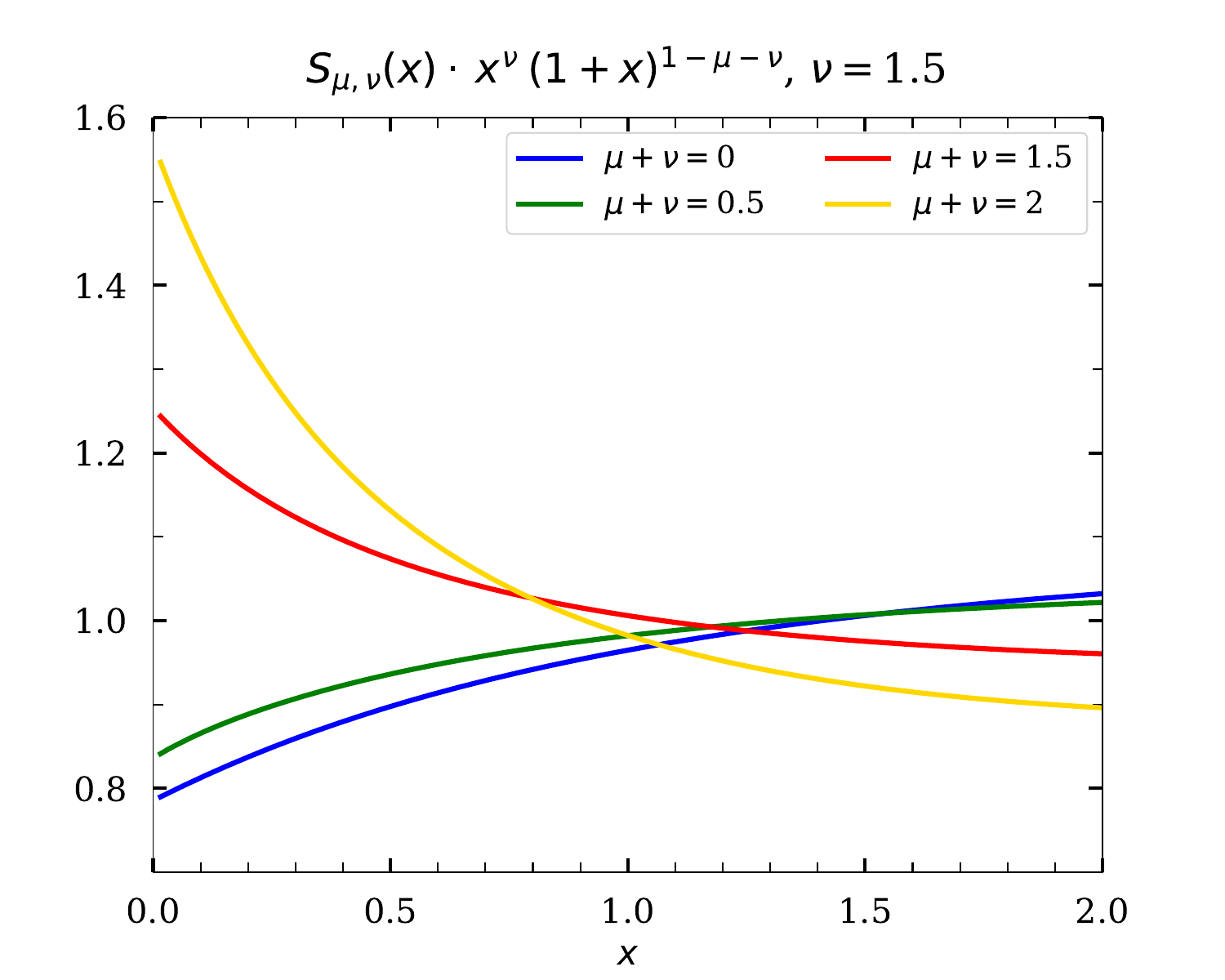}
}
\subfloat[$\nu = 3/2$, \ $x > 2$]{
   \includegraphics[width=0.47\textwidth,trim=0 0 40 57,clip]{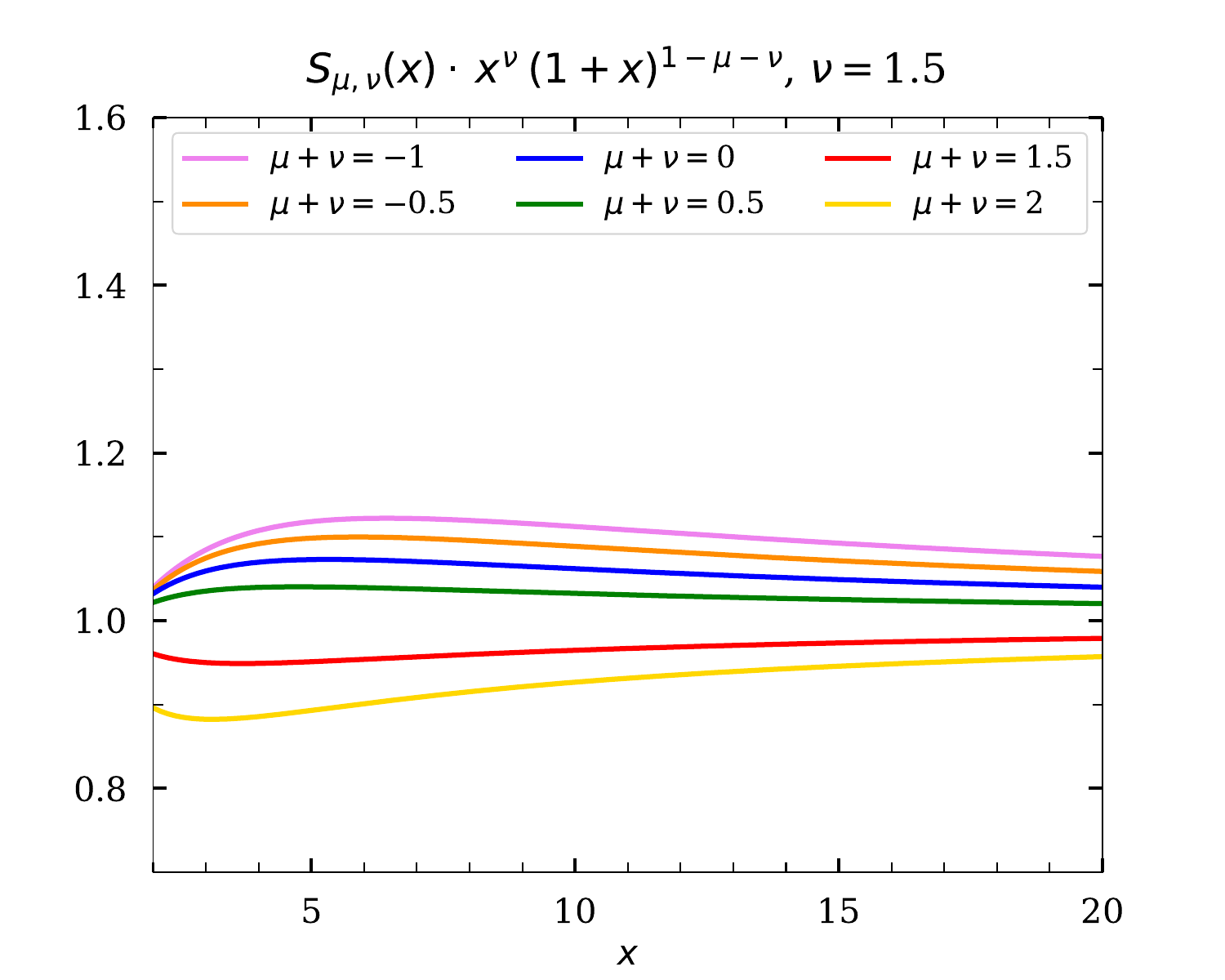}
}
\caption{\label{fig:Lommel-fcts} Lommel functions of the second kind, scaled as
specified in \protect\eqref{Lommel-scaled}.  The functions for $\mu + \nu = 1$
are not shown, since according to \protect\eqref{Lommel-simple} they are
identical to $1$.}
\end{figure}

It turns out that it is the Lommel functions of the second kind that are free of
fast oscillations in $z$.  This is illustrated for selected values of $\mu$ and
$\nu$ in \fig{\ref{fig:Lommel-fcts}}.  In the plots, we have scaled the
functions as
\begin{align}
   \label{Lommel-scaled}
   x^{\nu} \, (1+x)^{1 - \mu - \nu} \, S_{\mu, \nu}(x)
\end{align}
for clarity.  According to \eqref{Lommel-large-x} the scaled functions tend to
$1$ for $x\to \infty$, and according to \eqref{Lommel-gen-series} they have a
finite limit for $x\to 0$ if $\mu + \nu \ge 0$ and $\nu \ge 1$.

The corresponding solution for $h_1$ is readily obtained by inserting the one
for $h_3$ into \eqref{ODE-h3-lin-f2}.  Using the relations \eqref{Lommel-basic},
we obtain
\begin{align}
\label{pow-solution-f1}
   \begin{pmatrix} h_1 \\ h_3 \end{pmatrix} \Bigg|_{\text{$f_1$}}
   &=
   \tilde{c}_1
   \begin{pmatrix}
      (\mu-\nu-1)\, (q z)^{\nu+1} \, S_{\mu-1, \nu+1}(q z)
      \\
      q \, (q z)^{\nu} \, S_{\mu, \nu}(q z)
   \end{pmatrix}
\end{align}
with $\tilde{c}_1 = c_1 \ms q^{-\mu-\nu-1}$.

We now turn to the case where $f_1 = 0$ and
\begin{align}
   \label{pow-fct-2}
   f_2(z) &= c_2 \ms z^{\mu + \nu}
   \,,
\end{align}
which corresponds to the integral of $c_2 \ms z^{\mu} J_{\nu + 1}(q z)$ in
\eqref{Levin-int-resc}.
The r.h.s.\ of \eqref{ODE-h3-only} then equals $(\mu+\nu) \, c_2 \ms z^{\mu}$,
and the equation is again solved by Lommel functions, but with the first index
decreased by $1$ and with a different overall factor:
\begin{align}
\label{pow-solution-f2}
   \begin{pmatrix} h_1 \\ h_3 \end{pmatrix} \Bigg|_{\text{$f_2$}}
   &=
   (\mu+\nu) \, \tilde{c}_2
   \begin{pmatrix}
      (\mu-\nu-2)\, (q z)^{\nu+1} \, S_{\mu-2, \nu+1}(q z) \\
      q \, (q z)^{\nu} \, S_{\mu-1, \nu}(q z)
   \end{pmatrix}
   - \tilde{c}_2
   \begin{pmatrix}
      (q z)^{\mu+\nu} \\
      0
   \end{pmatrix}
\end{align}
where $\tilde{c}_2 = c_2 \, q^{-\mu-\nu-1}$.  The solution for $\mu + \nu = 0$
is trivial to verify.

By virtue of \eqref{Lommel-gen-series}, the functions in
\eqref{pow-solution-f1} and \eqref{pow-solution-f2} have series expansions
\begin{align}
   \label{pow-series-f1}
   \begin{pmatrix} h_1 \\ h_3 \end{pmatrix} \Bigg|_{\text{$f_1$}}
   &=
   \begin{pmatrix}
      \ser(z^2) + z^{\mu + \nu + 1} \, \ser(z^2)
         + z^{2 \nu + 2} \, \ser(z^2)
         + z^{2 \nu + 2} \ms \ln(z) \, \ser(z^2) \\
      \ser(z^2) + z^{\mu + \nu + 1} \, \ser(z^2)
         + z^{2 \nu} \, \ser(z^2)
         + z^{2 \nu} \ms \ln(z) \, \ser(z^2)
   \end{pmatrix}
\intertext{and}
   \label{pow-series-f2}
   \begin{pmatrix} h_1 \\ h_3 \end{pmatrix} \Bigg|_{\text{$f_2$}}
   &=
   \begin{pmatrix}
      \ser(z^2) + z^{\mu + \nu} \, \ser(z^2)
         + z^{2 \nu + 2} \, \ser(z^2)
         + z^{2 \nu + 2} \ms \ln(z) \, \ser(z^2) \\
      \ser(z^2) + z^{\mu + \nu} \, \ser(z^2)
         + z^{2 \nu} \, \ser(z^2)
         + z^{2 \nu} \ms \ln(z) \, \ser(z^2)
   \end{pmatrix}
   \,,
\end{align}
respectively.  In addition to terms already present in the homogeneous solution
\eqref{hom-solution-small-z}, we thus have a further power series in each
function $h_1$ and $h_3$, with a prefactor $z^{\mu + \nu + 1}$ for $f_1$ and a
prefactor $z^{\mu + \nu}$ for $f_2$.

This implies that the power $\mu + \nu$ of $z$ in \eqref{pow-fct-1} or
\eqref{pow-fct-2} must be large enough if $h_1$ and $h_3$ are to be sufficiently
well approximated by basis functions that have a Taylor expansion around $z=0$.
In particular, a finite first derivative of $h_1$ and $h_3$ at $z=0$ requires $\mu + \nu \ge 0$ for $f_1$ and $\mu + \nu \ge 1$ for $f_2$.


\subsubsection{Behaviour of solutions at large \texorpdfstring{$z$}{z}}
\label{sec:rescale_lin_large_z}

We now turn to the behaviour of the functions $h_1$ and $h_3$ in the limit $z\to
\infty$.  If $f_1(z)$ or $f_2(z)$ is proportional to $z^{\mu + \nu}$, we can
use the the solutions \eqref{pow-solution-f1} or \eqref{pow-solution-f2} derived
in the previous subsection.  According to \eqref{Lommel-large-x}, they behave
like
\begin{align}
   \label{large-z-f1-f2-pow}
   \begin{pmatrix} h_1 \\ h_3 \end{pmatrix} \Bigg|_{\text{$f_1$}}
   &=
   \begin{pmatrix}
      z^{\mu+\nu-1} \; \ser(z^{-2})
      \\
      z^{\mu+\nu-1} \; \ser(z^{-2})
   \end{pmatrix}
   \,,
   &
   \begin{pmatrix} h_1 \\ h_3 \end{pmatrix} \Bigg|_{\text{$f_2$}}
   &=
   \begin{pmatrix}
      z^{\mu+\nu} \; \ser(z^{-2})
      \\
      z^{\mu+\nu-2} \; \ser(z^{-2})
   \end{pmatrix}
\end{align}
for $z\gg 1$.  We postpone the discussion of this result to
\sect{\ref{sec:final_scaling_z_behaviour}}.

Let us now consider functions with an exponential falloff in $z$, namely
\begin{align}
   f_1(z)
   &=
   c_1 \ms z^{\mu + \nu} \, \exp( - \kappa z )
   \,,
   &
   f_2(z)
   &=
   c_2 \ms z^{\mu + \nu} \, \exp( - \kappa z )
   \,.
\end{align}
with $\kappa > 0$.  To solve the system of ODEs in \eqref{ODE-h3-lin-f1} and
\eqref{ODE-h3-lin-f2}, we make the ansatz
\begin{align}
   h_i(z, q)
   &=
   \hat{h}_i(z, q) \, \exp( - \kappa z )
   &
   (i=1, 3).
\end{align}
This leads to the new system
\begin{align}
   \label{ODE-yukawa}
   \begin{pmatrix}
      c_1 \ms z^{\mu + \nu} \\
      c_2 \ms z^{\mu + \nu}
   \end{pmatrix}
   &=
   - \begin{pmatrix}
         \kappa & - q \\ q & \kappa
     \end{pmatrix}
   \cdot
   \begin{pmatrix}
      \hat{h}_1 \\ z \hat{h}_3
   \end{pmatrix}
   +
   \begin{pmatrix}
      d\ms \hat{h}_1 / dz \\
      z\ms d\ms \hat{h}_3 / dz - 2\nu \ms \hat{h}_3
   \end{pmatrix}
   \,,
\end{align}
which can be solved with the ansatz
\begin{align}
   \hat{h}_1(z, q)
   &=
   z^{\mu + \nu} \, \sum_{k=0}^\infty b_{1,k}(q) \ms z^{-k}
   \,,
   \notag \\
   \hat{h}_3(z, q)
   &=
   z^{\mu + \nu - 1} \, \sum_{k=0}^\infty b_{3,k}(q) \ms z^{-k}
   \,.
\end{align}
For every single $k$ in the sums, the first term on the r.h.s.\ of
\eqref{ODE-yukawa} is leading at large $z$, whereas the second term on the
r.h.s.\ is down by one power of $z$.
For the leading coefficients, we readily get
\begin{align}
   \begin{pmatrix} b_{1,0} \\ b_{3,0} \end{pmatrix}
   &=
   - \frac{1}{\kappa^2 + q^2}
   \begin{pmatrix} \kappa & q \\ -q & \kappa \end{pmatrix}
   \cdot
   \begin{pmatrix} c_{1} \\ c_{3} \end{pmatrix}
   \,,
\end{align}
whereas the subleading coefficients can be obtained recursively as
\begin{align}
   \begin{pmatrix} b_{1,k+1} \\ b_{3,k+1} \end{pmatrix}
   &=
   \frac{1}{\kappa^2 + q^2}
   \begin{pmatrix} \kappa & q \\ -q & \kappa \end{pmatrix}
   \cdot
   \begin{pmatrix}
      (\mu + \nu - k) \, b_{1,k} \\
      (\mu - \nu - 1 - k) \, b_{3, k}
   \end{pmatrix}
   \,,
\end{align}
starting with $k=0$.

The upshot of this calculation is that if the functions $f_1$ and $f_2$ are of
the form ``power times exponential'' at large $z$, then the solutions at large
$z$ have the same form, with the same damping factor $\kappa$ in the
exponential.

In \app{\ref{sec:gaussian}} we show that a similar statement holds if $f_1$
and $f_2$ have a Gaussian decrease in $z$ (possibly modified by a power and an
exponential in $z$).

\section{Final form of the method}
\label{sec:method-final}

In the previous section, we have shown that the rescaling function $r(z) = z$
leads to well behaved solutions $h_1$ and $h_3$ of the differential equations
\eqref{ODE-h3} in the limit $z=0$.  This choice is however not optimal for the
opposite limit $z \to \infty$, where the requirement that the functions $f_1$
and $f_2$ in \eqref{Levin-int-resc} must remain finite at the end points of
the integration interval would reduce the class of integrals that can be
evaluated.  Our final choice of rescaling functions is therefore
\begin{align}
   \label{final-r}
   r(z) &= \frac{z}{1 + z}
   \,,
\end{align}
which interpolates smoothly between the choice $r(z) = z$ at small $z$ and no
rescaling as $z\to \infty$.  A more general version would be $r(z) = z / (1 +
\alpha z)$ with $\alpha > 0$, but we shall not explore this option here.

We can then compute the integral
\begin{align}
   \label{int-final-def}
   I(q)
   &=
   \int_{z_a}^{z_b} dz \;
   \biggl( \frac{1+z}{z} \biggr)^{\nu} \,
   \biggl[ J_{\nu}(q z) \, f_1(z) + J_{\nu+1}(q z) \, f_2(z) \biggr]
\end{align}
as
\begin{align}
   \label{Levin-int-final}
   I(q)
   &=
   \biggl[ J_{\nu}(q z) \, \biggl( \frac{1+z}{z} \biggr)^{\nu}\, h_1(z, q)
   + J_{\nu+1}(q z) \, \biggl( \frac{1+z}{z} \biggr)^{\nu-1}\, h_3(z, q)
   \biggr]_{z_a}^{z_b}
   \;,
\end{align}
where $h_1$ and $h_3$ solve the system
\begin{align}
   \label{ODE-final}
   f_1(z)
   &=
   \frac{d}{dz} \, h_1(z, q)
   + q \, \frac{z}{1+z}\, h_3(z, q)
   + \frac{\nu}{1+z}\, h_1(z, q)
   \,,
   \notag\\
   f_2(z)
   &=
   \frac{z}{1+z}\, \frac{d}{dz} \, h_3(z, q)
   - q h_1(z, q)
   - \biggl[\ms \frac{\nu-1}{(1+z)^2} + \frac{\nu+1}{1+z} \ms\biggr] \, h_3(z, q)
\end{align}
of ODEs.  For convenience, we will call \eqref{ODE-final} the \emph{Levin
equations} in the following.

For the reasons given in \sect{\ref{sec:rescale_lin}}, the relations
\eqref{Levin-int-final} and \eqref{ODE-final} are used for $\nu \ge 1$. To
evaluate integrals with Bessel functions of order below $1$, we instead use the
relation \eqref{Bessel-diff} and subsequent integration by parts, which gives
\begin{align}
   \label{IBP}
   &
   \int_{z_a}^{z_b} dz \; J_{\nu-1}(q z) \,
   \biggl( \frac{1+z}{z} \biggr)^{\nu-1}\, f_0(z)
   \notag \\[0.3em]
   & \qquad
   =
   \frac{1}{q} \int_{z_a}^{z_b} dz \;
   \frac{d}{dz} \ms \biggl[ z^{\nu} \ms J_{\nu}(q z) \biggr] \;
   \frac{1}{z^{\nu}}\, \biggl( \frac{1+z}{z} \biggr)^{\nu-1}\, f_0(z)
   \notag \\[0.3em]
   & \qquad
   =
   \frac{1}{q} \biggl[ J_{\nu}(q z) \,
   \biggl( \frac{1+z}{z} \biggr)^{\nu-1}\, f_0(z) \biggr]_{z_a}^{z_b}
   - \frac{1}{q} \int_{z_a}^{z_b} dz \; J_{\nu}(q z) \,
     \biggl( \frac{1+z}{z} \biggr)^{\nu} \, f_1(z)
\end{align}
with
\begin{align}
   \label{IBP-f1-f0}
   f_1(z)
   &=
   \frac{z}{1+z} \ms \frac{d}{dz} \ms f_0(z)
   - \biggl[ \frac{\nu-1}{(1+z)^2} + \frac{\nu}{1+z} \biggr] \ms f_0(z)
   \,.
\end{align}
\rev{Since we require $f_0(z)$ to be finite at the integration limits, the
boundary terms on the r.h.s.\ of \eqref{IBP} vanish for $z_a = 0$ and $z_b =
\infty$.  The derivative in \eqref{IBP-f1-f0} is computed numerically in our method, as specified in \eqn{\eqref{IBP-f1-f0-discrete}}.}


\subsection{Behaviour of solutions at small and large
\texorpdfstring{$z$}{z}}
\label{sec:final_scaling_z_behaviour}

From \eqn{\eqref{Levin-int-resc}} it is clear that one obtains the same
integral with different rescaling factors by a suitable transformation of the
functions $f_{i}$ and $h_{i}$ in the system of ODEs.  For the choices made in
the present and in the previous section, we have
\begin{align}
   f_{i}(z) \ms \big|_{r = z/(1+z)}
   &=
   (1+z)^{-\nu} \; f_{i}(z) \ms \big|_{r = z}
   &&
   (i = 1, 2),
   \notag \\[0.3em]
   h_{1}(z, q) \ms \big|_{r = z/(1+z)}
   &=
   (1+z)^{-\nu} \; h_{1}(z, q) \ms \big|_{r = z}
   \,,
   \notag \\[0.3em]
   h_{3}(z, q) \ms \big|_{r = z/(1+z)}
   &=
   (1+z)^{-\nu + 1} \; h_{3}(z, q) \ms \big|_{r = z}
   \,.
\end{align}
We thus readily obtain the small and large $z$ behaviour of $h_{i}$ with our final rescaling choice from our results in \sect{\ref{sec:rescale_lin}}.

For the limit $z\to 0$, we deduce from \eqref{pow-series-f1} and
\eqref{pow-series-f2} that
\begin{align}
   \label{final-small-z-pow-f1}
   f_{1}(z)
   &
   \sim z^{\mu + \nu} \, \ser(z)
   \notag \\
   \Rightarrow \;
   h_{3}(z, q)
   &
   \sim
   \ser(z) + z^{\mu + \nu + 1} \, \ser(z) + z^{2 \nu} \, \ser(z)
           + z^{2 \nu} \ms \ln(z) \, \ser(z)
\intertext{and}
   \label{final-small-z-pow-f2}
   f_{2}(z)
   &
   \sim z^{\mu + \nu} \, \ser(z)
   \notag \\
   \Rightarrow \;
   h_{3}(z, q)
   &
   \sim
   \ser(z) + z^{\mu + \nu} \, \ser(z) +  z^{2 \nu} \, \ser(z)
           + z^{2 \nu} \ms \ln(z) \, \ser(z)
   \,,
\end{align}
where we recall that $\ser(z)$ denotes a power series in $z$ in the generic
sense explained after \eqn{\eqref{gen-series-def}}.
The behaviour of $h_{1}(z, q)$ is as the one of $h_{3}(z, q)$ in both
\eqref{final-small-z-pow-f1} and \eqref{final-small-z-pow-f2}, except that the
power $z^{2\nu}$ is replaced with $z^{2\nu + 2}$.  It is understood that $\mu +
\nu \ge 0$, which ensures that $f_1$ and $f_2$ remain finite at $z=0$ and that
the integral \eqref{int-final-def} with $z_a = 0$ actually exists.

We recall from our discussion in \sect{\ref{sec:f1-or-f2-power}} that small
non-integer values of $\mu + \nu$ lead to a non-analytic behaviour of $h_1$ and
$h_3$ at $z=0$ that is difficult to approximate by functions having a Taylor
expansion around that point.  We therefore expect our method to become less
accurate in this case.  The comparison between \eqref{final-small-z-pow-f1}
and \eqref{final-small-z-pow-f2} reveals that for an integrand behaving like
$z^{\mu} J_\rho(q z)$ around $z=0$, one should use \eqref{int-final-def} with
$\nu = \rho$ and $f_2=0$ if $1 < \mu + \rho < 2$, rather than with $\nu = \rho -
1$ and $f_1=0$.  The same holds if $0 \le \mu + \rho < 1$.  For $\mu + \rho = 1$
or $\mu + \rho \ge 2$, both options are possible.

For the large-$z$ limit, we can discuss several cases.  From
\eqref{large-z-f1-f2-pow} we deduce that for a power-law behaviour of $f_{i}$ we
have
\begin{align}
   \label{final-large-z-pow}
   f_{1}(z)
   &
   \sim z^{\mu} \, \ser(z^{-1})
   \quad
   \Rightarrow \;
   \begin{pmatrix} h_{1}(z, q) \\ h_{3}(z, q) \end{pmatrix}
   \sim
   \begin{pmatrix}
      z^{\mu - 1} \ms \ser(z^{-1}) \\
      z^{\mu} \ms \ser(z^{-1})
   \end{pmatrix}
   \,,
   \notag \\[0.4em]
   f_{2}(z)
   &
   \sim z^{\mu} \, \ser(z^{-1})
   \quad
   \Rightarrow \;
   \begin{pmatrix} h_{1}(z, q) \\ h_{3}(z, q) \end{pmatrix}
   \sim
   \begin{pmatrix}
      z^{\mu} \ms \ser(z^{-1}) \\
      z^{\mu - 1} \ms \ser(z^{-1})
   \end{pmatrix}
   \,.
\end{align}
Both the functions $f_i$ and the solutions $h_i$ remain finite for $z\to
\infty$ provided that $\mu \le 0$.  Our method thus allows the computation of
Fourier-Bessel integrals not only for functions $f_{i}$ with a power-law
decrease but also for functions that tend to a finite value at $z\to \infty$.

For functions with an exponential falloff at large $z$, we get
\begin{align}
   \label{final-large-z-exp}
   f_{1}(z) \text{ or } f_{2}(z)
   &
   \sim
   z^{\mu} \, \ser(z^{-1}) \ms \exp(- \kappa z)
   \quad
   \Rightarrow \;
   \begin{pmatrix} h_{1}(z, q) \\ h_{3}(z, q) \end{pmatrix}
   \sim
   \begin{pmatrix}
      z^{\mu} \ms \ser(z^{-1}) \\
      z^{\mu} \ms \ser(z^{-1})
   \end{pmatrix}
   \exp(- \kappa z)
\end{align}
for $\kappa > 0$ according to \sect{\ref{sec:rescale_lin_large_z}}, whereas for
functions with a Gaussian falloff, the results of \app{\ref{sec:gaussian}} imply
that
\begin{align}
   \label{final-large-z-gauss}
   f_{1}(z)
   &
   \sim
   z^{\mu} \, \ser(z^{-1}) \ms
   \exp( - \lambda^2 z^2 - \kappa z \ms\bigr) \;
   \quad
   \Rightarrow \;
   \begin{pmatrix} h_{1}(z, q) \\ h_{3}(z, q) \end{pmatrix}
   \sim
   \begin{pmatrix}
      z^{\mu - 1} \ms \ser(z^{-1}) \\
      z^{\mu - 2} \ms \ser(z^{-1})
   \end{pmatrix}
   \exp( - \lambda^2 z^2 - \kappa z \ms\bigr) \;
   \,,
   \notag \\[0.4em]
   f_{2}(z)
   &
   \sim
   z^{\mu} \, \ser(z^{-1}) \ms
   \exp( - \lambda^2 z^2 - \kappa z \ms\bigr) \;
   \quad
   \Rightarrow \;
   \begin{pmatrix} h_{1}(z, q) \\ h_{3}(z, q) \end{pmatrix}
   \sim
   \begin{pmatrix}
      z^{\mu - 2} \ms \ser(z^{-1}) \\
      z^{\mu - 1} \ms \ser(z^{-1})
   \end{pmatrix}
   \exp( - \lambda^2 z^2 - \kappa z \ms\bigr) \;
\end{align}
for $\lambda > 0$.
Both in the exponential and in the Gaussian case, the solutions $h_1$ and $h_3$
thus have the same leading behaviour for $z\to 0$ as the functions $f_1$ or
$f_2$, except for a possible downward shift of the power $z^{\mu}$ by one or
two units.


\paragraph{Using integration by parts.}

It remains to discuss the case where one uses integration by parts \eqref{IBP}
and then evaluates the integral on its r.h.s.\ by Levin's method.

With \eqref{IBP-f1-f0} we find
\begin{align}
   f_0(z)
   &
   \sim
   z^{\mu + \nu - 1} \, \ser(z)
   \quad
   \Rightarrow \;
   f_1(z)
   \sim
   z^{\mu + \nu - 1} \, \ser(z)
\end{align}
for $z\to 0$, so that our method can be applied to integrands behaving like
$z^{\mu} \ms J_\rho(q z)$ with $\mu + \rho \ge 0$ for all $\rho \ge 0$.

At large $z$ we have
\begin{align}
   f_0(z)
   &
   \sim z^{\mu} \, \ser(z^{-1}) \ms
      \exp( - \lambda^2 z^2 - \kappa z \ms\bigr) \;
   \quad
   \Rightarrow \;
   f_1(z)
   \sim
   z^{\rho} \, \ser(z^{-1}) \,
   \exp( - \lambda^2 z^2 - \kappa z \ms\bigr) \;
\end{align}
with
\begin{align}
   \rho
   &=
   \begin{cases}
      \mu - 1  & \text{ if } \lambda = \kappa = 0 \,,
      \\
      \mu      & \text{ if } \lambda = 0 \text{ and } \kappa > 0 \,,
      \\
      \mu + 1  & \text{ if } \lambda > 0 \,.
   \end{cases}
\end{align}
One thus gets the same leading exponential or Gaussian behaviour for $f_0$ and
$f_1$.  A pure power law $f_0 \sim z^{\mu}$ with $\mu \le 0$ implies a power-law
decrease $f_1 \sim z^{\mu-1}$.


\paragraph{Grand summary.}  As an upshot of our detailed discussion, we retain
that if a set of basis functions is good for approximating the functions $f_1$
or $f_2$ in \eqref{int-final-def}, or $f_0$ in \eqref{IBP}, it can be expected
to be good for approximating the solutions $h_1$ and $h_3$ of the Levin
equations \eqref{ODE-final} and hence allow for an accurate computation of the
integrals.  A small-$z$ behaviour like $z^{1 - \delta}$ with $\delta < 1$ is tolerable for $f_0$ or $f_1$, because the corresponding power of $z$
for $h_1$ and $h_3$ is larger by one unit.


\subsection{Discretisation on Chebyshev grids}
\label{sec:cheb}

We now discuss in detail how we implement the method of collocation for solving
the Levin equations \eqref{ODE-final}.

In most examples of Levin's original work \cite{Levin:1982, Levin:1996}, the
chosen basis functions are polynomials $b_i(z) = (z - z_0)^{i}$, and the
system of ODEs is evaluated at \emph{equispaced} points between $z_a$ and $z_b$.
 It was found that the accuracy of the method could be improved by dividing the
integration domain into several subintervals.  Other options for the basis
functions are mentioned in \cite{Levin:1982}.

It is well known that polynomial approximation on equispaced grids suffers from
Runge's phenomenon: the approximation becomes poor in the vicinity of the
interval end points when one increases the number of grid points (i.e.\ the
polynomial order).  This problem is avoided by using suitable grids that are not
equispaced.  A prominent example is \emph{Chebyshev interpolation}, which is
described in detail in \cite{Trefethen:2012}.
%
This was indeed used in the context of Levin's method in \cite{Li:2010,
Li:2011}.

Motivated by our experience with interpolating (single and double) parton
distributions on Chebyshev grids \cite{Diehl:2021gvs, Diehl:2023cth}, we adopt
Chebyshev interpolation to our present case.  Our method involves three steps.


\paragraph{Variable transform.}
We first define a transformation from $z$ to the variable $u$ that will be used
for interpolation with Chebyshev polynomials.  The values $u_a = u(z_a)$ and
$u_b = u(z_b)$ must both be finite.  We require $u(z)$ to be an analytic
function with $du / dz > 0$ for all $z \in [z_a, z_b]$, except for the point
$z_b = \infty$, where necessarily $du / dz = 0$.

A primary purpose of this transformation is to map an infinite integration
interval $[z_a, \infty]$ onto a finite interval in $u$.  On a finite interval in
$z$, one may take the trivial transformation $u = z$.  However, even in this
case a non-trivial variable transformation can be beneficial, for instance when
$f_i(z)$ rises or decreases very steeply.

Depending on the behaviour of $f_i(z)$, different variable transformations
$u(z)$ lead to good results of our method.  A selection of such transformations
is presented in \sect{\ref{sec:var-transforms}}.


\paragraph{Subintervals.}
We find that it is often useful to split $[z_a, z_b]$ into $k$ subintervals
$[z_0, z_1]$, $[z_1, z_2]$, \ldots, $[z_{k-1}, z_{k}]$ with $z_0 = z_a$ and
$z_{k} = z_b$ and to evaluate the integral on each of them separately.  We
will see that this benefits both the integration accuracy and the computation
time.  Following the notation introduced in \cite{Diehl:2021gvs, Diehl:2023cth},
we write
\begin{align}
   \label{grid-notation}
   [z_0, z_1, \ldots, z_{k}]_{(n_1, \ldots, n_k)}
\end{align}
to characterise a grid and its subgrids, where $z_j$ are the interval
boundaries, and $n_j$ is the number of points on each subgrid.  The total number
of grid points is then $\sum_{j=1}^{k} n_j - (k-1)$, because adjacent subgrids
share their end points.  In the notation \eqref{grid-notation} it is understood
that one and the same variable transform is used on the full interval from $z_0$
to $z_k$.

For simplicity, we will continue to present our method for a single integration
interval $[z_a, z_b]$, bearing in mind that this may be a subinterval in the
sense just described.


\paragraph{Chebyshev interpolation.}
On each subinterval, we approximate the functions $h_1(z, q)$ and $h_3(z, q)$
using Chebyshev interpolation.  A compendium of this method with the formulae
needed for our method is given in \app{\ref{sec:cheb-details}}.  In brief, we
approximate
\begin{align}
   h_i(z, q)
   & \approx
   p_i\bigl( u(z), q \bigr)
   \,,
\end{align}
where $p_i(u, q)$ is the unique polynomial of order $N$ that reproduces the
function
\begin{align}
   h_i(z_j, q)
   &=
   p_i\bigl( u_j, q \bigr)
   \,,
   &
   u_j &= u(z_j)
\end{align}
at the $n = N+1$ points $z_j$ that correspond to the Chebyshev points $u_j$ on
the interval $[u_a, u_b]$, with $u_j$ given in \eqref{u-Cheb-points}.  Note that
these points include the interval boundaries, $z_a = z_0$ and $z_b = z_N$,
where the functions $h_i$ are required for evaluating the Fourier-Bessel
integral with the master formula \eqref{Levin-int-final} of our method.

The derivatives of $d h_i /dz$ appearing in the Levin equations
\eqref{ODE-final} can be approximated by the derivatives of the interpolants $d
p_i / dz$.  When evaluating the equations at the points $z_j$, these derivatives
can be evaluated via a matrix multiplication.  In summary, we discretise
\eqref{ODE-final} by making the following replacements:
\begin{align}
   f_i(z)
   &\to f_i(z_j)
   && (i = 1, 2)
   \,,
   \notag \\[0.5em]
   h_i(z, q)
   &\to p_i(u_j, q)
   && (i = 1, 3)
   \,,
   \notag \\
   \frac{d}{d z} \ms h_i(z, q)
   &\to
   \frac{d u}{d z} (z_j) \sum_{k} D^u_{j k} \, p_i(u_k, q)
   && (i = 1, 3)
   \,,
\end{align}
where $D^{u}_{j k}$ is given in \eqref{diff-mat-u} and $du /dz$ is the Jacobian of the variable transform.
The Levin equations are thus replaced by a system of linear equations, which can
be written as
\begin{align}
   \label{lin-system}
   B(q) \cdot \vect{P}(q)
   &=
   \vect{F}
\end{align}
with the $2 n$-dimensional vectors
\begin{align}
   \vect{F}
   &=
   \bigl( f_1(z_0), f_1(z_1), \ldots, f_1(z_N),\;
          f_2(z_0), f_2(z_1), \ldots, f_2(z_N) \bigr)^T
   \,,
   \notag \\
   \vect{P}(q)
   &=
   \bigl( p_1(z_0, q), p_1(z_1, q), \ldots, p_1(z_N, q),\;
          p_3(z_0, q), p_3(z_1, q), \ldots, p_3(z_N, q) \bigr)^T
\end{align}
and the $2 n$-dimensional square matrix
\begin{align}
   B(q)
   &=
   \begin{pmatrix}
   B^{a a} & B^{a b}(q) \\ B^{b a}(q) & B^{b b}
   \end{pmatrix}
\end{align}
with diagonal blocks
\begin{align}
   \label{B-diag-blocks}
   B^{a a}_{j k}
   &=
   \frac{du}{dz} (z_j) \, D^u_{j k} + \frac{\nu}{1 + z_j} \, \delta_{j k}
   \,,
   \notag \\
   B^{b b}_{j k}
   &=
   \frac{z_j}{1 + z_j} \, \frac{du}{dz} (z_j) \, D^u_{j k}
   - \biggl[\ms \frac{\nu-1}{(1 + z_j)^2} + \frac{\nu+1}{1 + z_j} \ms\biggr] \ms
     \delta_{j k}
\end{align}
and off-diagonal blocks
\begin{align}
   \label{B-off-diag-blocks}
   B^{a b}_{j k}(q)
   &=
   q \, \frac{z_j}{1 + z_j} \, \delta_{j k}
   \,,
   &
   B^{b a}_{j k}(q)
   &=
   - q \ms \delta_{j k}
   \,,
\end{align}
where $j$ and $k$ run from $0$ to $N$ in all cases.

Note that up to this point, the only approximation we have made was to replace
the derivatives of $h_i$ at the points $z_j$ by the derivatives of their
approximants $p_i$.
If we use the integration-by-parts identity \eqref{IBP}, then we also
\rev{invoke Chebyshev interpolation to approximately compute the derivative
$d f_0 / d z$ in the definition \eqref{IBP-f1-f0} of $f_1$.  The values of $f_1$
at the grid points are thus obtained from the corresponding values of $f_0$ as}
\begin{align}
   \label{IBP-f1-f0-discrete}
   f_1(z_j)
   & \approx
   \sum_{k} C_{j k} \, f_0(z_k)
   \,,
   \notag \\
   C_{j k}
   &=
   \frac{z_j}{1 + z_j} \ms \frac{du}{dz} (z_j) D^u_{j k}
   - \biggl[ \frac{\nu-1}{(1 + z_i)^2} + \frac{\nu}{1 + z_i} \biggr] \ms
     \delta_{j k}
   \,.
\end{align}

In \sect{\ref{sec:final_scaling_z_behaviour}} we investigated in detail how the
small-$z$ behaviour of $h_1$ and $h_3$ is related to the small-$z$ behaviour of
$f_1$ or $f_2$.  Since we require the variable transformation between $u$ and
$z$ to be analytic and to have a finite derivative at $z=0$, the interpolation
polynomials $p_i\bigl( u(z), q \bigr)$ in $u$ have a Taylor expansion around
$z=0$.  To obtain a good approximation when discretising the Levin equations, we
hence need that $h_1$ and $h_3$ can be sufficiently well approximated by a
Taylor series around $z=0$, with the first $n$ Taylor coefficients being free
and the remaining ones fixed by the variable transform.

A corresponding statement holds for the small-$z$ behaviour of $f_0(z)$ when
using integration by parts.  Notice that if one has $f_0 \sim z^{1-\delta}$
with $0 < \delta < 1$, one gets a divergent first derivative of $f_0$ at $z =
0$, which is however multiplied with $z$ in \eqref{IBP-f1-f0} and thus can be
accommodated within our setup.


\paragraph{Using quadrature.}
We will see that the matrix $B$ in \eqref{lin-system} becomes ill-conditioned
when $q$ becomes small.  As described in the next subsection, this can in part
be dealt with by linear algebra methods.  On the other hand, there is no
compelling need to use the Levin method when the integrand in
\eqref{int-final-def} is not rapidly oscillating in the integration interval
$[z_a, z_b]$.  In our algorithm, we therefore replace the Levin method with
Clenshaw-Curtis quadrature if for a given $q$ one has
\begin{align}
   \label{CC-use-criterion}
   q z_{b} & \le j_\nu
   \,,
\end{align}
where $j_\nu$ is the first zero of the Bessel function $J_\nu(x)$.  We find that
this yields a high integration accuracy and also saves computing time.  The
quadrature result is obtained as the sum
\begin{align}
   \label{CC-final}
   I(q)
   & \approx
   \sum_j w^z_j \;
   \biggl( \frac{1 + z_j}{z_j} \biggr)^{\nu} \,
   \biggl[
      J_{\nu}(q z_j) \, f_1(z_j) + J_{\nu+1}(q z_j) \, f_2(z_j)
   \biggr]
   \,,
\end{align}
where $I(q)$ is defined in \eqref{int-final-def} and the weights are given by
\begin{align}
   \label{CC-weights-final}
   w^z_{j}
   &=
   \biggl( \frac{d u}{d z} (z_j) \biggr)^{-1} \; w^u_j
\end{align}
with $w^u_j$ from \eqref{CC-quadrature-u}.


\subsubsection{Solving the linear equation system}
\label{sec:linalg}

The system \eqref{lin-system} has a unique solution if the matrix $B$ is
nonsingular.  By taking increasingly fine interpolation grids, i.e.\
increasingly high order of the interpolation polynomials $p_i(u, q)$, one should
eventually approach the limit of the original Levin equations, which has an
infinity of solutions.  One can hence expect that for sufficiently fine
interpolation grids, the matrix $B$ is close to singular.  This poses a
challenge to the numerical solution of \eqref{lin-system}.

The work in \cite{Li:2010, Li:2011} contains a detailed analysis of the linear
equation system for a different set of integrals, along with a strategy for its
solution. We follow a rather similar strategy in the present work, selecting one
out of two algorithms to solve \eqref{lin-system}.

Our default algorithm uses the LU decomposition (see e.g.\ \cite[Chapter
2.3]{NumericalRecipes}), which represents $B$ in the form
\begin{align}
   \label{LU-decomp}
   B &= \Pi \cdot L \cdot U
   \,,
\end{align}
where $\Pi$ is a permutation matrix, $L$ is a lower diagonal matrix with all
diagonal elements equal to $1$, and $U$ an upper diagonal matrix.  The product
of the diagonal elements $U_{i i}$ is equal to $\det (B)$, so that at least one
of these elements is zero when $B$ is singular.  If this is not the case, the
solution of $B \cdot \vect{P} = \vect{F}$ is obtained by successively solving
\begin{align}
   L \cdot \vect{G} &= \Pi^{-1} \cdot \vect{F}
   \,,
   &
   U \cdot \vect{P} &= \tvec{G}
   \,,
\end{align}
which can be done by forward and backward substitution, respectively.  The
second step involves the inverse diagonal elements $1/U_{i i}$ of $U$ and
becomes numerically unstable when $B$ is close to singular.  As criterion for
this case, we use
\begin{align}
   \label{SV-use-crit}
   \min_i \bigl( |U_{i i}| \bigr)
   & \;\le\;
   r_{LU}^{\text{max}} \, \max_i \bigl( |U_{i i}| \bigr)
   \,,
\end{align}
where $r_{LU}^{\text{max}}$ can be chosen by the user.

If \eqref{SV-use-crit} is satisfied for a given matrix $B$, we use the
singular value (SV) decomposition.  In this case, $B$ is represented as
\begin{align}
   \label{SV-decomp}
   B &= U \cdot S \cdot V^T
   \,,
\end{align}
where $U$ and $V$ are orthogonal\footnote{%
   Note that $U$ denotes different matrices in \eqref{SV-decomp} and
   \eqref{LU-decomp}.}
and $S = \operatorname{diag}(S_{i i})$ is diagonal and positive semidefinite.

For nonsingular $B$, the solution of $B \cdot \vect{P} = \vect{F}$ is then
readily given by
\begin{align}
   \label{SV-solution}
   \vect{P}
   &= V \cdot \operatorname{diag}( 1/S_{i i} ) \cdot U^T \cdot \vect{F}
   \,.
\end{align}
If $B$ is singular, then some of the $S_{i i}$ vanish, and one replaces the
corresponding entries $1/S_{i i}$ in \eqref{SV-solution} with zero.  As shown
e.g.\ in \cite[Chapter 2.6]{NumericalRecipes}, this yields an approximate
solution of \eqref{lin-system} in the sense that $\vect{P}$ obtained in this way
minimises $| B \cdot \vect{P} - \vect{F}|$, i.e.\ the distance between the
vectors on the two sides of the equation.  The numerical stability of this
procedure is greatly enhanced by also replacing $1/S_{i i}$ with zero when $S_{i
i}$ is very small; this is called a \emph{truncated} SV decomposition.  We
choose to replace an entry $1/S_{i i}$ with zero if it satisfies
\begin{align}
   \label{SV-trunc-crit}
   S_{i i}
   & \; < \;
   r_{SV}^{\text{max}} \, \max_j(S_{j j})
   \,,
\end{align}
with $r_{SV}^{\text{max}}$ to be chosen by the user.

A numerical computation is typically much faster for the LU decomposition
\eqref{LU-decomp} than for the SV decomposition \eqref{SV-decomp}.  One should
hence take $r_{LU}^{\text{max}}$ as small as can be done without compromising
the accuracy of the result.    Our choice of the control parameters
$r_{LU}^{\text{max}}$ and $r_{SV}^{\text{max}}$ is discussed in
\sect{\ref{sec:condition-numbers}}.

The computation time for both the LU and the SV decomposition grows very quickly
with the dimension of the matrix $B$.  From this point of view, it is beneficial
to split the overall integration region $[z_a, z_b]$ into subintervals, since
this allows one to work with smaller matrices $B_k$ on each subinterval $k$.


\subsubsection{Examples of variable transforms}
\label{sec:var-transforms}

A selection of variable transformations that are suitable for integration over
$z$ from $0$ to $\infty$ is presented in \tab{\ref{tab:var-trf}}. The table
has been adapted from \tab{1} in \cite{Diehl:2023cth}, where the same
transformations were used to describe the dependence of double parton
distributions on the transverse distance between the two partons.

The following brief characterisation of the different transforms is the result
of detailed numerical investigations, part of which will be presented in the
next sections.
\begin{itemize}
\item The inverse power law transformation in the first row of the table is
   suitable for integrals with a large-$z$ behaviour like $f(z) \sim z^{\mu}$
   with $\mu \le 0$.
\item \rev{The transformation in the second row behaves like
   $u \approx - \ln^{\alpha} (z_{\text{hi}} / z)$ for
   $z_{\text{lo}} \ll z \ll z_{\text{hi}}$ and like an inverse power law
   $u \approx {}- \mathit{const} \; z^{-\alpha}$ for $z \gg z_{\text{hi}}$.
   This is useful for functions that involve high powers of $\ln z$, as we will
   see in \sect{\ref{sec:special-int}}.}
\item The exponential transform in the \rev{third} row works well for functions that
   have a Gaussian behaviour $f(z) \sim \exp(- \lambda^2 z^2)$ at large $z$,
   possibly modified by an exponential $\exp(- \kappa z)$ times a power of $z$.
\item The exponential transform with a square root in the \rev{fourth} row works well
   for $f(z) \sim \exp(- \kappa z)$ at large $z$, possibly modified by a power of
   $z$.
\end{itemize}
In our study \cite{Diehl:2023cth} of double parton distributions, we have shown
that the Gaussian transformation in the last row yields good
\emph{interpolation} accuracy for functions with a Gaussian behaviour at large
$z$.  Likewise, we found that the exponential transform in the \rev{third} row allows
for an accurate interpolation of functions with an exponential falloff. As we
will see in \sect{\ref{sec:tmd-numerics}}, the same combinations of variable
transforms and integrands work with Levin's method but are less powerful than
the combinations specified above.

\begin{table}
\centering
\begin{tabular}{c c c c}
\toprule
   shorthand &
      ${}-u(z)$ &
      $z(u)$ &
      $du / dz$
\\ \midrule
   \texttt{inv pow} &
      $(z + z_0)^{-\alpha}$ &
      $|u|^{- 1 / \alpha} - z_0$ &
      $\alpha \, (z + z_0)^{-1-\alpha}$
\\[0.35em]
   \texttt{log pow} &
   $\Bigl[ \ln \frac{z + z_{\text{hi}}}{z + z_{\text{lo}}} \Bigr]^\alpha$ &
   $\frac{z_{\text{hi}} - z_{\text{lo}} \exp(|u|^{1/\alpha})}{
      \exp(|u|^{1/\alpha}) - 1}$ &
   $\frac{\alpha \, (z_{\text{hi}} - z_{\text{lo}})}{(z + z_{\text{hi}}) \,
      (z + z_{\text{lo}})} \, |u|^{(\alpha - 1)/\alpha}$
\\[0.35em]
   \texttt{exp} &
      $\exp\Bigl[-\frac{1}{4} \ms m z \Bigr]$ &
      $\frac{4}{m} \ms L(u)$ &
      $\frac{1}{4} \ms m |u|$
\\[0.35em]
   \texttt{exp sqrt} &
      $\exp\Bigl[\ms 1 - \sqrt{1 + m z / 2} \,\Bigr]$ &
      $\frac{2}{m} \ms \Bigl[ L^2(u) + 2 L(u) \ms\Bigr]$ &
      $\frac{1}{4} \ms m |u| \, \Bigl[ L(u) + 1 \ms\Bigr]^{-1}$
\\[0.35em]
   \texttt{Gauss} &
   $\exp\Bigl[- \frac{1}{4} \ms (m^2 z^2 + m z) \Bigr]$ &
   $\frac{1}{2 m} \, \bigl[ \sqrt{16 L(u) + 1} - 1 \ms\bigr]$ &
   $\frac{1}{4} \ms m  |u| \, \sqrt{16 L(u) + 1}$
\\ \bottomrule
\end{tabular}
\caption{\label{tab:var-trf}Variable transformations suitable for
integration over $z$ from $0$ to $\infty$.  It is understood that $\alpha > 0$,
$z_0 > 0$, \rev{$z_{\text{hi}} > z_{\text{lo}} > 0$,} and $m > 0$.  All
transformations satisfy $u \le 0$ and $du / dz > 0$
for $u < 0$. At $u=0$ one has $z = \infty$ and $du / dz = 0$.  We abbreviate
$L(u) = \ln\bigl( 1 / |u| \bigr)$.}
\end{table}


\subsection{Summary of the algorithm}
\label{sec:algorithm}

Our complete algorithm naturally proceeds in several steps.
\begin{enumerate}
\item The user must specify an interpolation grid, i.e.\ a variable transform
   $u(z)$, a subinterval division in $z$, and the number of grid points per
   subinterval.  Then
   \begin{enumerate}
   \item \label{step:grid-points} compute the grid points $z_j$,
   \item \label{step:discr-functions} evaluate the values at the grid points
      for the function(s) to be integrated over, i.e.\ for $f_1(z), f_2(z)$ or
      $f_0(z)$ in \eqref{int-final-def} or \eqref{IBP},
   \item perform steps \ref{step:fixed-nu} to \ref{step:integrate} for each
      subinterval and add up the integrals from these subintervals.
   \end{enumerate}
\item \label{step:fixed-nu} For a given value of $\nu$
   \begin{enumerate}
   \item compute the values of $\bigl[ z_j / (1 + z_j) \bigr]^{\rho}$ with
      $\rho = 1$, $\nu$,  and $\nu-1$ for each grid point,
   \item compute the diagonal blocks \eqref{B-diag-blocks} of the matrix $B$, as
      well as the matrix $C$  in \eqref{IBP-f1-f0-discrete},
   \item compute the Clenshaw-Curtis weights $w^z_j$ in
      \eqref{CC-weights-final}.
   \end{enumerate}
   We recall that for a given $\nu$, the method allows the computation of
   integrals with $J_{\nu-1}(q z)$, $J_{\nu}(q z)$, and $J_{\nu+1}(q z)$.
\item \label{step:fixed-q} For a given value of $q$
   \begin{enumerate}
   \item determine from \eqref{CC-use-criterion} whether to use quadrature or
      the Levin method,
   \item \label{step:comp-Js} compute the values of
      $\bigl[ (1+z)/z \ms\bigr]^{\nu} \, J_{\nu}(q z)$ and
      $\bigl[ (1+z)/z \ms\bigr]^{\nu} \, J_{\nu + 1}(q z)$, for all grid
      points~$z_j$ if quadrature is used, and for the interval boundaries otherwise,
   \item if the Levin method is used: compute the off-diagonal blocks
      \eqref{B-off-diag-blocks} of the matrix~$B$,
   \item \label{step:setup-lin-solv} if the Levin method is used: set up the
      linear equation solver by computing the LU and --- if needed --- the SV
      decomposition of $B$.
   \end{enumerate}
\item \label{step:integrate} For given values $f_i(z_j)$ on the interpolation grid ($i=0,1$, or $2$)
   \begin{enumerate}
   \item for integrals with $J_{\nu-1}(q z) \ms$:
      evaluate the boundary terms at $z_a$ and $z_b$ in \eqref{IBP} and compute
      $f_1(z_j)$ from $f_0(z_j)$ using \eqref{IBP-f1-f0-discrete},
   \item if the Levin method is used: solve the matrix equation
      \eqref{lin-system},
   \item evaluate the integral from the master formula \eqref{Levin-int-final}
      or \eqref{IBP}.
   \end{enumerate}
\end{enumerate}

The time required for step \ref{step:discr-functions} crucially depends on the
functions to be integrated; an important feature of our method is that the
functions need not be evaluated separately for different values of $q$.  The
evaluation of Bessel functions is limited to step \ref{step:comp-Js} and
required only once for a given~$q$.

The operations in step \ref{step:fixed-nu} need to be done only once for a given
choice of $\nu$ and the interpolation grid -- they should be irrelevant for the
overall computing time in a typical workflow.  We find step
\ref{step:setup-lin-solv} to be by far the most time consuming one.  By
comparison, all operations in step \ref{step:integrate} are fast.

\rev{Recalling the use cases mentioned in the introduction, we see that our
method (as well as the Levin method in its original form) is particularly
efficient for a workflow in which one needs the Fourier-Bessel transform of
\emph{many} similar functions at a given value of $q$.  With functions of
sufficiently similar shape, one can take a common grid for discretisation.  The
computationally expensive step \ref{step:setup-lin-solv} has to be carried out
once for a given $q$, after which the much faster steps in \ref{step:integrate}
can be repeated for each discretised function in turn.}


\subsection{Implementation in BestLime}
\label{sec:code}

We have implemented this algorithm in the C++ library \bestlime, which is
available on \cite{BestLime:release}.  It \rev{offers} a number of variable
transforms $u(z)$, including those in \tab{\ref{tab:var-trf}}, as well as
the possibility to work with user-defined transformations.  \rev{Routines are
provided to compute the integral in \eqref{int-final-def} for given $f_1$ or
$f_2$, or the integral in \eqref{IBP} for given $f_0$, taking as arguments the
value of $q$ and the values of the relevant function $f_i$ on the points of the
interpolation grid.  There are also calls for computing
$\int dz \, J_{\rho}(q z) \, \ftil(z)$} with $\rho = \nu$, $\nu+1$, or
$\nu-1$. These calls simply multiply $\ftil(z)$ with $[ z / (1+z) ]^{\nu}$ or $[
z / (1+z) ]^{\nu - 1}$ and then \rev{evaluate the integral in}
\eqref{int-final-def} or \eqref{IBP}.

\rev{Step \ref{step:grid-points} and the steps in \ref{step:fixed-nu} are
carried out once when initialising an integration setup, which is characterised
by a discretisation grid and a value $\nu \ge 1$.  An integration call invokes
the steps in \ref{step:fixed-q} if and only if the requested $q$ differs from
its value in the previous call.  The results of these steps are stored and
reused in subsequent calls at the same $q$.  This enables the workflow described
at the end of the previous subsection.}

For linear algebra operations, \bestlime\ uses the Eigen library
\cite{Eigen:home}, specifically the part for dense matrices in Eigen version
3.4.0.  In particular, the linear equations are solved using the methods
\texttt{PartialPivLU} (LU decomposition with partial pivoting) or
\texttt{BDCSVD} (Bidiagonal Divide and Conquer SVD).

For the computation of Bessel functions (and of the hypergeometric function
${}_{2} F_{1}$ in the benchmark comparison of \sect{\ref{sec:benchmarks}}), the
GNU Scientific Library (GSL) \cite{GSL:doc} is used.

The methods of \bestlime\ that deal with Chebyshev interpolation have been taken
form the \chilipdf\ library \cite{Diehl:2021gvs, Diehl:2023cth}, which is under
development.

\section{Numerical studies}
\label{sec:numerics}

In this section, we investigate several numerical and computational aspects of
our integration algorithm.  To characterise the interpolation grids, we use the
shorthand names for variable transforms in the first column of
\tab{\ref{tab:var-trf}}, along with the notation for grid boundaries and grid
points in \eqn{\eqref{grid-notation}}.

All results in this and the next sections have been obtained with the library
\bestlime.  We begin with a brief discussion of the parameters that control the
algorithm used for solving the discretised Levin equations \eqref{lin-system}.
After this, we present benchmarks for the integration accuracy.  Finally, we
give an indication of the computation time required by \bestlime\ for
representative integration setups.


\subsection{Control parameters for solving the linear equation system}
\label{sec:condition-numbers}

As explained in \sect{\ref{sec:linalg}}, we use the LU decomposition
\eqref{LU-decomp} to solve the discretised Levin equations, provided that the
matrix $B$ is sufficiently far from being singular.  A measure for this is the
ratio
\begin{align}
   \label{r-LU-def}
   r_{LU}
   &=
   \frac{\min_i \bigl( |U_{i i}| \bigr)}{\max_i \bigl( |U_{i i}| \bigr)}
   \,,
\end{align}
where $U_{i i}$ denotes the diagonal elements of the matrix $U$ in
\eqref{LU-decomp}.  Let us see when this ratio becomes very small.

\begin{figure}[t]
\centering
\subfloat[\mbox{$[0, z_1]$} with $n_1 = 16$]{
   \includegraphics[width=0.47\textwidth]{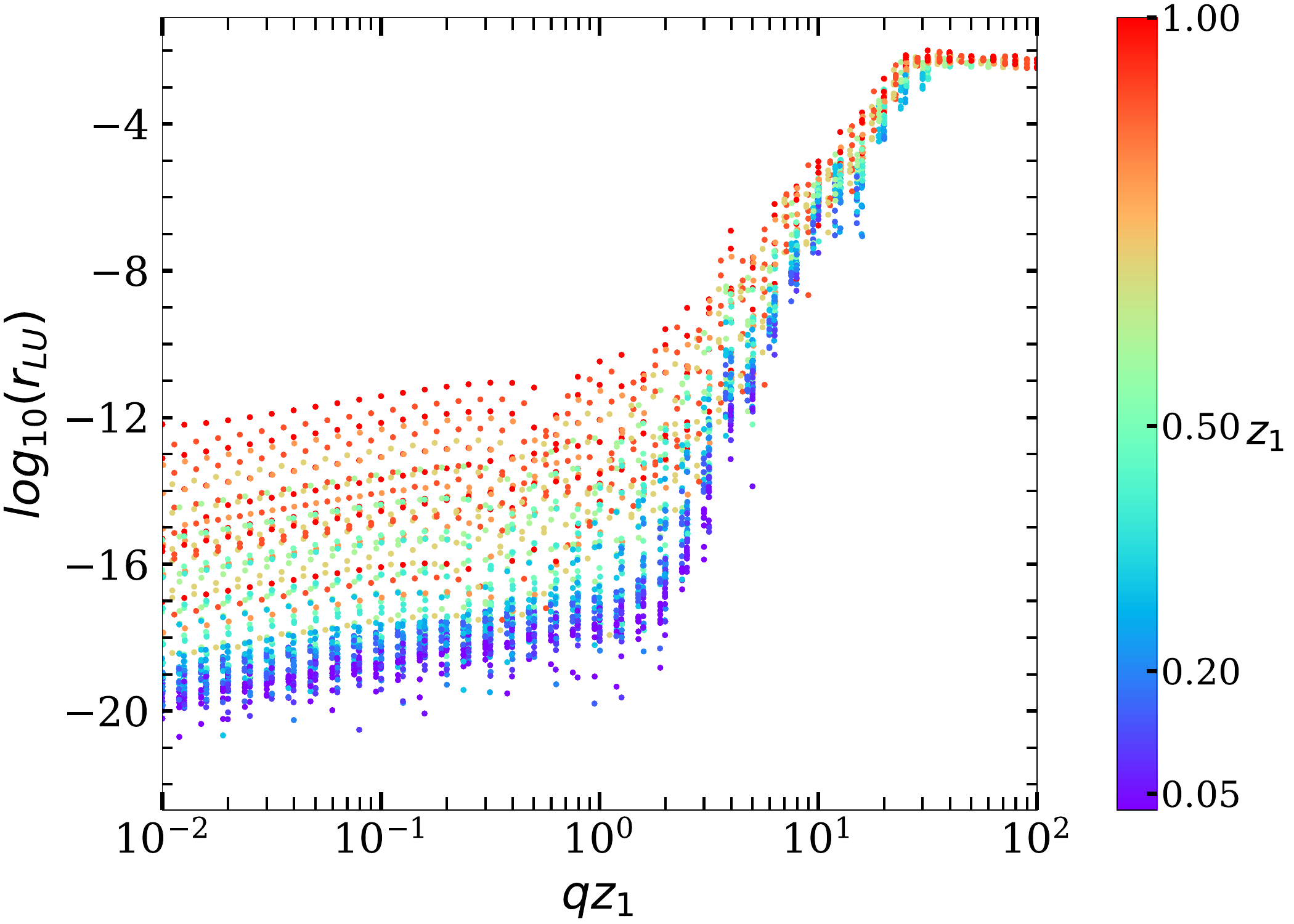}
}
\subfloat[\mbox{$[0, z_1]$} with $n_1 = 32$]{
   \includegraphics[width=0.47\textwidth]{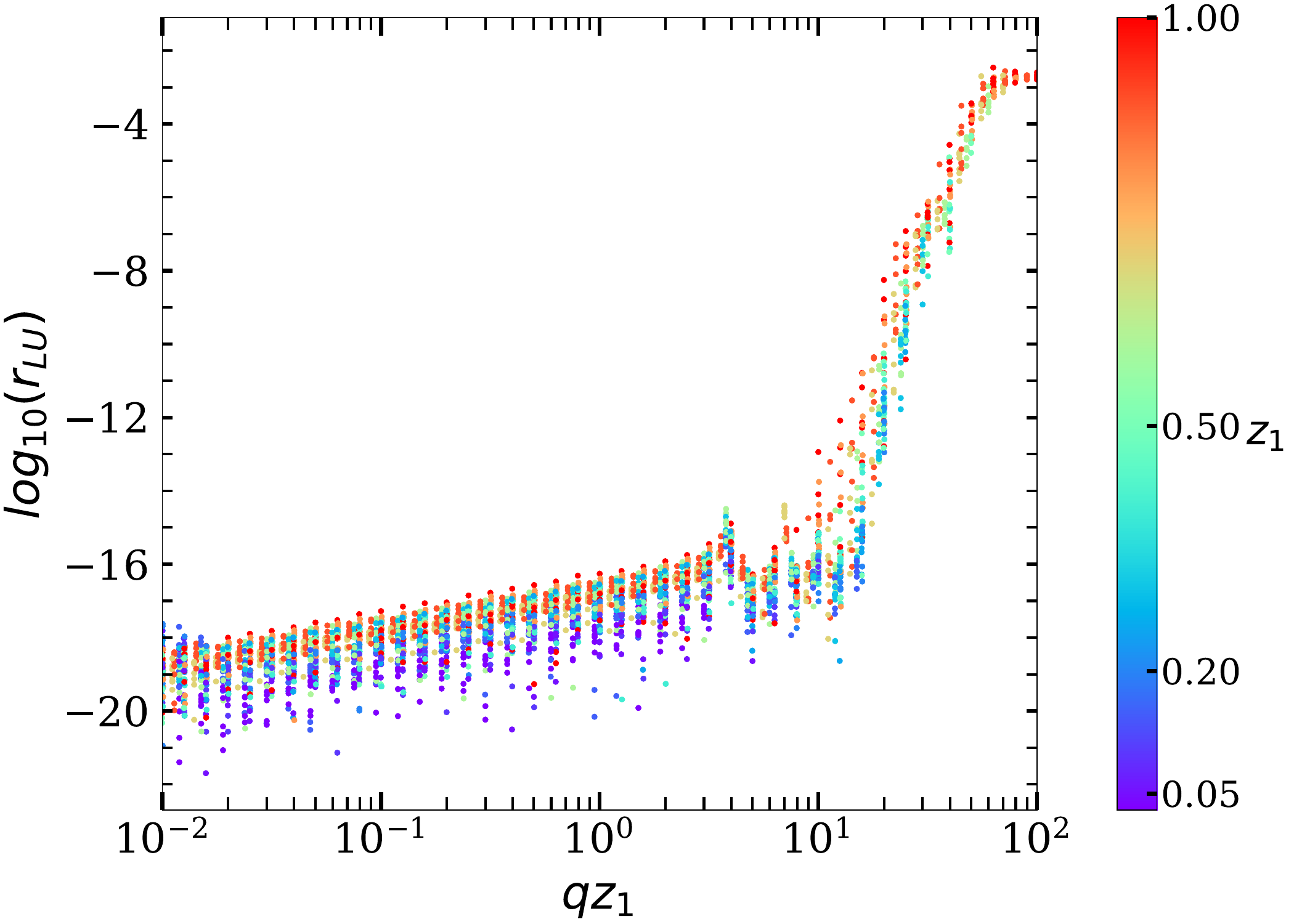}
}
\\[1.5em]
\subfloat[\mbox{$[z_1, \infty]$} with $n_2 = 32$]{
   \includegraphics[width=0.47\textwidth]{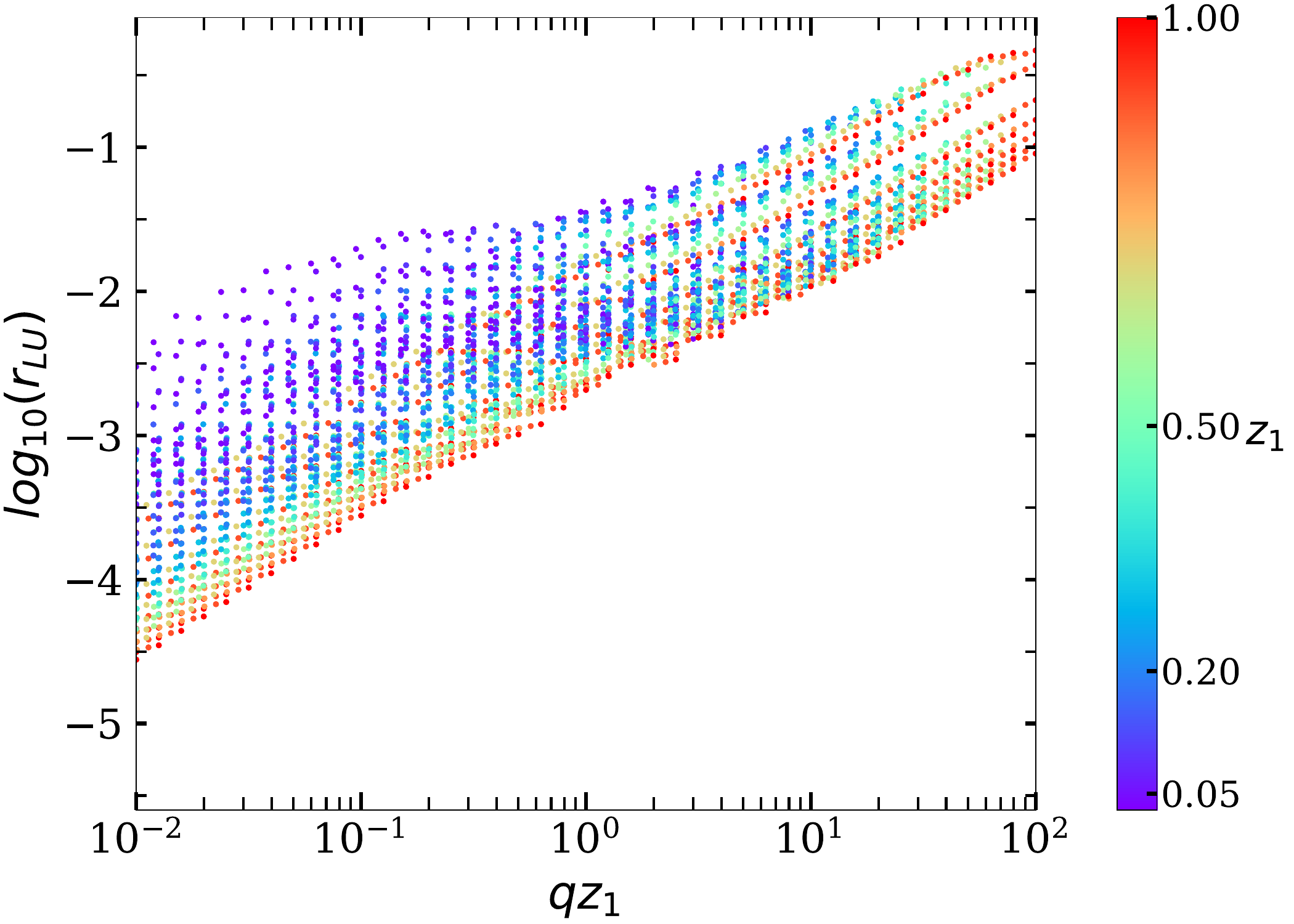}
}
\subfloat[\mbox{$[z_1, \infty]$} with $n_2 = 64$]{
   \includegraphics[width=0.47\textwidth]{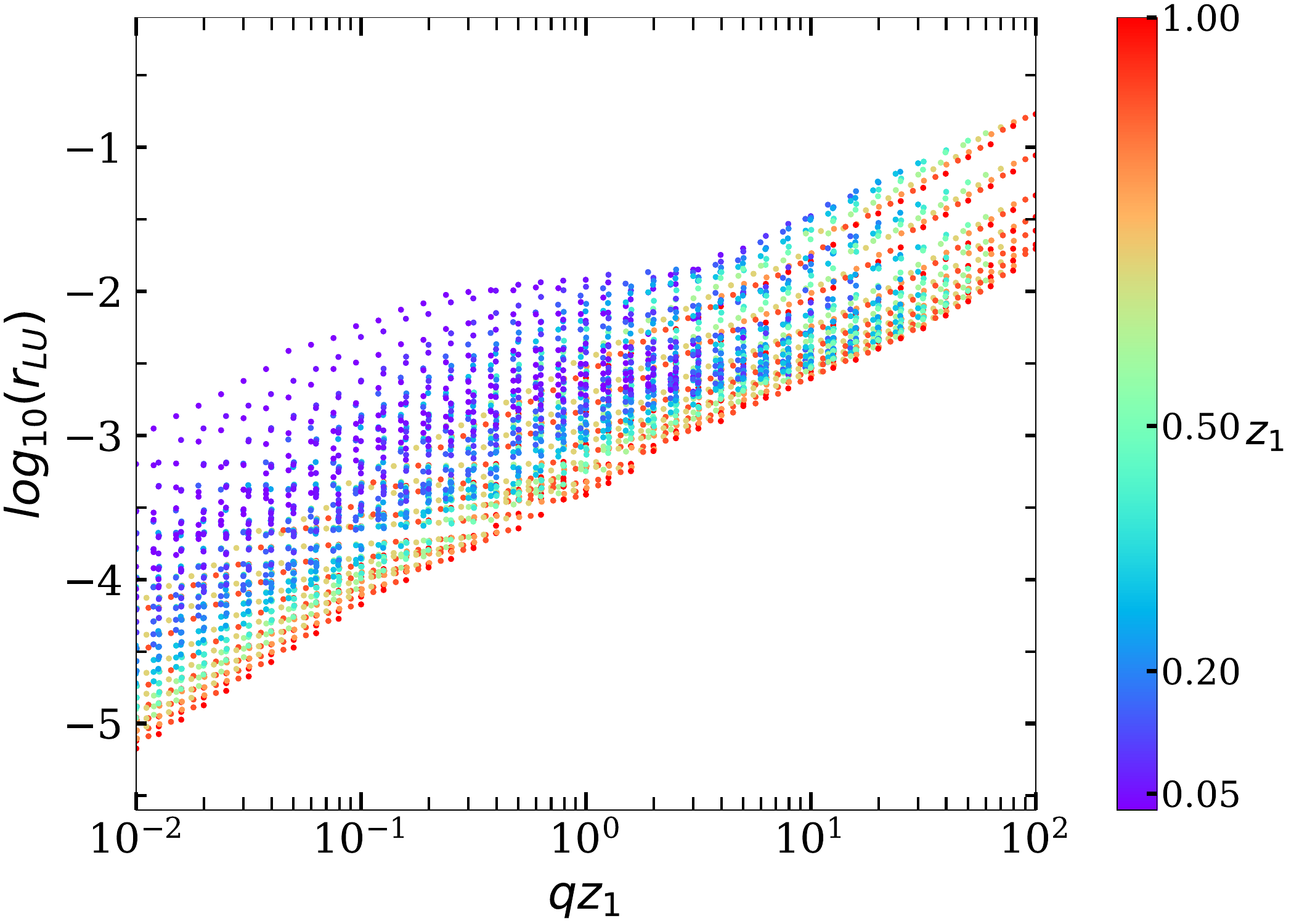}
}
\caption{\label{fig:mat-cond-nu-1} The ratio $r_{LU}$ defined in
\protect\eqref{r-LU-def} for the matrix $B$ on a given subgrid.  The plots in
this figure are for $\nu=1$.  See the text for details of the variable transform
$u(z)$.}
\end{figure}

\begin{figure}[t]
\centering
\subfloat[\mbox{$[0, z_1]$} with $n_1 = 16$]{
   \includegraphics[width=0.47\textwidth]{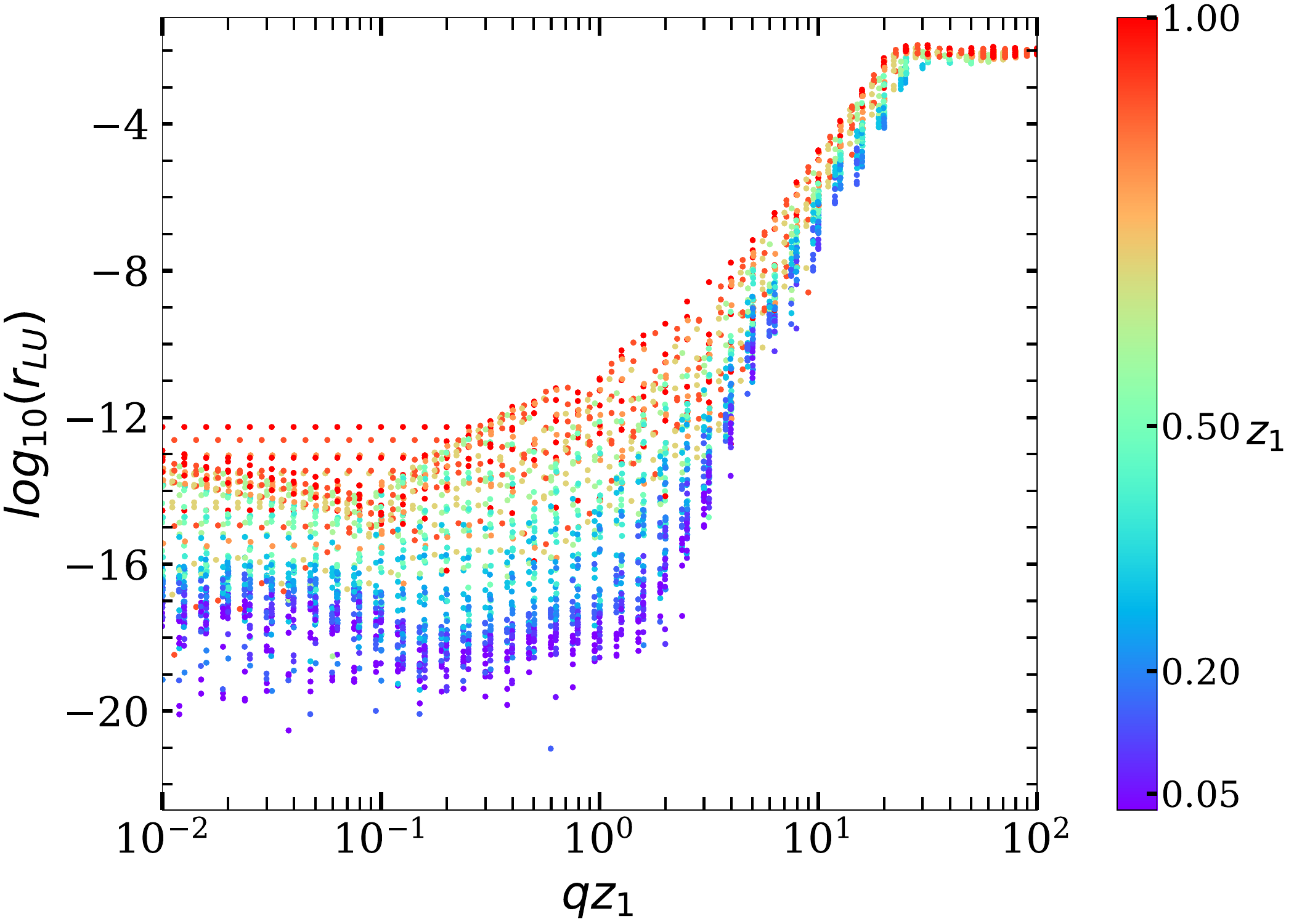}
}
\subfloat[\mbox{$[0, z_1]$} with $n_1 = 32$]{
   \includegraphics[width=0.47\textwidth]{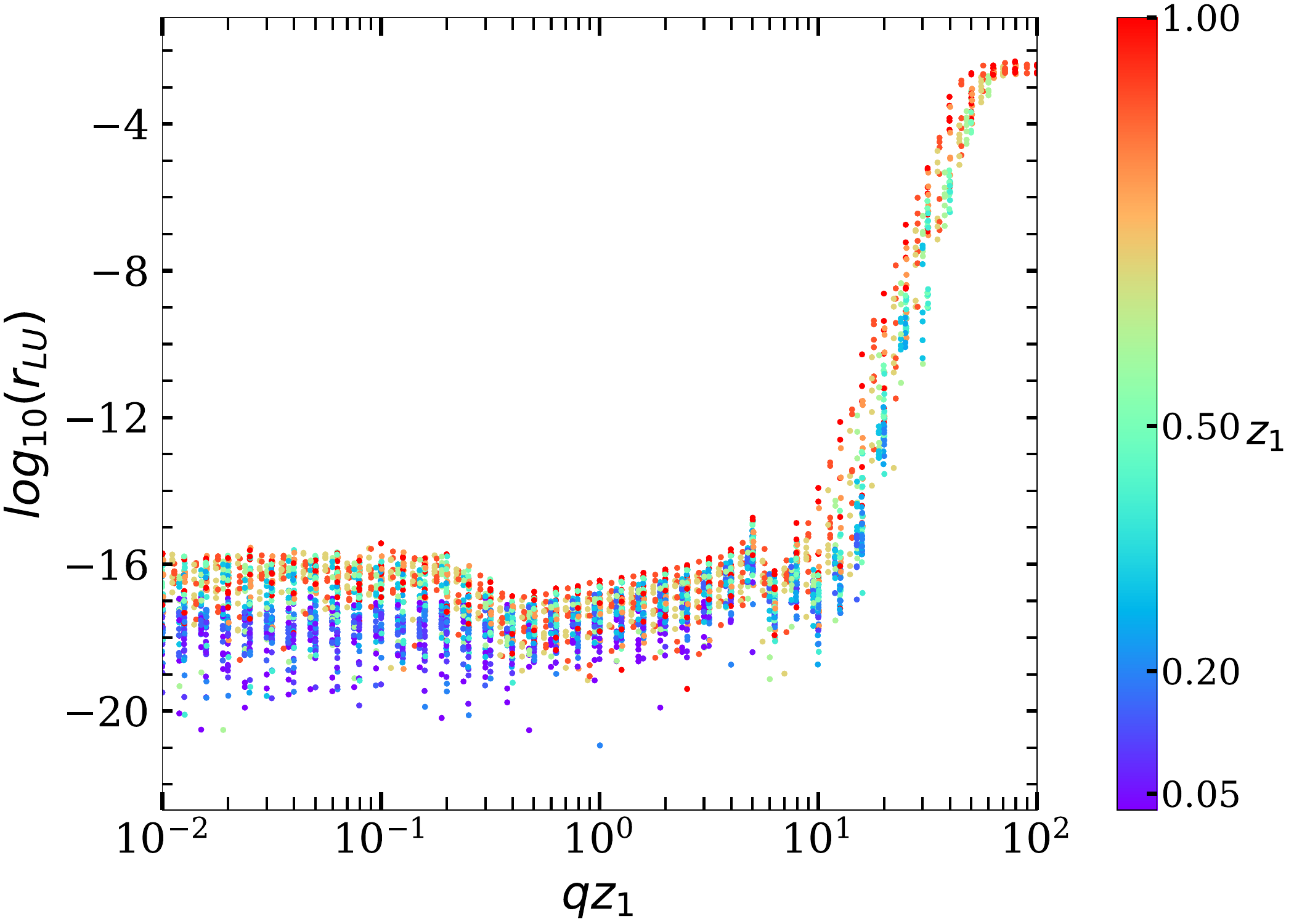}
}
\\[1.5em]
\subfloat[\mbox{$[z_1, \infty]$} with $n_2 = 32$]{
   \includegraphics[width=0.47\textwidth]{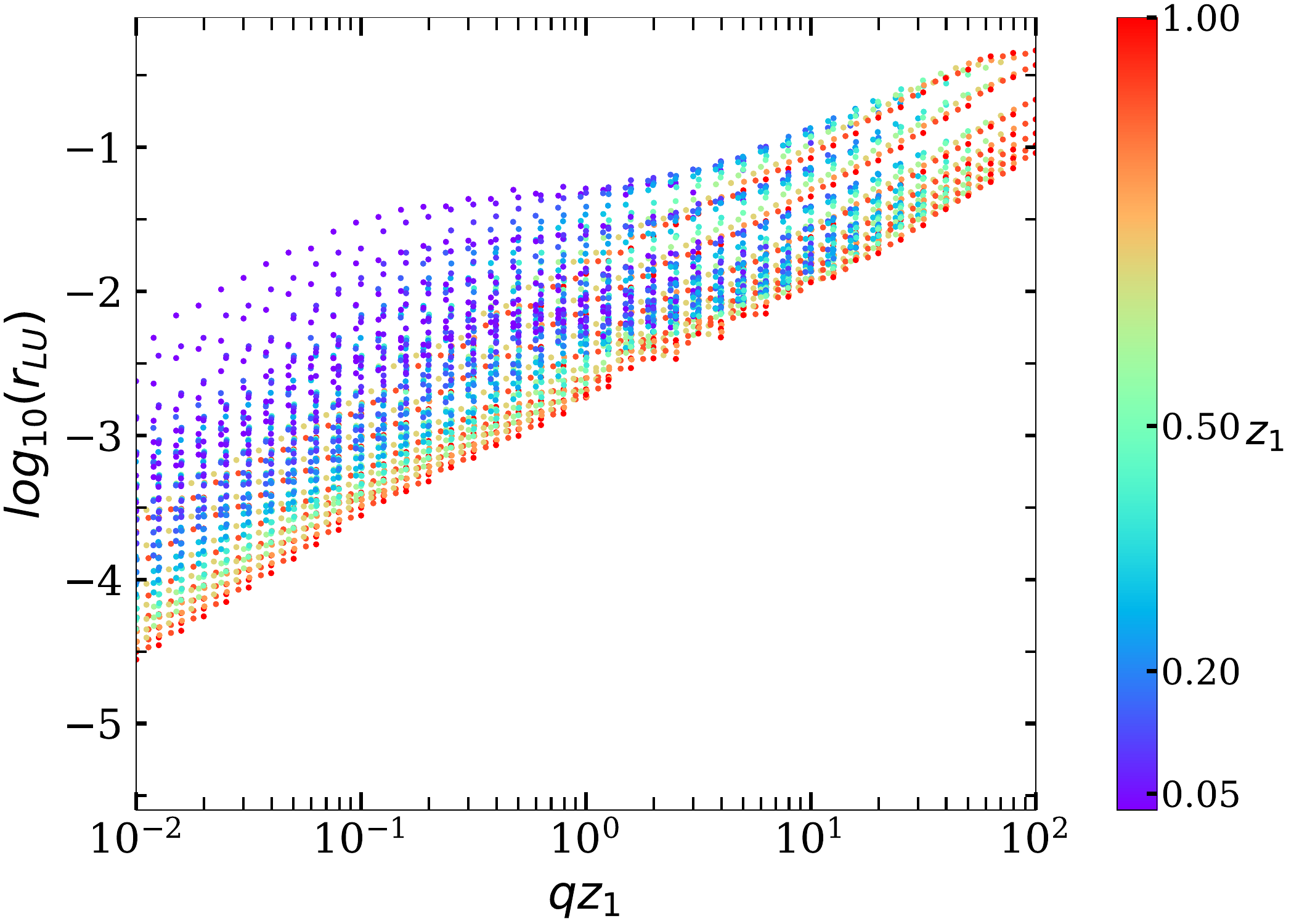}
}
\subfloat[\mbox{$[z_1, \infty]$} with $n_2 = 64$]{
   \includegraphics[width=0.47\textwidth]{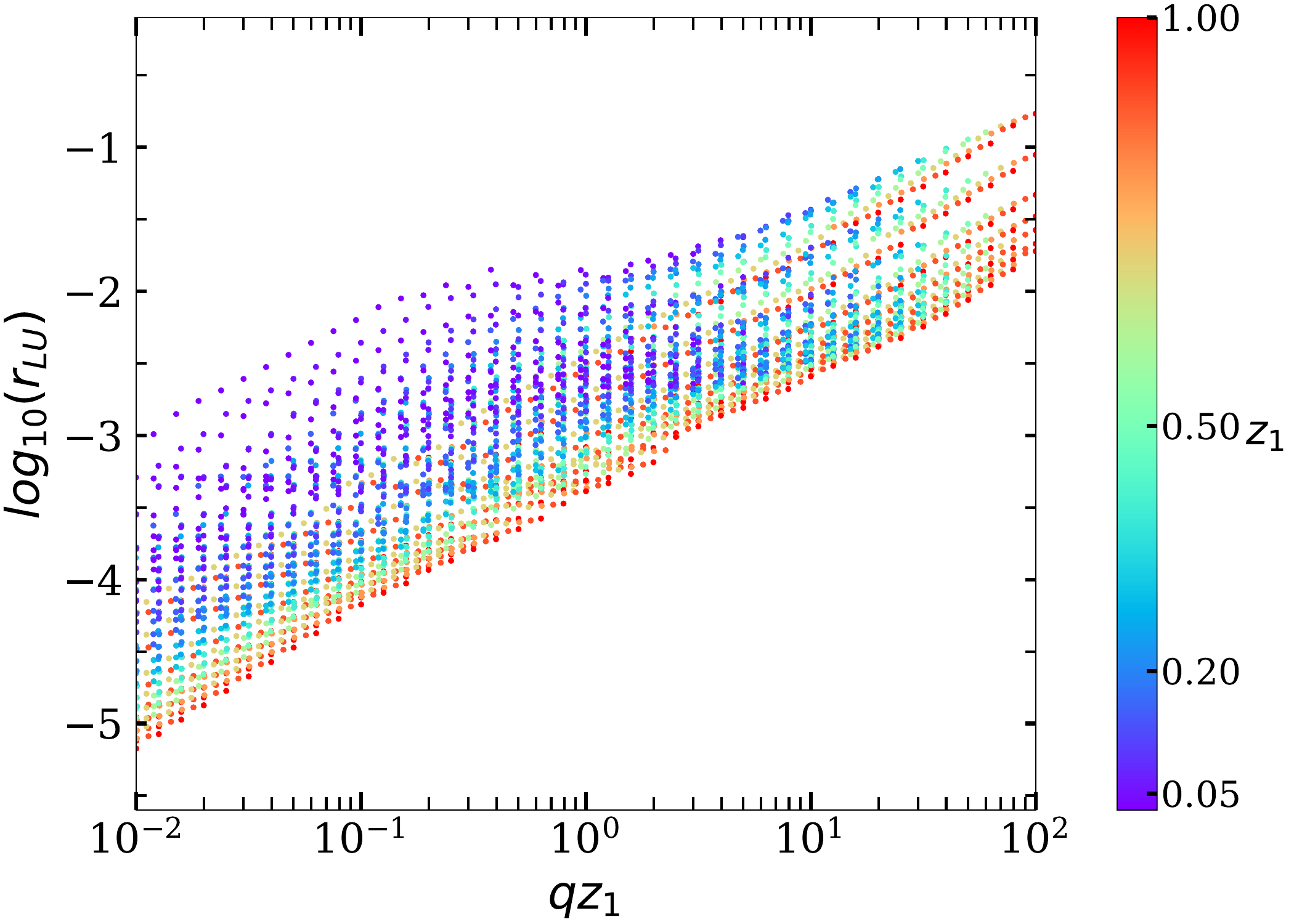}
}
\caption{\label{fig:mat-cond-nu-2} As \fig{\protect\ref{fig:mat-cond-nu-1}}, but
for $\nu=2$.}
\end{figure}

In the following exercise, we consider grids of the form $[0, z_1,
\infty]_{(n_1, n_2)}$ with $(n_1, n_2)$ equal to $(16, 32)$ or $(32, 64)$ .  We
use either the \texttt{exp sqrt} transform specified in \tab{\ref{tab:var-trf}}
with the parameter $m$ ranging from $0.32$ to $3.85$, or the \texttt{exp}
transform with $m$ ranging from $0.19$ to $2.24$. (These settings correspond to
the parameter scans presented in \sect{\ref{sec:params}}.)
Apart from some details, our results are very similar for different values of
$\nu$, and in \figs{\ref{fig:mat-cond-nu-1}} and \ref{fig:mat-cond-nu-2} we show
them for $\nu=1$ and $\nu=2$, respectively.

We compute $B$ for a wide range of $z_1$ and $q$ values, with linearly spaced
$z_1$ values between $0.03$ and $1$ and logarithmically spaced $q$ values
between $0.01$ and $100$.  We find $r_{LU}$ to be rather strongly correlated
with the product $q z_1$ of these variables, as shown for the lower subinterval
$[0, z_1]$ in the top panels of \figs{\ref{fig:mat-cond-nu-1}} and
\ref{fig:mat-cond-nu-2} and for the upper subinterval $[z_1, \infty]$ in the
bottom panels of these figures.

On the upper subinterval, $r_{LU}$ is always relatively large, decreasing only
slightly when the number of grid points is increased.  One can expect the LU
decomposition to give reliable numerical results in this range, and our
integration benchmarks confirm that this is indeed the case.

The situation is very different in the lower subinterval.  Here $r_{LU}$ drops
sharply as $q z_1$ decreases to values of order $1$, and it remains very small
when $q z_1$ decreases even further.  This confirms our expectation given in
\sect{\ref{sec:linalg}}: in a $z$ interval where the Bessel functions $J_{\nu}(q
z)$ and $J_{\nu+1}(q z)$ have only few or no oscillations, the discretised
version of the Levin equations approaches the ``continuum limit'', in which
these equations have an infinite number of solutions.

We checked separately that with a single subinterval $[0, \infty]$ the value of
$r_{LU}$ stays above $10^{-6}$ for $n \le 96$ points with the variable
transformations and $q$ values specified above.  For a single interval from $0$
to $\infty$, one can hence expect the LU method to work well throughout.

When $r_{LU}^{} \le r_{LU}^{\text{max}}$, we no longer use the LU decomposition
to solve the linear equations and resort to the computationally more stable (and
more expensive) SV decomposition.  When the latter is used, we also need to
choose the threshold $r_{SV}^{\text{max}}$ for truncating small singular values
as specified in \eqref{SV-trunc-crit}.

To set these two control parameters of our method, we disabled the use of
quadrature in intervals with $q z_1 \le j_{\nu}$ (see the discussion around
\eqref{CC-use-criterion}) and monitored how the integration accuracy depends on
$r_{LU}^{\text{max}}$ and $r_{SV}^{\text{max}}$.  We performed this monitoring
for the integrals in \sects{\ref{sec:benchmarks}} and \ref{sec:tmd-numerics} and
found
\begin{align}
   \label{r-values-default}
   r_{LU}^{\text{max}}
   &=
   r_{SV}^{\text{max}}
   =
   10^{-12}
\end{align}
to be a satisfactory setting.  Our choice for $r_{LU}^{\text{max}}$ is
conservative in the sense that decreasing it by two orders of magnitude does not
degrade the accuracy in the test cases just mentioned.

When the use of quadrature in intervals with $q z_1 \le j_{\nu}$ is enabled, the
integration accuracy is even less dependent on the precise settings of
$r_{LU}^{\text{max}}$ and $r_{SV}^{\text{max}}$.  This is because small values
of $r_{LU}$ or $r_{SV}$ tend to appear only for small values of $q z$,
when quadrature can be used.

We find that the only settings where the SV decomposition is needed involve very
fine subgrids and intermediate values of $q$ (small enough to give a small
$r_{LU}$ but large enough to fail our criterion for using quadrature). An
example is given in the rightmost column of \tab{\ref{tab:timing}}.


\subsection{Benchmarks: grids and precision}
\label{sec:benchmarks}

To assess the performance of our algorithm, we compute a number of
Fourier-Bessel integrals whose exact form is known analytically.  The integrals
considered are
\begin{align}
   \label{test-integrals}
   I(q)
   &=
   \int_{0}^{z_b} dz\, J_{\nu}(q z) \, \ftil(z)
   &
   \text{ with }
   z_b =
   \begin{cases}
      \infty & \text{ for cases 1a to 7a,}
      \\
      10     & \text{ for cases 7b and 8,}
   \end{cases}
\end{align}
where $\ftil(z)$ and $I(q)$ for the different cases are given in
\tab{\ref{tab:functions}}.  Unless specified otherwise in
\tab{\ref{tab:int-details}}, the integrals are evaluated for
\begin{align}
   \label{nu-values}
   \nu
   &=
   \begin{cases}
      0,\, 0.5,\, 1,\, 1.5,\, 2 & \text{ with $J_{\nu-1}$ calls,} \\
      1,\, 1.5,\, 2,\, 2.5,\, 3 & \text{ with $J_{\nu}$ calls,}  \\
      2,\, 2.5,\, 3             & \text{ with $J_{\nu+1}$ calls,}
   \end{cases}
   \\
   \label{q-values}
   q
   &=
   0.01,\, 0.1,\, 1,\, 2,\, 3,\, 5,\, 10,\, 15,\, 20,\, 25,\, 30
\end{align}
The different ``calls'' specified in \eqref{nu-values} refer to the master
equations \eqref{int-final-def} and \eqref{IBP} of our method, i.e.\ the
$J_{\nu-1}$, $J_{\nu}$, and $J_{\nu+1}$ calls respectively correspond to a
nonzero $f_0$, $f_1$, and $f_2$ in these equations.

\begin{table}
\centering
\renewcommand{\arraystretch}{1.5}
\begin{tabular}{c l l}
\toprule
   case  & $\ftil(z)$ & $I(q)$
\\
\midrule
   1a &
   $z^{\nu + 1} \, K_0(\kappa z)$
   &
   $\Gamma(\nu+1) \; (2 q)^{\nu} \,
      \big/ (\kappa^2 + q^2)^{\nu+1}$
\\[0.25em]
   1b &
   $z^{\mu + \nu + 1} \, K_{\mu}(\kappa z)$
   &
   $\Gamma(\mu+\nu+1) \; (2 \kappa)^{\mu} \, (2 q)^{\nu} \,
      \big/ (\kappa^2 + q^2)^{\mu+\nu+1}$
\\[0.25em]
   2 &
   $z^\mu \, e^{- \kappa z}$
   &
   $\frac{\Gamma(\mu+\nu+1)}{\Gamma(\nu+1)} \,
   \bigl( \frac{q}{2} \bigr)^\nu \, \kappa^{-\mu -\nu -1} \,
   {}_2 F_{1}\Bigl( \frac{\mu+\nu+1}{2}, \frac{\mu+\nu+2}{2};
      \nu+1; - \frac{q^2}{\kappa^2} \Bigr)$
\\[0.25em]
   3 &
   $e^{- \lambda^2 z^2}$
   &
   $\frac{\sqrt{\pi}}{2 \lambda}\,
   \exp\Bigl( - \frac{q^2}{8 \lambda^2} \Bigr) \,
   I_{\nu / 2}\Bigl( \frac{q^2}{8 \lambda^2} \Bigr)$
\\[0.25em]
   4 &
   $z^{\nu + 1} \, e^{- \lambda^2 z^2}$
   &
   $\frac{1}{q} \, \bigl( \frac{q}{2 \lambda^2} \bigr)^{\nu+1} \,
   \exp\Bigl( - \frac{q^2}{4 \lambda^2} \Bigr)$
\\[0.25em]
   5a &
   $\bigl( \frac{z}{z^2 + b^2} \bigr)^{\nu+1}$
   &
   $\frac{1}{\Gamma(\nu+1)} \,
   \bigl( \frac{q}{2} \bigr)^{\nu} \,
   K_{0}(q b)$
\\[0.25em]
   5b &
   $z^{\nu+1} \big/ (z^2 + b^2)^{\mu+\nu+1}$
   &
   $\frac{1}{\Gamma(\mu+\nu+1)} \,
   \bigl( \frac{q}{2} \bigr)^{\nu} \, \bigl( \frac{q}{2 b} \bigr)^{\mu} \,
   K_{\mu}(q b)$
\\[0.25em]
   6a &
   $1$
   &
   $\frac{1}{q}$
\\[0.25em]
   6b &
   $z^{-\nu}$
   &
   $\frac{\Gamma(1/2)}{2 \ms \Gamma(\nu + 1/2)} \,
   \bigl( \frac{q}{2} \bigr)^{\nu-1}$
\\[0.25em]
   7a &
   $z^{-\nu+1}$
   &
   $\frac{1}{2 \ms \Gamma(\nu)} \,
   \bigl( \frac{q}{2} \bigr)^{\nu-2}$
\\[0.25em]
   7b &
   $z^{-\nu+1}$
   &
   $\frac{1}{2 \ms \Gamma(\nu)} \,
   \bigl( \frac{q}{2} \bigr)^{\nu-2} \,
   - \frac{1}{q} \, z_b^{-\nu+1} J_{\nu-1}(q z_b)$
\\[0.25em]
   8 &
   $z^{\nu+1}$
   &
   $\frac{1}{q} \, z_b^{\nu+1} \,
   J_{\nu+1}(q z_b)$
\\
\bottomrule
\end{tabular}
\caption{\label{tab:functions} The functions $\ftil(z)$ used for our benchmark
exercise and their Fourier-Bessel transforms $I(q)$ specified in
\eqref{test-integrals}.  \rev{Cases 1a, 5a, 6a, and 7a are special cases of 1b,
5b, 6b, and 7b, respectively, but are handled with different integration
settings (see \eqn{\eqref{high-acc-grids}} and
\tab{\protect\ref{tab:int-details}}).}  Here $I_\nu$ and $K_\nu$ are the modified
Bessel functions of the first and second kind, and ${}_2 F_{1}$ is the Gaussian
hypergeometric function.}
\end{table}

\begin{table}
\centering
\renewcommand{\arraystretch}{1.3}
\begin{tabular}{c l l l}
\toprule
   case & parameters & variable transform & remarks
\\
\midrule
   1a &
   $\kappa = 1.5$
   &
   \texttt{exp sqrt}, $m = 1.5 \ms \kappa$
   &
   $\nu \le 2.5$
\\[0.25em]
   1b &
   $\kappa = 1.5, \, \mu = 1$
   &
   \texttt{exp sqrt}, $m = \kappa$
   &
   $\nu \le 2.5$
\\[0.25em]
   2 &
   $\kappa = 1.5, \, \mu = 2.5$
   &
   \texttt{exp sqrt}, $m = 1.5 \ms \kappa$
\\[0.25em]
   3 &
   $\lambda = 2$
   &
   \texttt{exp}, $m = 4\lambda$
   &
   not $\nu = 0.5$
 \\[0.25em]
   4 &
   $\lambda = 2$
   &
   ------------ " ------------
   &
   $q \le 10$
\\[0.25em]
   5a &
   $b=1.2$
   &
   \texttt{inv pow}, $z_0 = 1, \alpha = 0.5$
   &
   $q \le 10$
\\[0.25em]
   5b &
   $b=1.2, \, \mu=2.5$
   &
   \texttt{inv pow}, $z_0 = 1, \alpha = 1$
   &
   $q \le 10$
\\[0.25em]
   6a &
   &
   \texttt{inv pow}, $z_0 = 1, \alpha = 0.5$
   &
   not $\nu = 0.5$
\\[0.25em]
   6b &
   &
   ------------ " ------------
   &
   only $\nu \ge 0.5$, not using $J_{\nu+1}$ call
\\[0.25em]
   7a &
   &
   ------------ " ------------
   &
   only $\nu \ge 1$
\\[0.25em]
   7b &
   &
   ------------ " ------------
   &
   also for $q = 0.001,\, 50,\, 100,\, 200,\, 300$
\\[0.25em]
   8 &
   &
   trivial transform $z = u$
   &
   also for $q = 0.001,\, 50,\, 100,\, 200,\, 300$
\\
\bottomrule
\end{tabular}
\caption{\label{tab:int-details} Details for computing the integrals in
\tab{\protect\ref{tab:functions}}.  The variable transforms are specified in
\tab{\protect\ref{tab:var-trf}} and discussed in
\sect{\protect\ref{sec:var-transforms}}.}
\end{table}

We find grid settings both for intermediate or for high precision. Specifically,
we obtain a relative integration error $\varepsilon$ of
\rev{
\begin{align}
   \label{low-acc-grids}
   \varepsilon
   &
   \le 5 \times 10^{-4}
   &&
   \text{ with }
   \begin{cases}
      \, [0, 1, \infty]_{(20, 25)} & \text{ for cases 1a to 7a,} \\[0.1em]
      \, [0, \infty]_{(45)}        & \text{ for cases 1b to 7a,} \\[0.1em]
      \, [0, 10]_{(24)}            & \text{ for cases 7b and 8,}
   \end{cases}
\intertext{and}
   \label{high-acc-grids}
   \varepsilon
   &
   \le 10^{-6}
   &&
   \text{ with }
   \begin{cases}
      \, [0, 0.1, \infty]_{(30, 44)} & \text{ for case 1a,} \\[0.1em]
      \, [0, 1, \infty]_{(30, 44)}   & \text{ for cases 1b to 7a,} \\[0.1em]
      \, [0, 10]_{(34)}              & \text{ for cases 7b and 8.}
   \end{cases}
\end{align}
These grid settings and the parameters of variable transforms in
\tab{\ref{tab:int-details}} are not optimised for each individual integrand, but
have been chosen with as little variation as needed to achieve the accuracy goals of $5 \times 10^{-4}$ or $10^{-6}$.  In practical applications, one will often deal with less variation in the shape of integrands and in the value of $\nu$.  One can then perform a more systematic optimisation of grid settings, as we will do for TMD-like functions in \sect{\ref{sec:tmd-numerics}}.}

We note that in \rev{cases 5a and 5b} the integral $I(q)$ decreases exponentially
with $q$, whilst in \rev{case 4} it decreases like a Gaussian in $q$.  We limit the
benchmark to $q \le 10$ in these cases.  For larger value the accuracy degrades
rather quickly with our grid settings, which is not too surprising because the
integrals become extremely small as a result of numerical cancellations.

\rev{For $\nu=0$, the integrand of case 1a behaves like $-z \ln(z)$ at $z\to 0$
and hence has a divergent first derivative in that limit.  This puts higher
requirements on the discretisation.  In particular, the grid $[0, \infty]_{(45)}$
yields an accuracy of only $\varepsilon \approx 3 \times 10^{-3}$, which is
quite inferior to what we obtain with the same grid for cases 1b to 7a.  In such
a situation, splitting the integration interval into subgrids gives more
flexibility and allows one to achieve significantly better accuracy for a given
number of points. We will confirm this trend in
\sect{\ref{sec:subgrid-choice}}.}

Notice also that we omit the value $\nu = 0.5$ in our benchmark for \rev{cases 3 and
6a}.  In these cases, one has an integrand behaving like $z^\mu \ms J_{\nu}(q z)$
with $\mu + \nu = 0.5$ at small $z$, for which we expect our method to be less
accurate as discussed in \sect{\ref{sec:final_scaling_z_behaviour}}.  This is
borne out by our numerical study: the relative integration error for $\nu = 0.5$
in \rev{cases 3 and 6a is $\varepsilon \le 2 \times 10^{-3}$ for the grids $[0, 1,
\infty]_{(20, 25)}$ and $[0, \infty]_{(45)}$, and $\varepsilon \le 3 \times
10^{-4}$ for the grid $[0, 1, \infty]_{(30, 44)}$}. Although this is worse than
the benchmark accuracy \rev{in \eqref{low-acc-grids} and \eqref{high-acc-grids}}, it shows that our method
can handle these cases as well.


\subsection{Computation time}
\label{sec:timing}

Indicative computation times needed for integration calls under different
conditions are given in \tab{\ref{tab:timing}}.  The first two grids are those
in our benchmark exercise of the previous subsection; and the grid with $(n_1,
n_2) = (40, 50)$ may be used for estimating the integration error of the grid
with $(n_1, n_2) = (20, 25)$, as will be discussed in \sect{\ref{sec:error}}.
The times in the table do \emph{not} include the time required for computing
the integrand on the grid points (step \ref{step:discr-functions} in
\sect{\ref{sec:algorithm}}), since that time crucially depends on the function
for the integrand and not on our algorithm.

As discussed in \sect{\ref{sec:condition-numbers}}, the integration method used
in the first of two subintervals depends on $q$.  At sufficiently low $q$,
quadrature can be used, whilst for large $q$ the matrix $B$ becomes well
conditioned and the LU decomposition can be employed.  For intermediate $q$
values, the LU method can be used for less fine grids, but it will become
unstable for finer grids, when the SV decomposition must be taken instead.

\begin{table}
\centering
\begin{tabular}{l |c |c c |c c c}
\toprule
grid
       & $[0, \infty]_{(45)}$
       & \multicolumn{2}{c|}{$[0, 1, \infty]_{(20, 25)}$}
       & \multicolumn{3}{c }{$[0, 1, \infty]_{(40, 50)}$}
\\
method & LU
       & (CC, LU) & (LU, LU)
       & (CC, LU) & (LU, LU) & (SV, LU)
\\
$q$ range & low
          & low & intermediate
          & low & high & intermediate
\\
\midrule
new $q$ [$\microsec$]
       & 45
       & 12 & 18
       & 60 & 100 & 1500
\\
same $q$ [$\microsec$]
       & 1.9
       & 0.8 & 1.3
       & 2.5 & 3.6 & 4.1
\\
\bottomrule
\end{tabular}
\caption{\label{tab:timing} Indicative computation times of the integration call
for $J_{\nu - 1}$ (which includes integration by parts).  The second row
specifies the integration method used in each subinterval.  The $q$ ranges are
$0.095 \ldots 0.1$ (low), $14.25 \ldots 15$ (intermediate), and $38 \ldots 40$
(high).  Timings have been obtained on a recent laptop with an
Intel\textregistered\ Core\texttrademark\ i7-1355U processor, 32 GiB memory, and
code compiled with \texttt{g++} version 11.4.0.}
\end{table}

The \emph{absolute} times in the table should not be over-interpreted, since
they refer to a particular computer (where they fluctuate at the level of 20\%
between different runs).  Running the same code on a range of other laptops and
desktop PCs, we find that some times are larger than those in the table by
a factor up to 5, whilst in a few cases calls are faster by a factor up to
1.4.

Independently of the used computer, we observe a clear hierarchy for the timing
of calls under different conditions, and thus for the different elements of our
algorithm:
\begin{itemize}
\item integration is significantly faster if the $q$ value is the same as in the
   previous call,
\item times for integration at a new value of $q$ increase faster with the
   number of grid points than times for integration at the same value of $q$,
\item quadrature is faster than the Levin method, and within the Levin method,
   an SV decomposition takes more time than an LU decomposition (especially
   for the $q$ update),
\item for a comparable total number of grid points, using subgrids reduces the
   computation time.
\end{itemize}
These findings are in line with our expectations formulated in
\sects{\ref{sec:linalg}} and \ref{sec:algorithm}.  We also note that the time
for initialising the integration routine for a specific grid and $\nu$ value
(step \ref{step:fixed-nu} in \sect{\ref{sec:algorithm}}) is well below $100
\microsec$ in all cases.  As expected, this is negligible for the overall
computing time budget.

The times in the table are for $J_{\nu-1}$ calls, i.e.\ they require integration
by parts and hence include the matrix multiplication \eqref{IBP-f1-f0-discrete}.
 Correspondingly, the integration calls for $J_{\nu}$ or $J_{\nu+1}$ are
slightly faster, namely between 10\% to 20\% for integration without update
of~$q$.

\rev{The integration calls discussed so far are for a single discretised
function.  For convenience, \bestlime\ also provides calls that take one value
of $q$ but several discretised functions as arguments and return the vector of
integration results.  We find no significant difference in timing between $N$
single-function calls (at the same $q$) and a single call with $N$ functions at
once.}

\section{Numerical studies for TMD-like functions}
\label{sec:tmd-numerics}

In this section, we explore in some detail how our method performs for integrals
that appear in TMD cross sections.  Up to a global factor, such cross sections
are proportional to
\begin{align}
   \label{tmd-int}
   I(q) &=
   \int_0^\infty \! dz\, J_0(q z) \, z \ms W(z, Q)
\end{align}
where $W(z, Q)$ is product of TMDs evolved to the scale $Q$.  For brevity, we do
not display the additional dependence on longitudinal momentum fractions.  We
will also consider the cumulative $q$ spectrum
\rev{
\begin{align}
   \label{tmd-cumul}
   K(q)
   &=
   \frac{1}{2\pi} \int d^2\bs \tilde{q}\; \theta(q - \tilde{q}) \, I(\tilde{q})
   =
   \int_0^{q} d \tilde{q} \; \tilde{q} \, I(\tilde{q})
   =
   q \int_0^\infty \! dz\, J_1(q z) \, W(z, Q)
   \,.
\end{align}
Notice} that the integrands in \eqref{tmd-int} and \eqref{tmd-cumul} have the
same behaviour at $z\to 0$.

We recall that $q$ is a measured transverse momentum and $z$ is a distance
between parton fields in the plane transverse to the collision axis in the
process.  We will therefore give $q$ in units of $\gev$ and $z$ in units in
$\gev^{-1}$ (using a system in which the speed of light is unity).


\subsection{Integrands}
\label{sec:tmd-integrands}

In this subsection, we specify different forms of $W(z,Q)$ used in our numerical
study.  A reader not familiar with TMDs (but interested in Fourier-Bessel
transforms) may skip over the details, but should note two important features of
the integrand: $(i)$ it is non-analytic at $z=0$, and $(ii)$ its leading
behaviour at large $z$ is either given by an exponential $\exp(- \kappa z)$ or
by a Gaussian $\exp(- \lambda^2 z^2)$.

Our purpose is not a detailed phenomenological investigation of TMDs, but to
understand the interplay between their qualitative behaviour and our integration
algorithm.  We therefore use  leading-order evolution and tree-level
short-distance matching for the TMDs.  For details and references, we refer to
\cite[chapter~2]{Boussarie:2023izj}.  The product of two evolved TMDs
has the form
\begin{align}
   \label{evolved-W}
   W(z, Q)
   &=
   \bigl[ f_{\text{np}}(z) \bigr]^2 \, \exp\bigl[ - 2 S(z, Q) \bigr]
   \,,
\end{align}
where
\begin{align}
   S(z, Q)
   &=
   \int_{\mu_z}^{Q} \frac{d \mu}{\mu} \,
   \biggl[
      4 C_F \, \frac{\als(\mu)}{2\pi} \, \ln \frac{Q}{\mu}
      - 3 C_F \, \frac{\als(\mu)}{2\pi}
   \biggr]
   \notag \\[0.2em]
   &=
   \biggl[\, \frac{c_1}{\als(Q)} - c_2 \,\biggr] \,
      \ln \frac{\als(\mu_z)}{\als(Q)}
   - c_1 \biggl[\, \frac{1}{\als(Q)} - \frac{1}{\als(\mu_z)} \,\biggr]
\end{align}
with coefficients
\begin{align}
   c_1 &= 8 \pi C_F / \beta_0^2 \,,
   &
   c_2 &= 3 C_F / \beta_0 \,,
   &
   \beta_0 &= 11 - 2 n_f / 3 \,,
   &
   C_F = 4/3
\end{align}
and with one-loop running of the strong coupling:
\begin{align}
   \alpha_s(\mu)
   &=
   \frac{2 \pi}{\beta_0 \ln( \mu / \Lambda )}
   \,.
\end{align}
The scale parameter $\Lambda$ is chosen such that $\alpha_s(M_Z) = 0.13$ at the
mass of the $Z$ boson.  For simplicity, we take $n_f=5$ flavours everywhere; a
more realistic treatment would not affect the qualitative behaviour of our
integrands.
The initial conditions for evolution are formulated at the scale
\begin{align}
   \label{b0-def}
   \mu_z
   &=
   b_0 \ms \sqrt{1/z_{}^{2} +  1/z^2_{\text{max}} \rule{0pt}{1.78ex}}
   \,,
   &
   b_0 &= 2 e^{-\gamma} \approx 1.12
   \,,
   &
   z_{\text{max}} &= 0.5 \gev^{-1}
   \,.
\end{align}
For $z\to 0$, we have a non-analytic behaviour in $z$, namely
\begin{align}
   \label{LO-sudakov}
   \exp\bigl[ - 2 S(z, Q) \bigr]
   & \;\sim\;
   \biggl( \frac{z \Lambda}{b_0} \biggr)^{8 C_F / \beta_0}
   \biggl( \ln \frac{b_0}{z \Lambda} \biggr)^{2 c_1 / \als(Q)}
   \approx
   \biggl( \frac{z \Lambda}{b_0} \biggr)^{1.4} \;
   \biggl( \ln \frac{b_0}{z \Lambda} \biggr)^{1.14 / \als(Q)}
   \,.
\end{align}
Note that if $f_{\text{np}}(z)$ is sufficiently regular at $z=0$, the function
$z \ms W(z)$ appearing in the integral \eqref{tmd-int} has a finite first
derivative at that point.

We performed all numerical studies for three values of $Q$, namely
\begin{align}
   \label{Q-values}
   Q &= 2 \gev, \, 20 \gev, \, 100 \gev
   \,,
\end{align}
with corresponding values $\alpha_s(2 \gev) \approx 0.33$, $\alpha_s(20 \gev)
\approx 0.17$, and $\alpha_s(100 \gev) \approx 0.13$.  The resulting power of
$\ln (1/z)$ in \eqref{LO-sudakov} is about $3.5$, $6.7$, and $8.9$, respectively
and thus rather high for the two larger $Q$ values.

For the TMDs at scale $\mu_z$, we consider two different forms that have been
used in phenomenological analyses.  The first one is
%
%
\begin{align}
   \label{yukawa-tmd}
   f_{\text{np, Yukawa}}(z)
   &=
   \cosh\biggl[
      \biggl( \frac{2 c^2}{\kappa} - \frac{\kappa}{4} \biggr) z
   \,\biggr]
   \, \bigg/
   \cosh\biggl[
      \biggl( \frac{2 c^2}{\kappa} + \frac{\kappa}{4} \biggr) z
   \,\biggr]
   \,,
   &
   \kappa &= 0.642 \gev
   \,,
   \notag \\
   &&
   c &= 0.521 \gev
\end{align}
and corresponds to \eqn{(3.7)} and the first line of \tab{12} in
\cite{Scimemi:2017etj}.  At large $z$ this has a Yukawa-type behaviour like
$\exp(- \kappa z/2)$.  The second form has a Gaussian behaviour and reads
\begin{align}
   \label{pavia-tmd}
   f_{\text{np, Gauss}}(z)
   &=
   \bigl( 1 - c^2 z^2 \bigr) \exp\bigl( - \lambda^2 z^2 /2 \bigr)
   \,,
   &
   \lambda &= 0.374 \gev \,,
   &
   c &= 0.117 \gev
   \,.
\end{align}
This was used in the study \cite{Bacchetta:2017gcc}, see \eqn{(34)} and \tab{XI}
in that paper.
Notice that we define the parameters $\kappa$ and $\lambda$ such that the
\emph{squared} TMDs and hence the integrands in \eqref{tmd-int} and
\eqref{tmd-cumul} decrease like $\exp(-\kappa z)$ or $\exp(-\lambda^2 z^2)$.

In addition to these ``Yukawa'' and ``Gauss'' TMDs, we also consider the ``toy
TMD'' investigated in \cite{Kang:2019ctl}, which corresponds to
\begin{align}
   \label{toy-TMD-W}
   z \ms W_{\text{toy}}(z)
   &=
   \frac{1}{\Gamma\bigl( \beta^2 / \sigma^2 \bigr)}  \,
   \biggl( \frac{\beta z}{\sigma^2} \biggr)^{\beta^2 / \sigma^2}
   \exp\biggl[ - \frac{\beta z}{\sigma^2} \biggr]
   \,.
\end{align}
This has the same form as the first entry in our \tab{\ref{tab:functions}}, with
an exponential decay parameter $\kappa = \beta /\sigma^2$.  The function
$W_{\text{toy}}(z)$ takes its maximum when $1/z$ is equal to
\begin{align}
   Q &= \frac{\beta}{\beta^2 - \sigma^2}
   \,,
\end{align}
which in \cite{Kang:2019ctl} was taken as a representative of the physical scale
$Q$ of a hard process in this simplified TMD setup.  We choose the parameters in
\eqref{toy-TMD-W} as
\begin{align}
   \label{toy-TMD-params}
   \beta / \sigma^2 &= \kappa = 0.642 \gev
   \,,
   &
   \beta &= 1/\kappa + 1/Q
   \,,
\end{align}
thus taking the same value of $\kappa$ as in \eqref{yukawa-tmd} for all $Q$.

Unless specified otherwise, the plots in the present section are obtained with
the following interpolation grids:
\begin{align}
   \label{tmd-grid-default}
   \text{ \texttt{exp sqrt} grid with }
   &
   [0, z_1, \infty]_{(16, 32)}
   &&
   \text{ for the toy and Yukawa TMDs,}
   \notag \\
   \text{ \texttt{exp} grid with }
   &
   [0, z_1, \infty]_{(16, 32)}
   &&
   \text{ for the Gauss TMD.}
\end{align}
It is natural to specify the mass parameters $m$ in the grid transformations
(see \tab{\ref{tab:var-trf}}) as multiples of the characteristic exponential
decay parameter of the integrands.  We therefore define
\begin{align}
   \label{r-kappa-lam-def}
   r_{\kappa} &= m / \kappa
   &&
   \text{ for the toy and Yukawa TMDs,}
   \notag \\
   r_{\lambda} &= m / \lambda
   &&
   \text{ for the Gauss TMD.}
\end{align}
Our default grid parameters are
\begin{align}
   \label{tmd-param-default}
   z_1 &= 0.05 \,,
   &
   r_\kappa &= 3
   &&
   \text{ for the toy and Yukawa TMDs,}
   \notag \\
   z_1 &= 0.05 \,,
   &
   r_\lambda &= 5
   &&
   \text{ for the Gauss TMD}
\end{align}
and have been chosen such that one obtains good integration accuracy for all
three $Q$ values in \eqref{Q-values}, see \tab{\ref{tab:best-params}} and the
plots in \sect{\ref{sec:params}}.

\begin{figure}
\centering
\subfloat[$z \ms W(z)$\,, \ toy TMD]{
   \includegraphics[width=0.47\textwidth,trim=20 0 40 60,clip]{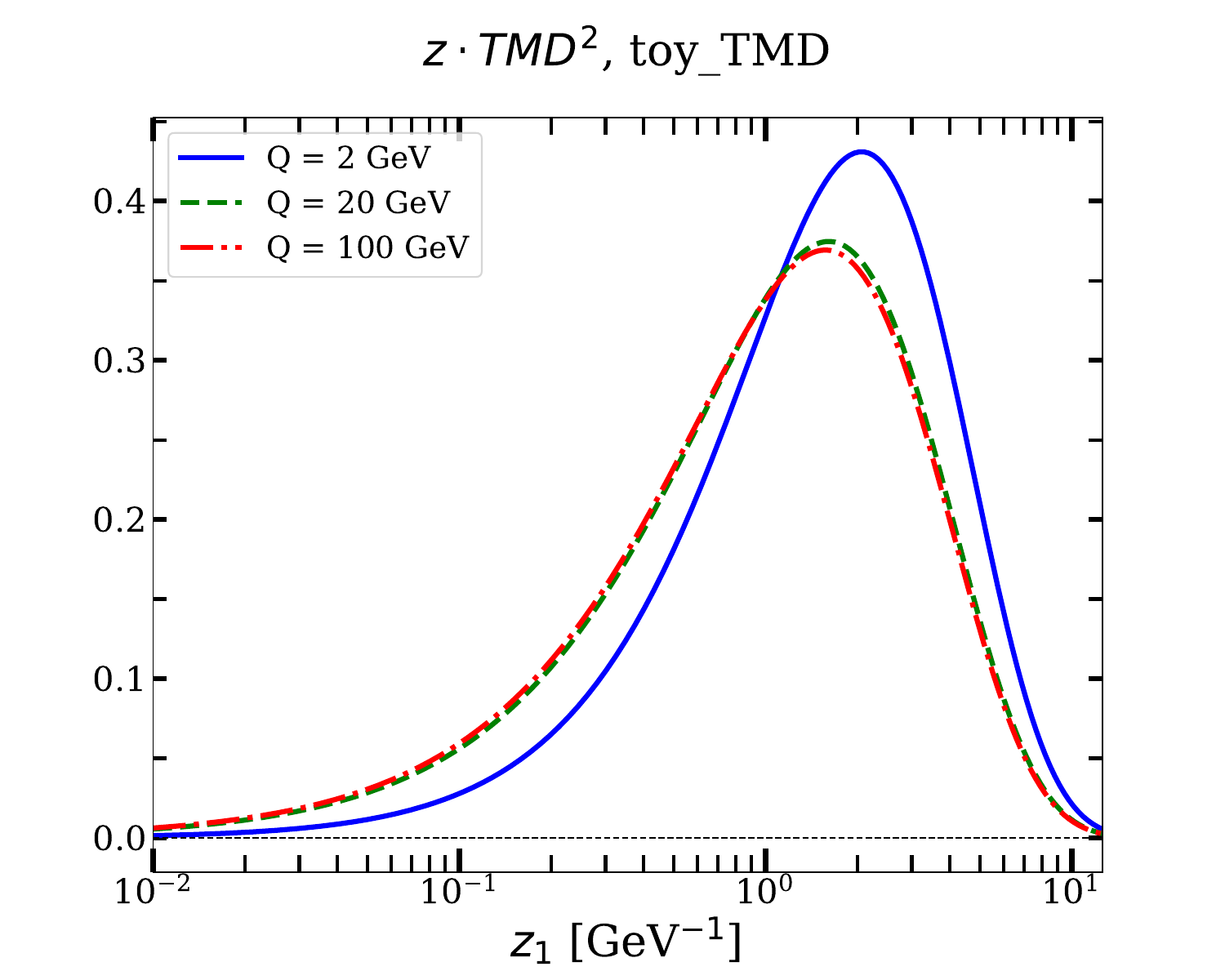}
}
\subfloat[$f_1(z)$\,, \ toy TMD]{
   \includegraphics[width=0.47\textwidth,trim=20 0 40 60,clip]{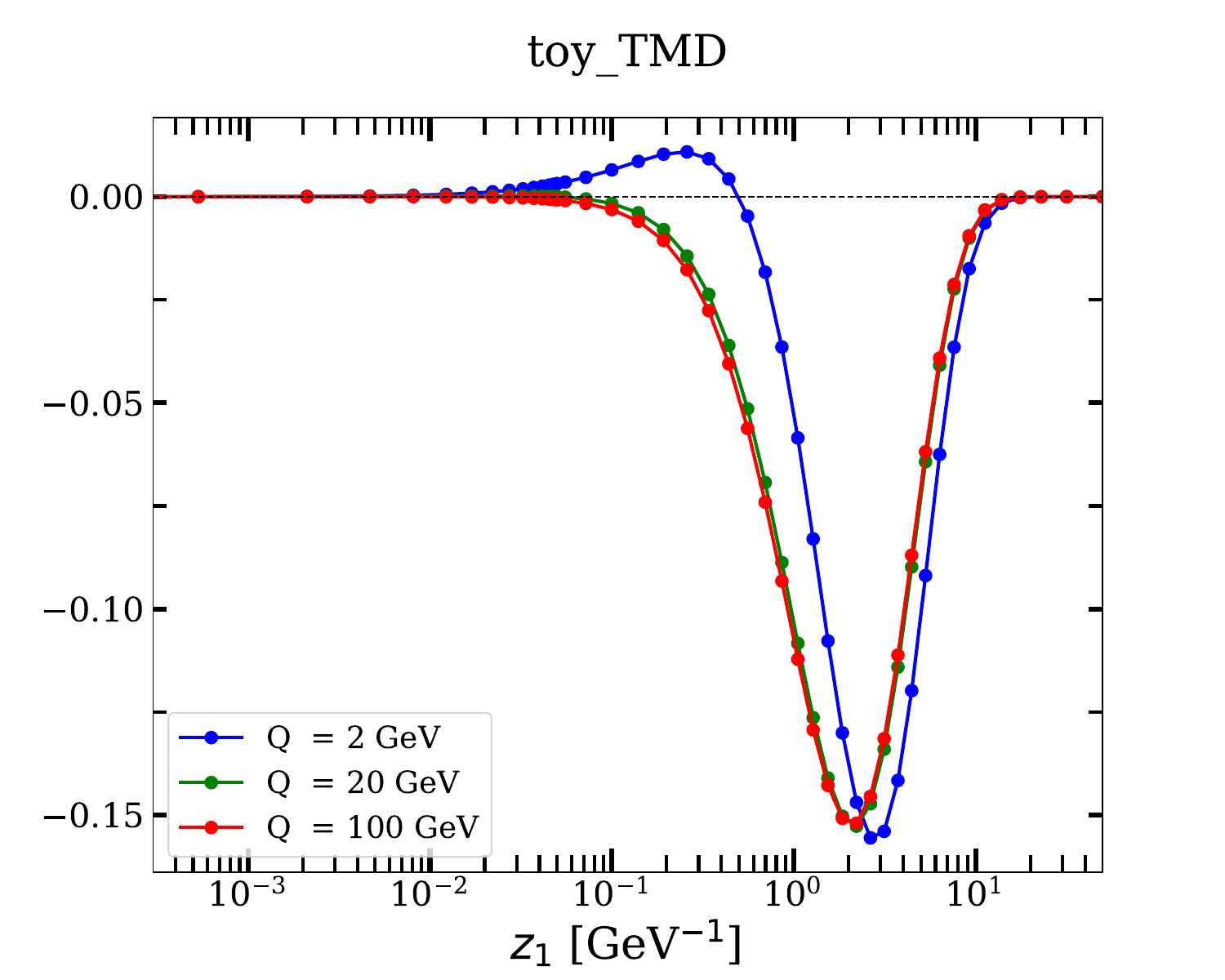}
}
\\[1.5em]
\subfloat[$z \ms W(z)$\,, \ Yukawa TMD]{
   \includegraphics[width=0.47\textwidth,trim=20 0 40 60,clip]{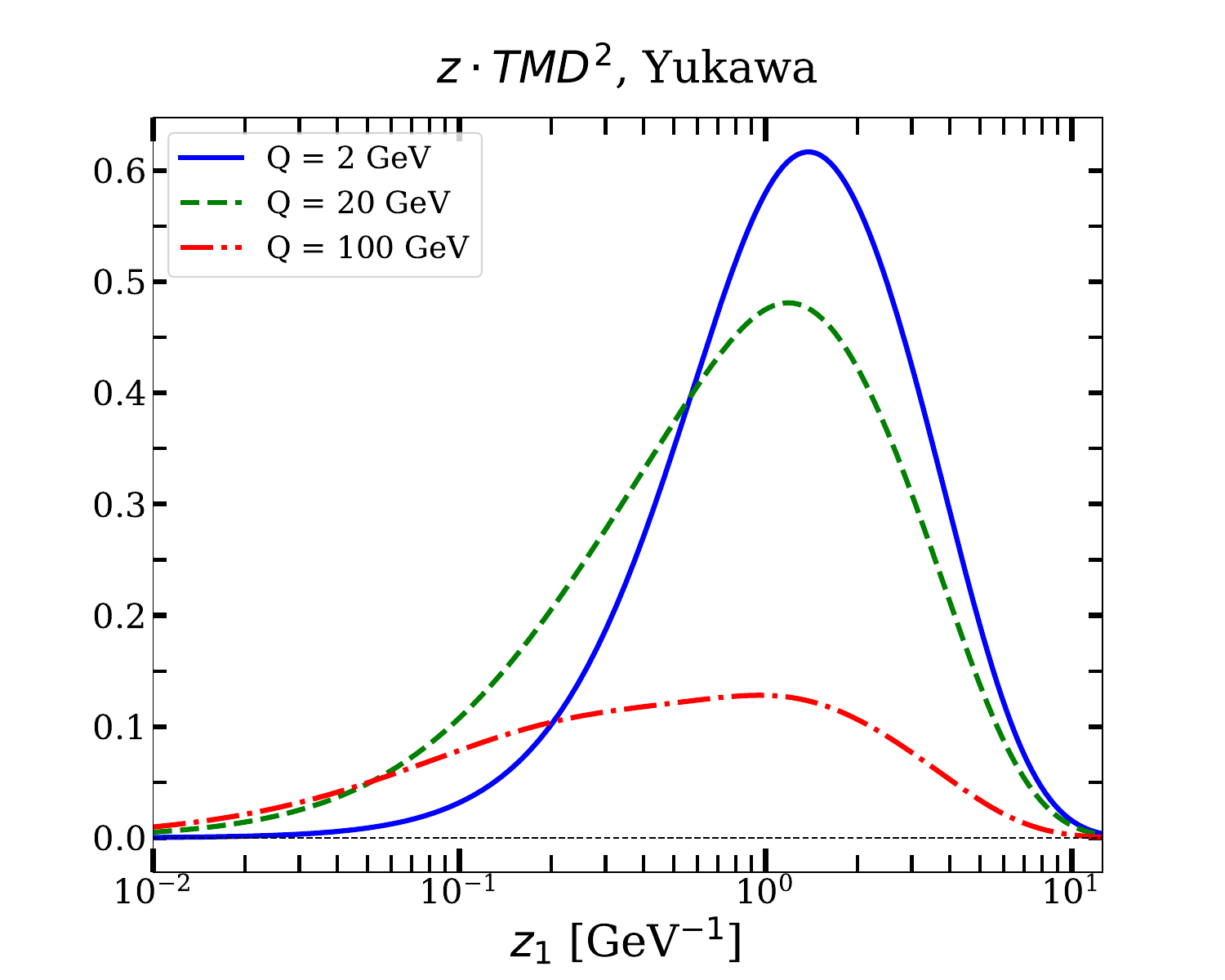}
}
\subfloat[$f_1(z)$\,, \ Yukawa TMD]{
   \includegraphics[width=0.47\textwidth,trim=20 0 40 60,clip]{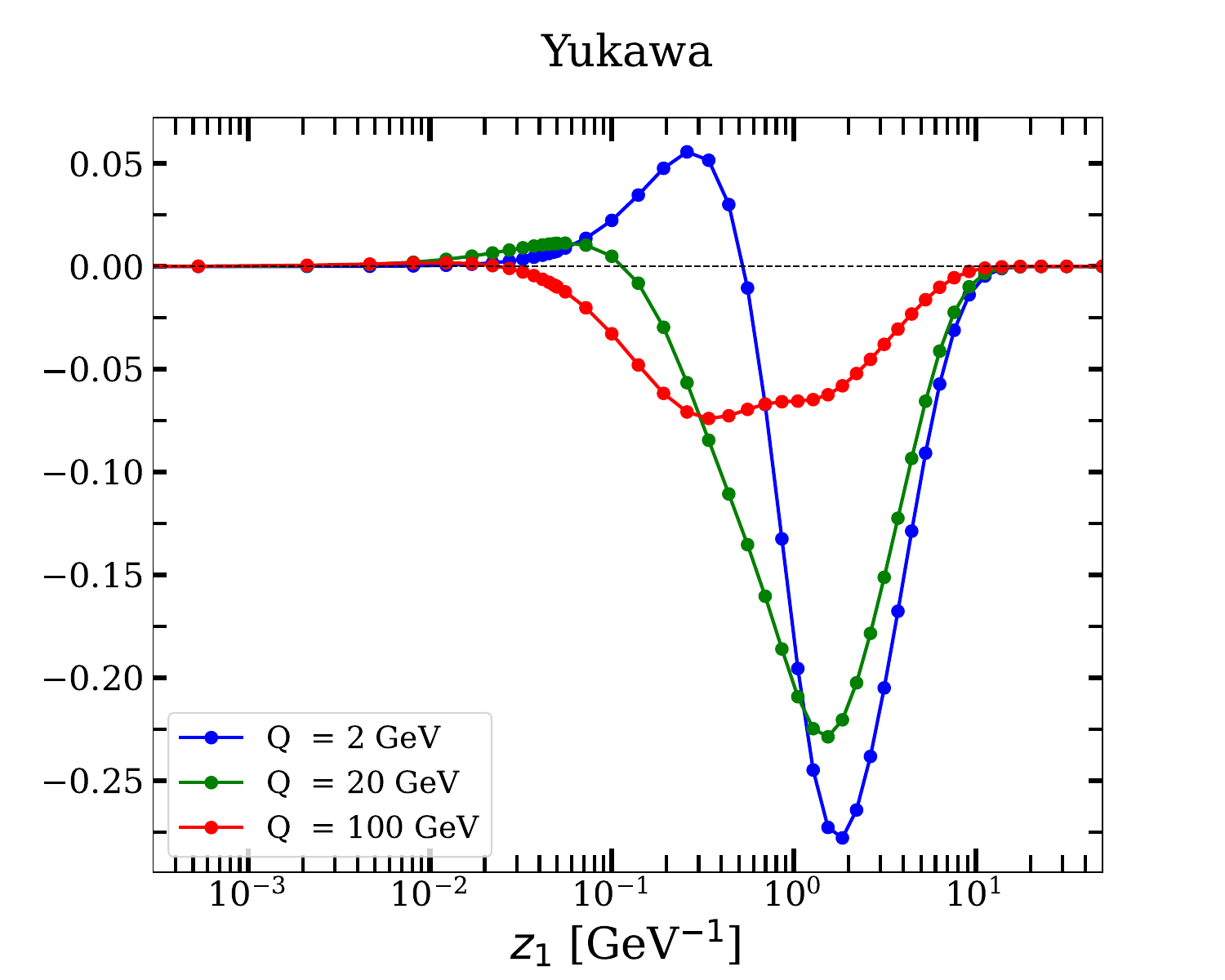}
}
\\[1.5em]
\subfloat[$z \ms W(z)$\,, \ Gauss TMD]{
   \includegraphics[width=0.47\textwidth,trim=20 0 40 60,clip]{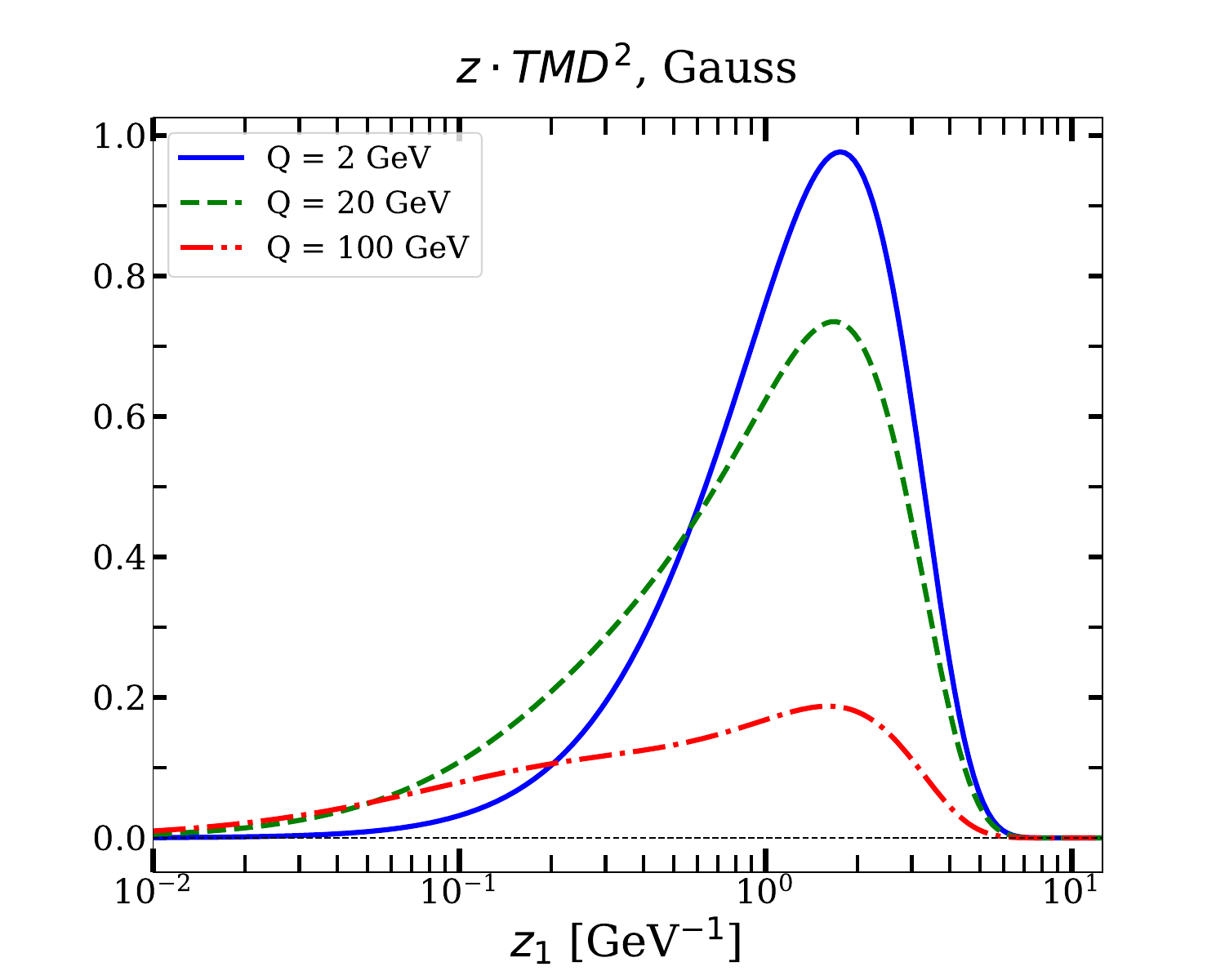}
}
\subfloat[$f_1(z)$\,, \ Gauss TMD]{
   \includegraphics[width=0.47\textwidth,trim=20 0 40 60,clip]{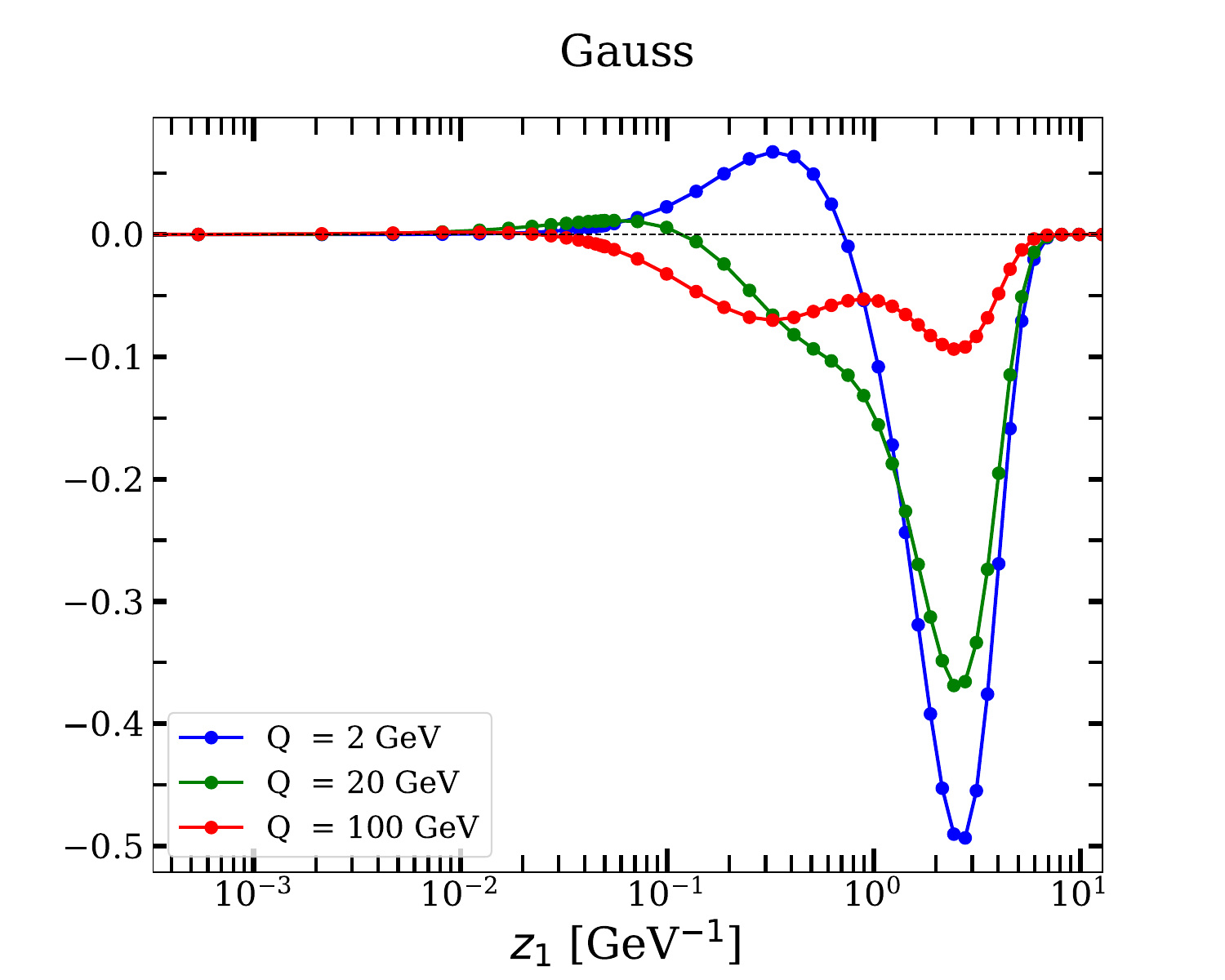}
}
\caption{\label{fig:integrands}  Left: the transformed function $z \ms W(z)$ in
\eqref{tmd-int} for different values of $Q$ and the three choices of TMDs
specified in the text. Right: the corresponding function $f_1(z)$ in our method,
see \eqs{\protect\eqref{int-final-def}} and \protect\eqref{IBP}.  The points on
the curves correspond to the interpolation grids specified by
\protect\eqref{tmd-grid-default} and \protect\eqref{tmd-param-default}.}
\end{figure}

The functions $z \ms W(z)$, which multiply $J_0(q z)$ in the integral $I(q)$, are
plotted on the l.h.s.\ of \fig{\ref{fig:integrands}}.  When computing $I(q)$,
integration by parts is used as specified in \eqref{IBP}, and the resulting
function $f_1(z)$ that appears in the master equation of the Levin method
\eqref{int-final-def} is shown on the r.h.s.\ of the figure.  We see non-trivial
structures of $f_1(z)$ appearing as a consequence of taking the derivative and
weighting with powers of $z/(1+z)$. The points on the curves for $f_1(z)$
correspond to the interpolation grids just specified.

\begin{figure}[t]
\centering
\subfloat[$|I(q)|$\,, \ toy TMD]{
   \includegraphics[width=0.48\textwidth,trim=40 0 40 75,clip]{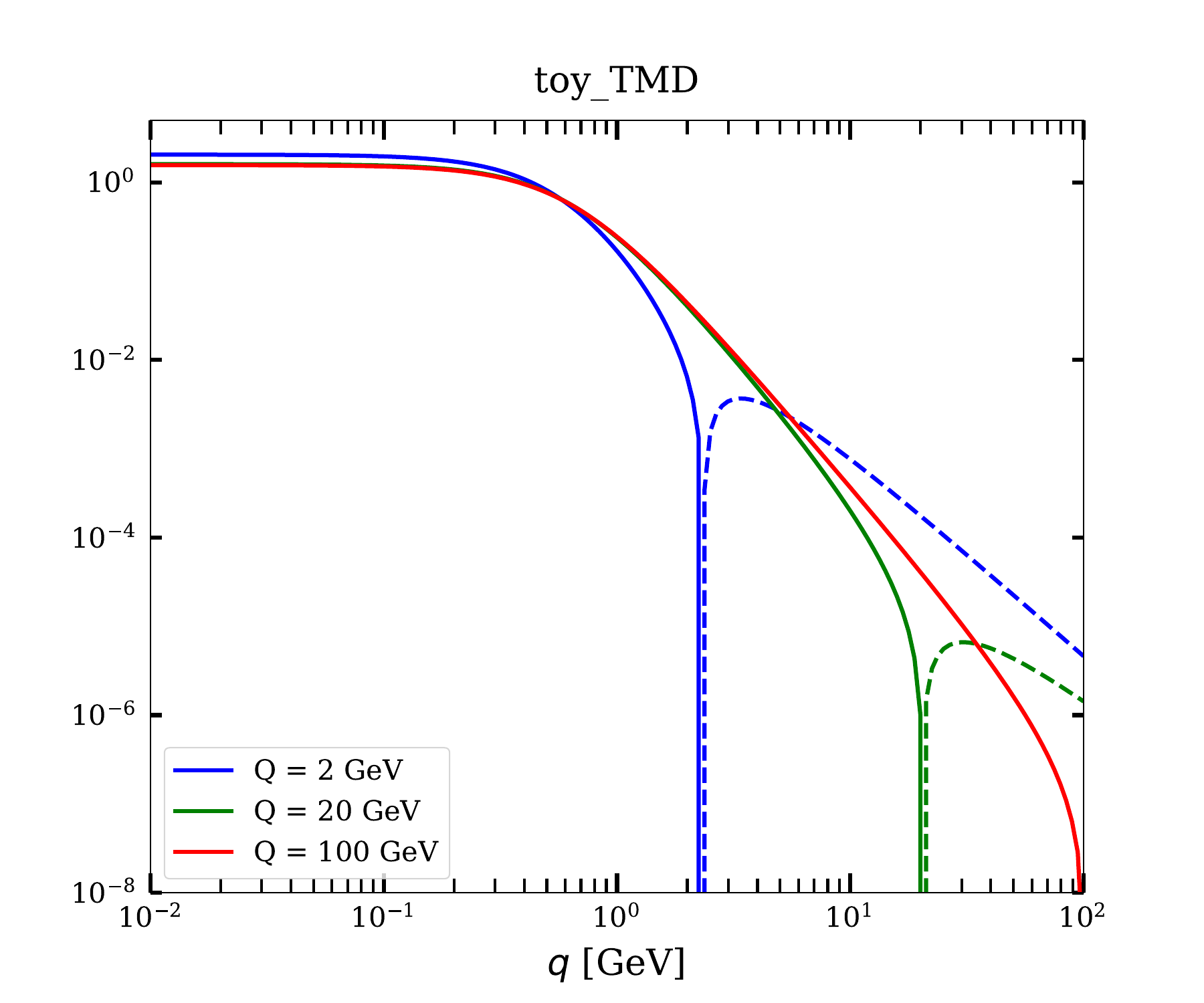}
}
\subfloat[$|I(q)|$\,, \ Yukawa TMD]{
   \includegraphics[width=0.48\textwidth,trim=40 0 40 75,clip]{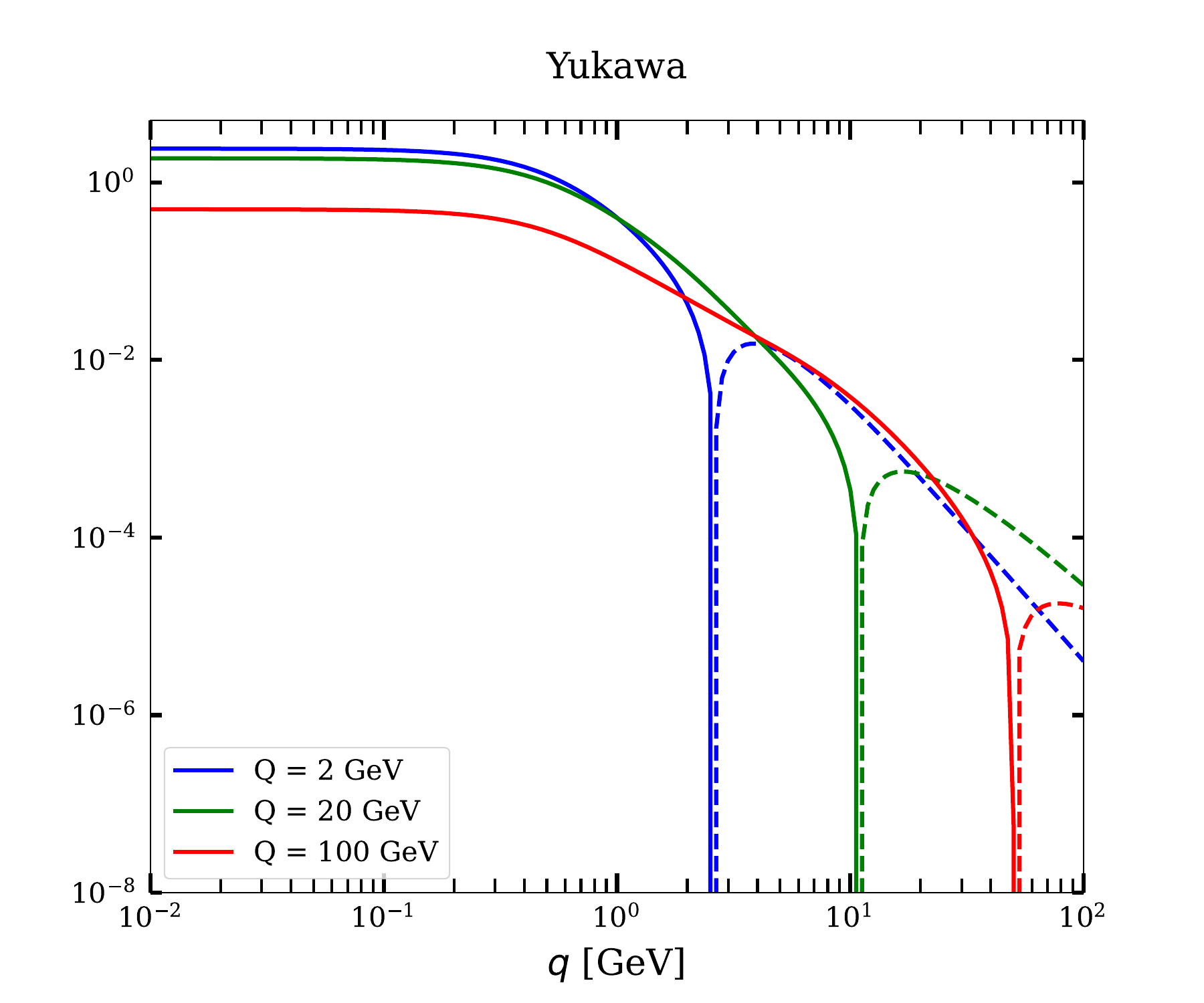}
}
\\[1.5em]
\subfloat[$|I(q)|$\,, \ Gauss TMD]{
   \includegraphics[width=0.48\textwidth,trim=40 0 40 75,clip]{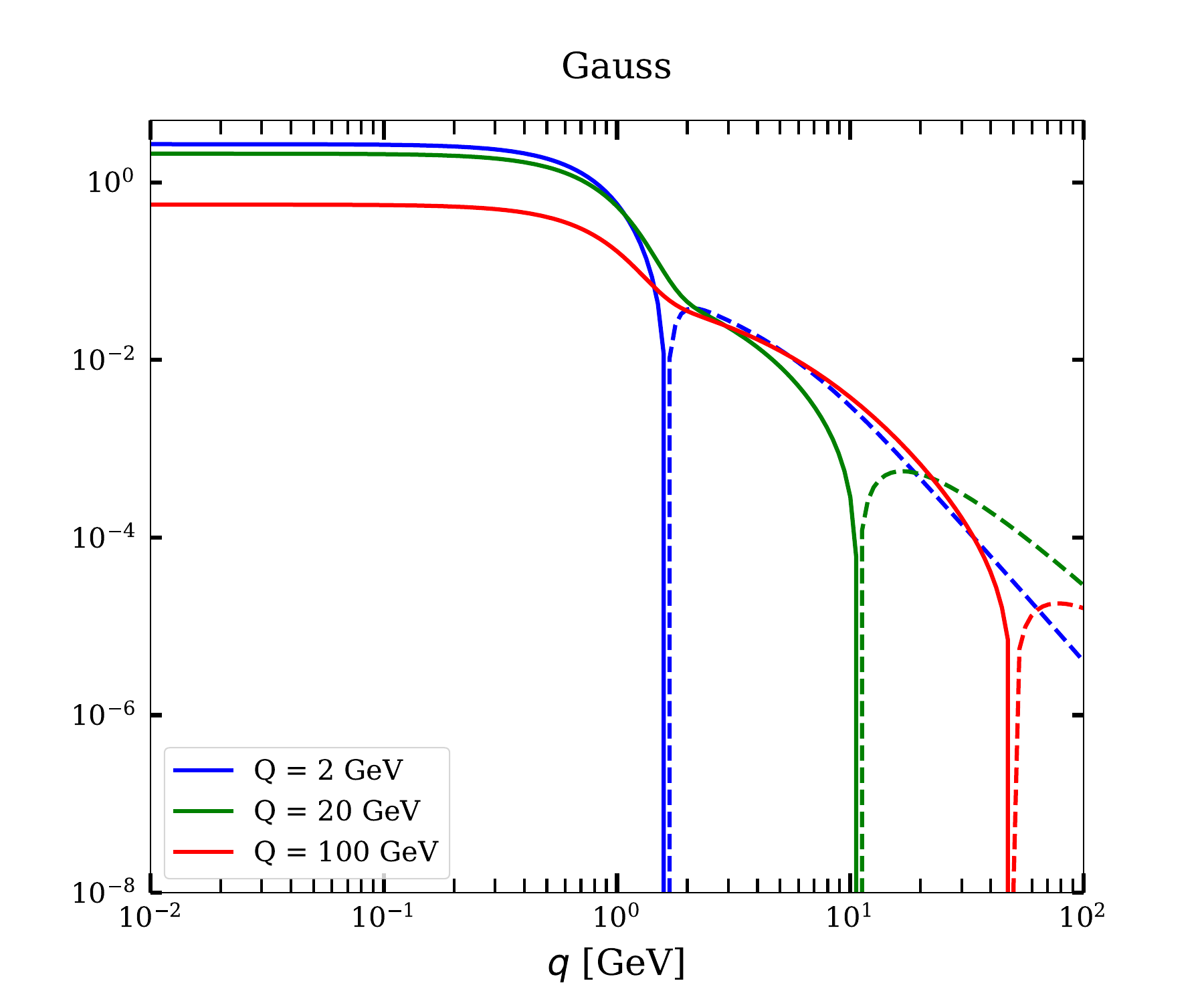}
}
\caption{\label{fig:bessel_transforms}  Absolute values of the Fourier-Bessel
transform $I(q)$ for the functions $z \ms W(z)$ shown on the l.h.s.\ of
\protect\fig{\ref{fig:integrands}}.  Dashed lines indicate that $I(q)$ is
negative.  The curve for $Q = 100 \gev$ in panel (a) has a zero crossing at $q$
between $100$ and $101 \gev$.}
\end{figure}

The Fourier-Bessel transform $I(q)$, which gives the transverse-momentum
spectrum of the TMD process, is shown in \fig{\ref{fig:bessel_transforms}}.  We
observe that the transform becomes negative at some value $q \sim Q$.  This is a
clear indication that TMD factorisation does not describe the cross section in
such a region.  We recall from the introduction that the evaluation of $I(q)$
may however be necessary even in that region, in order to achieve a smooth
transition between the TMD regime and the regime in which the cross section can
be computed in terms of collinear parton distributions and fixed-order
perturbative cross sections.

\subsection{Choosing interpolation grids}
\label{sec:grid-choice}

We now compare the performance of different settings for the interpolation grids
in our method.  We generally find that high accuracy is easier to reach for
$K(q)$ than for $I(q)$, which is why we mostly focus on $I(q)$ in the following.

To quantify the performance of a particular grid setting for a given choice of
$W(z)$, we evaluate the Fourier-Bessel integrals for a set of logarithmically
spaced values $q_i$ between $0.01 \gev$ and $100 \gev$.  We then determine the
root mean square of the relative integration error $\varepsilon$, averaging over
all $q_i$ that are not too close to the point $q_0$ where $I(q)$ has a zero
crossing.
Specifically, we define
\begin{align}
   \label{rms-err-def}
   \varepsilon^2_{\text{rms}}
   &=
   \langle \varepsilon^2 \rangle_{q_i \notin R}
   &
   \text{ with } R &= (q_0 - \Delta, q_0 + \Delta)
   \,,
\end{align}
where
\begin{align}
   \label{Delta-def}
   \Delta
   &=
   \begin{cases}
      1 \gev & \text{ for } Q = 2 \gev ,
      \\
      5 \gev & \text{ for } Q = 20 \gev ,
      \\
      10 \gev & \text{ for } Q = 100 \gev .
   \end{cases}
\end{align}
To compute the integration error, we take the analytic form of the
Fourier-Bessel transform for the toy TMD (see \tab{\ref{tab:functions}}) and the
result of an adaptive Gauss-Kronrod integrator for the Yukawa and Gauss
TMDs.\footnote{We use the routine \texttt{scipy.integrate.quad} of SciPy version
1.7.3, which employs the \texttt{QAGS} algorithm of the QUADPACK library
\protect\cite{QUADPACK}.  The relative target accuracy is set to $10^{-12}$.}


\subsubsection{Variable transformations}
\label{sec:transform-choice}

We begin with a comparison between different variable transformations, fixing
the grid structure to the form $[0, z_1, \infty]_{16, 32}$.  For each
transformation we need to choose parameters $(z_1, r_\kappa)$ or $(z_1,
r_\lambda)$.   We do this by performing a parameter scan in these values, with
$z_1 \in [0.01, 0.3]$ in steps of $0.01$ and with $r_\kappa$ or $r_\lambda$
varying from $0.25$ to $8$ in steps of $0.25$.  The parameters that give the
smallest value of $\varepsilon_{\text{rms}}$ are listed in
\tab{\ref{tab:best-params}}.

\begin{table}
\centering
\begin{tabular}{c | c | c c c}
\toprule
\multicolumn{5}{c}{toy TMD}
\\
\midrule
           & \texttt{exp}   & \multicolumn{3}{| c}{\texttt{exp sqrt}}
\\
$Q [\gev]$ & $(z_1, r_\kappa)$
           & $r_\kappa$ & $(z_1, r_\kappa)$ & $(z_1, z_2, r_\kappa)$
\\
\midrule
2   & (0.15, 1.0)
    & 7.0 & \phantom{1}(0.1, 2.0)  & \phantom{1}(0.1, 1.0, 1.5)
\\
20  & (0.15, 1.0)
    & 4.0 & \phantom{1}(0.1, 2.0)  & \phantom{1}(0.1, 1.5, 1.5)
\\
100 & (0.05, 1.5)
    & 6.0 & (0.05, 3.0)            & (0.05, 0.5, 1.0)
\\
\bottomrule
\end{tabular}

\vspace{2em}

\begin{tabular}{c | c | c c c}
\toprule
\multicolumn{5}{c}{Yukawa TMD}
\\
\midrule
           & \texttt{exp}   & \multicolumn{3}{| c}{\texttt{exp sqrt}}
\\
$Q [\gev]$ & $(z_1, r_\kappa)$
           & $r_\kappa$ & $(z_1, r_\kappa)$ & $(z_1, z_2, r_\kappa)$
\\
\midrule
2   & \phantom{1}(0.1,\ 1.5)
    & \phantom{1}8.0 & \phantom{1}(0.1,\ 2.0)  & \phantom{1}(0.1,\ 1.0,\ 1.5)
\\
20  & (0.15,\ 1.5)
    & \phantom{1}8.0 & \phantom{1}(0.1,\ 2.0)  & \phantom{1}(0.1, 0.8, 1.5)
\\
100 & (0.05,\ 1.0)
    & 10.0           & (0.05,\ 4.0)            & (0.05, 0.8, 6.0)
\\
\bottomrule
\end{tabular}

\vspace{2em}

\begin{tabular}{c | c | c c c}
\toprule
\multicolumn{5}{c}{Gauss TMD}
\\
\midrule
           & \texttt{Gauss} & \multicolumn{3}{| c}{\texttt{exp}}
\\
$Q [\gev]$ & $(z_1, r_\lambda)$
           & $r_\lambda$ & $(z_1, r_\lambda)$ & $(z_1, z_2, r_\lambda)$
\\
\midrule
2   & (0.15,\ 1.5)
    & 8.0 & \phantom{1}(0.1,\ 3.0) & \phantom{1}(0.2, 3.0, 5.0)
\\
20  & (0.15,\ 1.5)
    & 7.0 & \phantom{1}(0.1,\ 4.0) & \phantom{1}(0.1, 1.0, 4.0)
\\
100 & (0.05,\ 2.0)
    & 4.0 & (0.05,\ 5.0)           & (0.15, 1.2, 7.0)
\\
\bottomrule
\end{tabular}
\caption{\label{tab:best-params}  Grid parameter values used for the plots in
\figs{\protect\ref{fig:grid-transforms}} to \protect\ref{fig:2_vs_3_subgrids}.
These values give the best average integration error $\varepsilon_{\text{rms}}$
in a parameter scan.  The labels \texttt{exp}, \texttt{exp sqrt}, and
\texttt{Gauss} refer to the variable transformations in
\tab{\protect\ref{tab:var-trf}}. The corresponding grids have the structure $[0,
\infty]_{(48)}$ for one parameter, $[0, z_1, \infty]_{(16,32)}$ for two
parameters, and $[0, z_1, z_2, \infty]_{(16,16,16)}$ for three parameters.
$z_1$ and $z_2$ are given in units of $\gev^{-1}$.}
\end{table}

\begin{figure}
\centering
\subfloat[Toy TMD]{
   \includegraphics[width=0.98\textwidth,trim=10 0 0 0,clip]{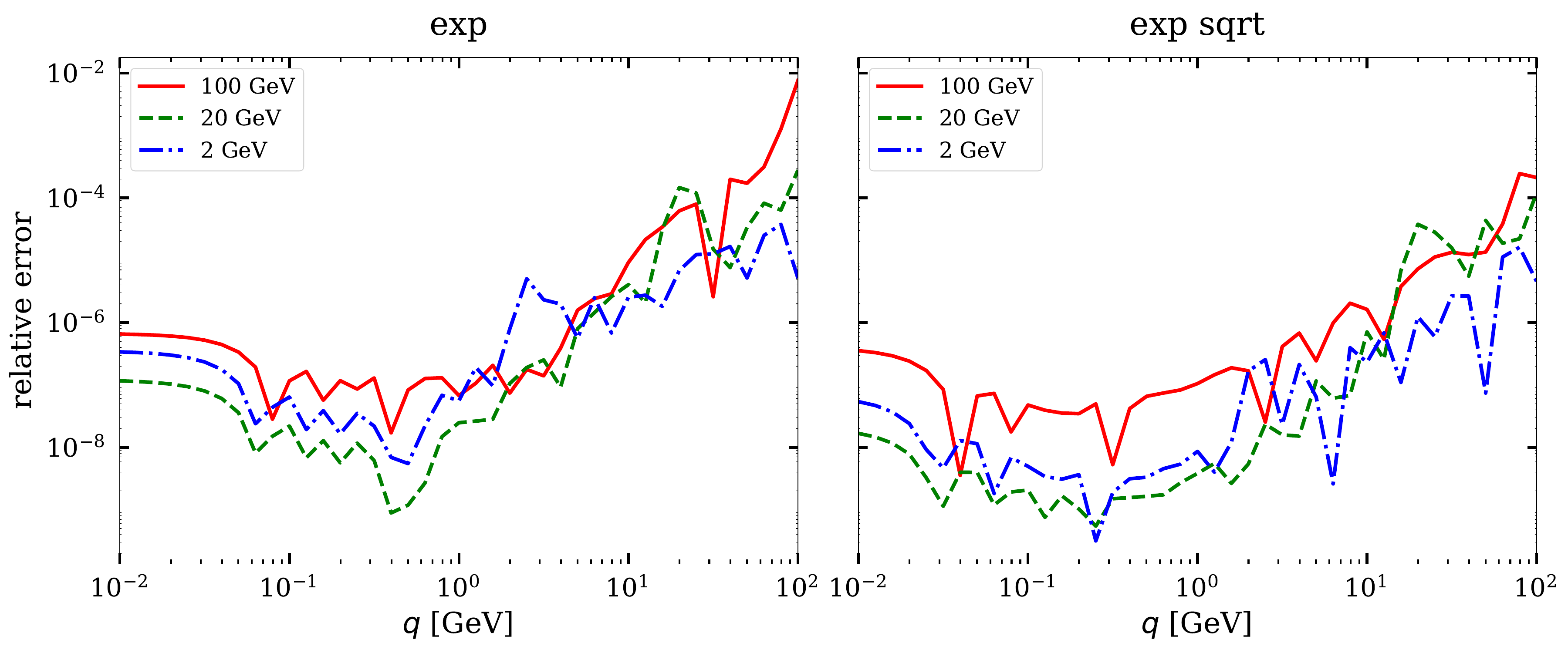}
}
\\[1.5em]
\subfloat[Yukawa TMD]{
   \includegraphics[width=0.98\textwidth,trim=10 0 0 0,clip]{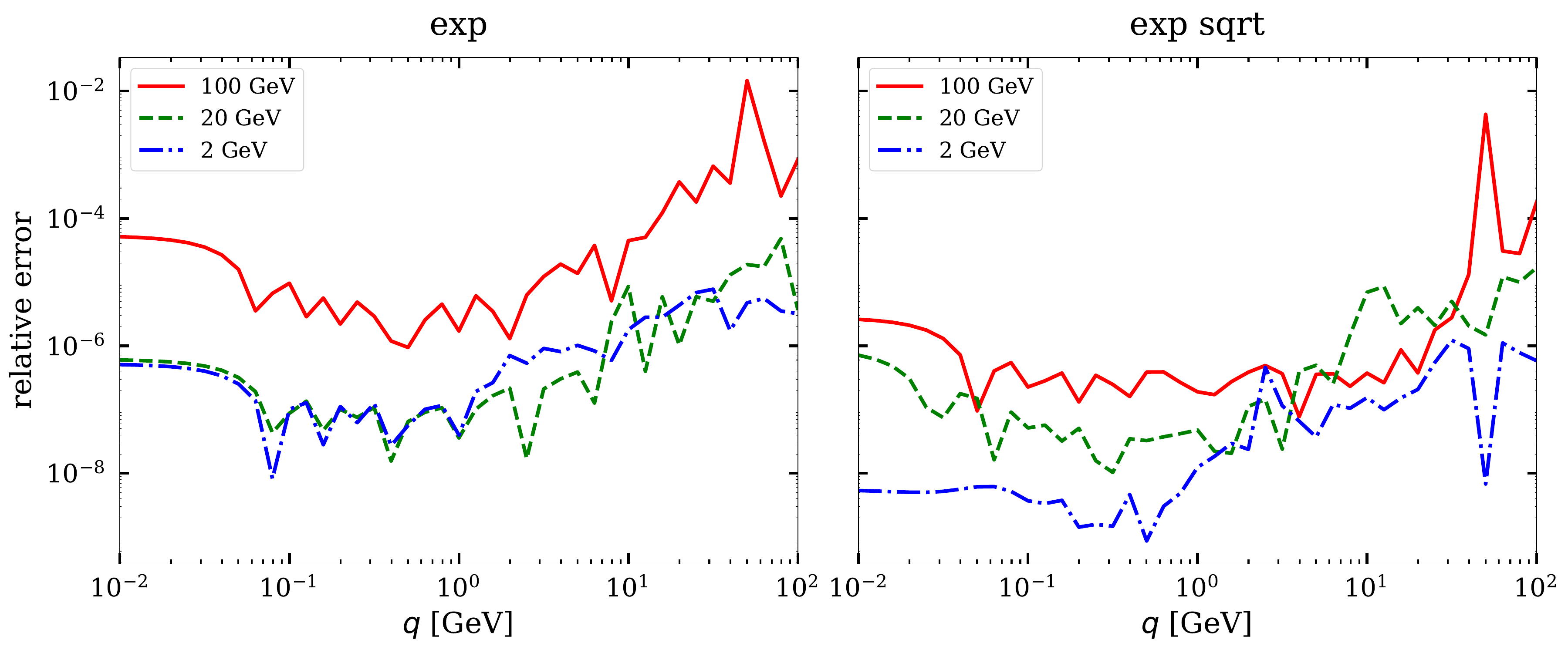}
}
\\[1.5em]
\subfloat[Gauss TMD]{
   \includegraphics[width=0.98\textwidth,trim=10 0 0 0,clip]{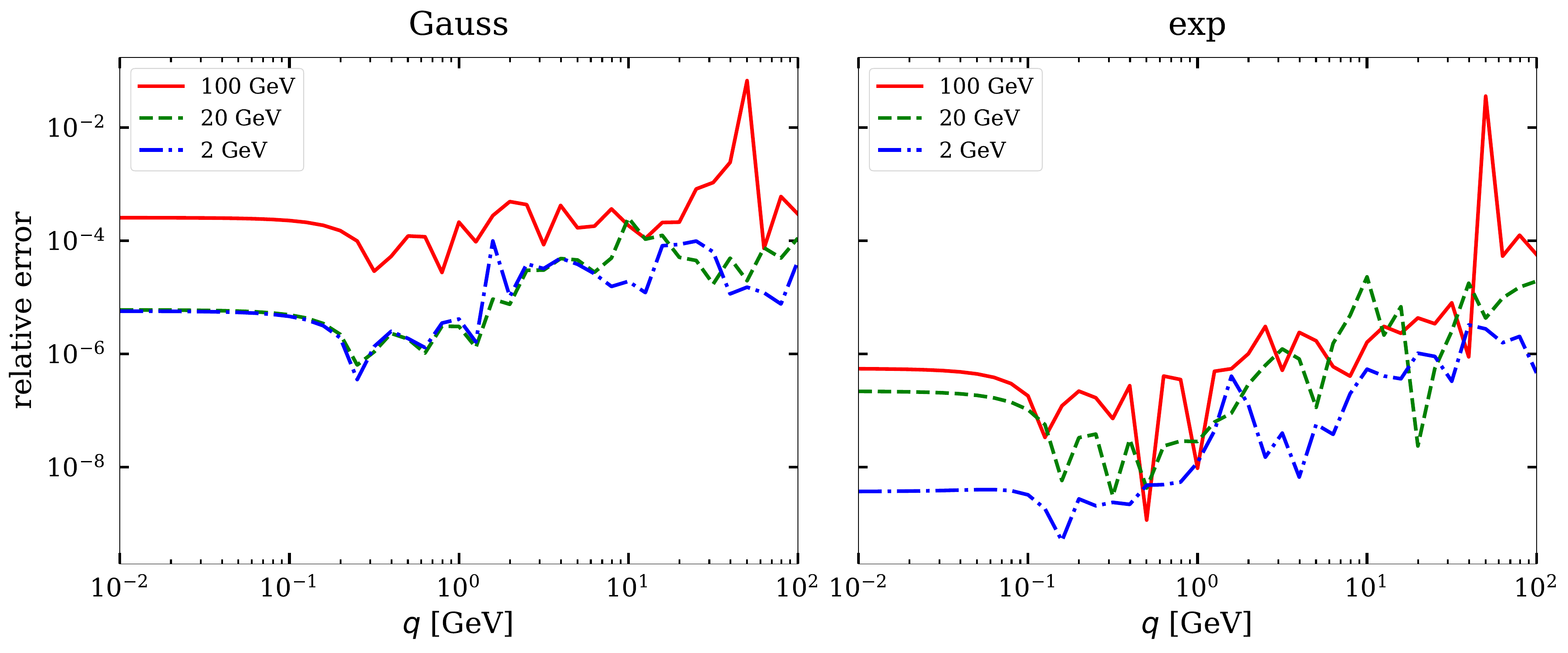}
}
\caption{\label{fig:grid-transforms}  Relative integration error
achieved with different variable transforms on grids of the form $[0, z_1,
\infty]_{(16, 32)}$.  Details are given in the text.}
\end{figure}

The relative integration error for the best parameter setting is shown in
\fig{\ref{fig:grid-transforms}} as a function of $q$.  We see that for the TMDs
with an exponential falloff (toy and Yukawa TMD) the \texttt{exp sqrt} transform
performs somewhat better than the \texttt{exp} transform.  In turn, for the
Gauss TMD , the \texttt{exp} transform performs much better than the
\texttt{Gauss} transform.  We note that in both cases, the preferred
transformation is the one that leads to a stronger decrease of the integrand for
$u \to 0$ (which corresponds to $z \to \infty$).  We use this preferred
transformation in all subsequent comparisons.

\subsubsection{Number of subgrids}
\label{sec:subgrid-choice}

Let us now compare the accuracy of our algorithm with one, two, or three
subgrids for a comparable total number of points.  As in the previous
subsection, we determine the best parameter settings for each grid by a
parameter scan; the resulting values are given in \tab{\ref{tab:best-params}}.
We note in passing that for a single subgrid, the preferred values of $r_\kappa$
or $r_\lambda$ are much larger than with two or more subgrids.

\begin{figure}
\centering
\subfloat[toy TMD]{
   \includegraphics[width=0.97\textwidth,trim=10 0 0 0,clip]{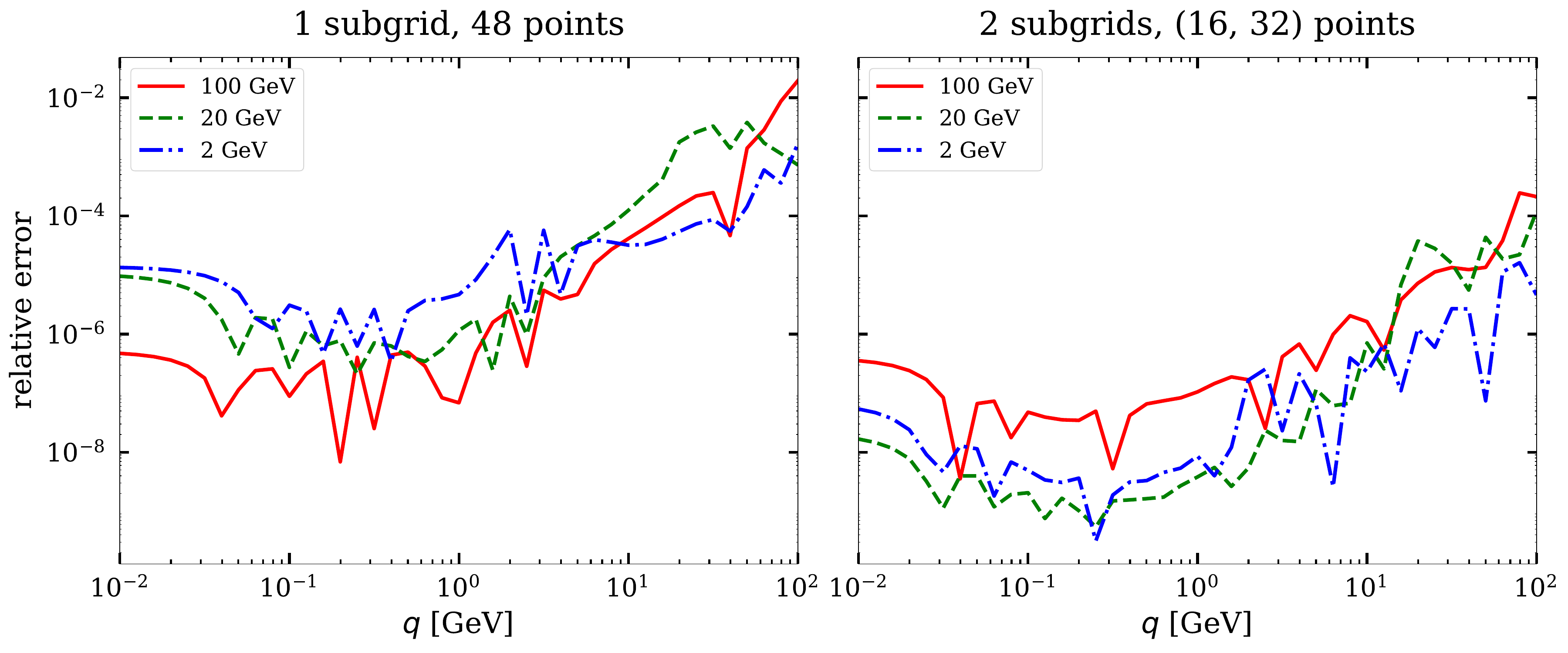}
}
\\[1.5em]
\subfloat[Yukawa TMD]{
   \includegraphics[width=0.97\textwidth,trim=10 0 0 0,clip]{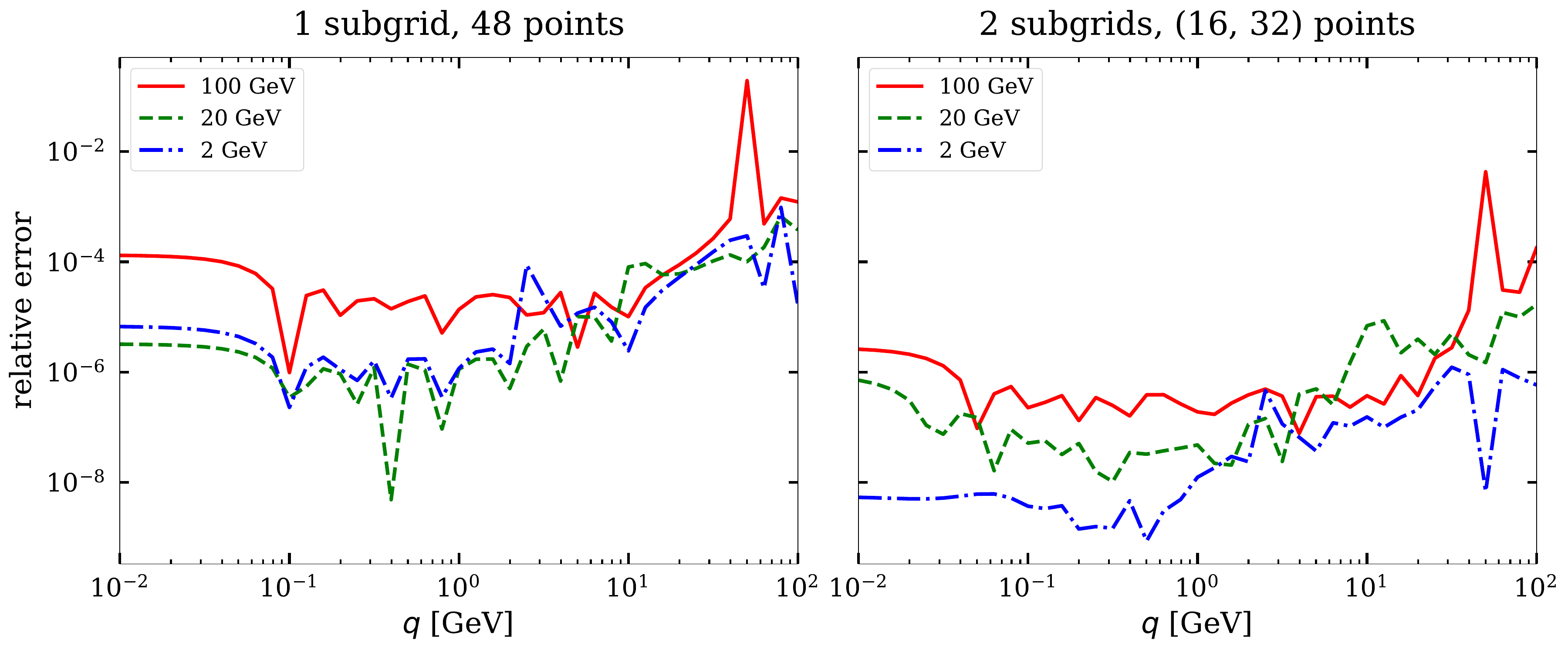}
}
\\[1.5em]
\subfloat[Gauss TMD]{
   \includegraphics[width=0.97\textwidth,trim=10 0 0 0,clip]{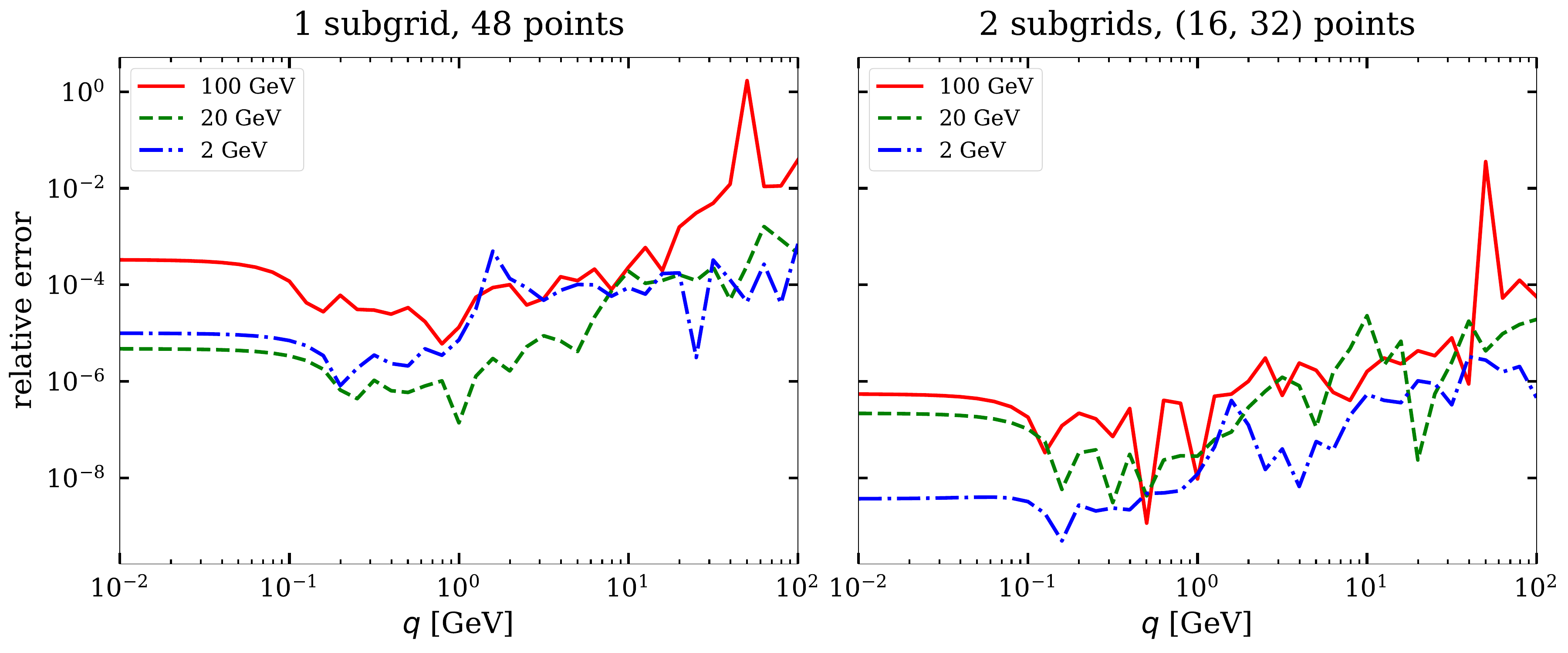}
}
\caption{\label{fig:1_vs_2_subgrids}  Relative integration error for grids of
the form $[0, \infty]_{(48)}$ (left) and $[0, z_1, \infty]_{(16, 32)}$ (right).
The variable transform for a given TMD is as in the right panels of
\fig{\protect{\ref{fig:grid-transforms}}}.}
\end{figure}

\begin{figure}
\centering
\subfloat[toy TMD]{
   \includegraphics[width=0.97\textwidth,trim=10 0 0 0,clip]{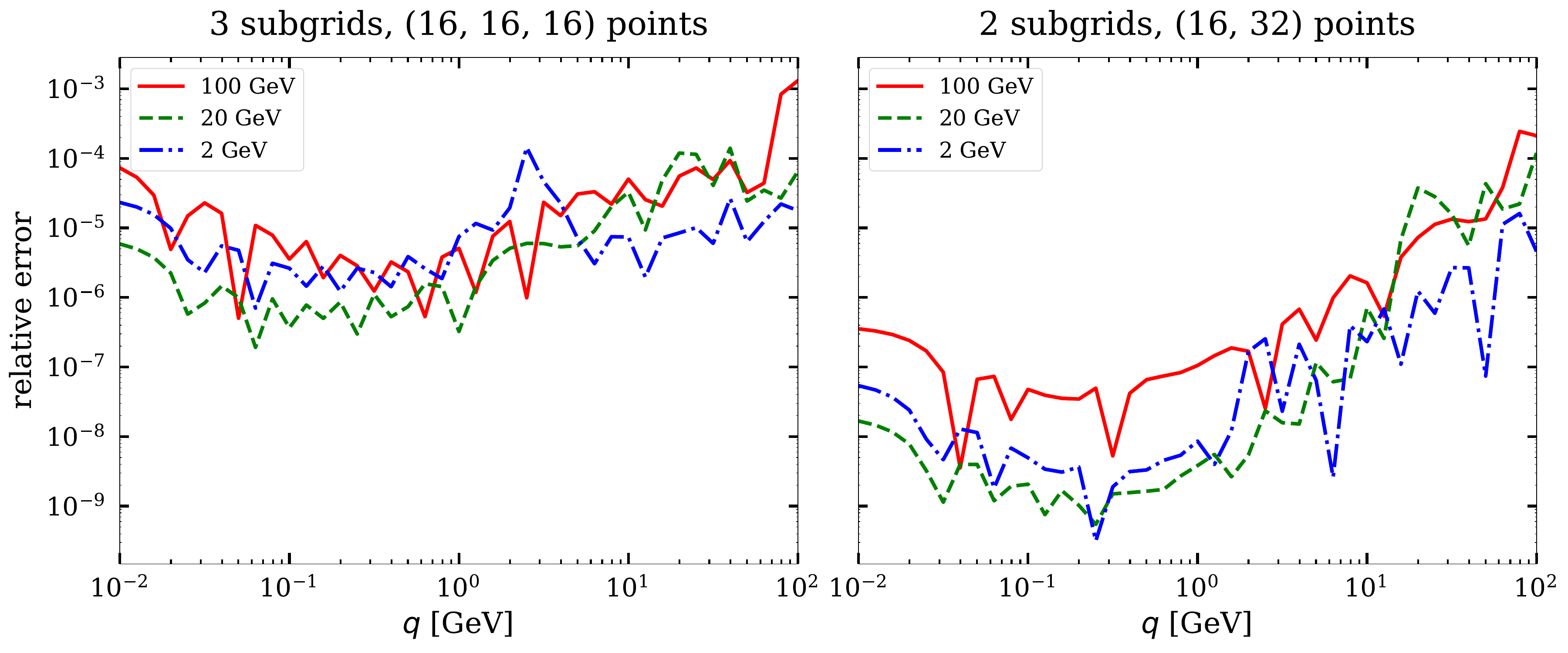}
}
\\[1.5em]
\subfloat[Yukawa TMD]{
   \includegraphics[width=0.97\textwidth,trim=10 0 0 0,clip]{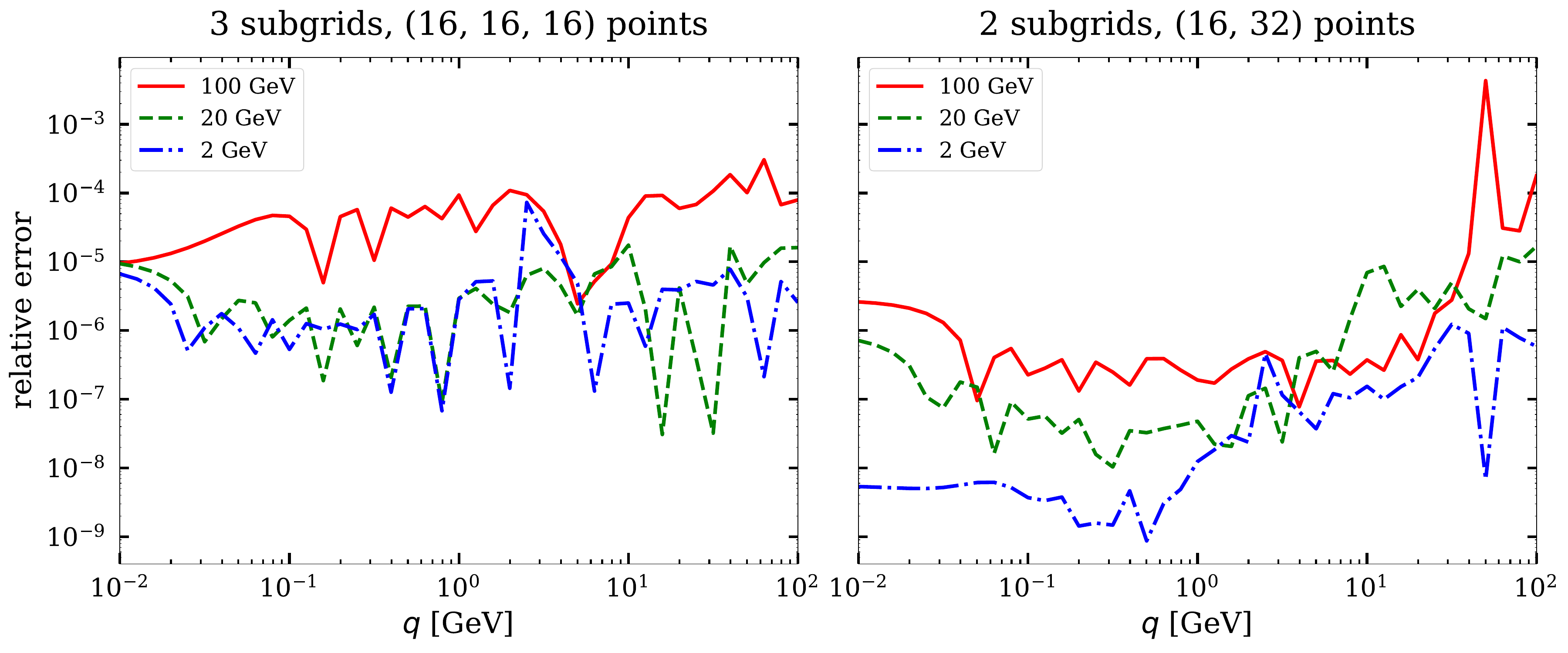}
}
\\[1.5em]
\subfloat[Gauss TMD]{
   \includegraphics[width=0.97\textwidth,trim=10 0 0 0,clip]{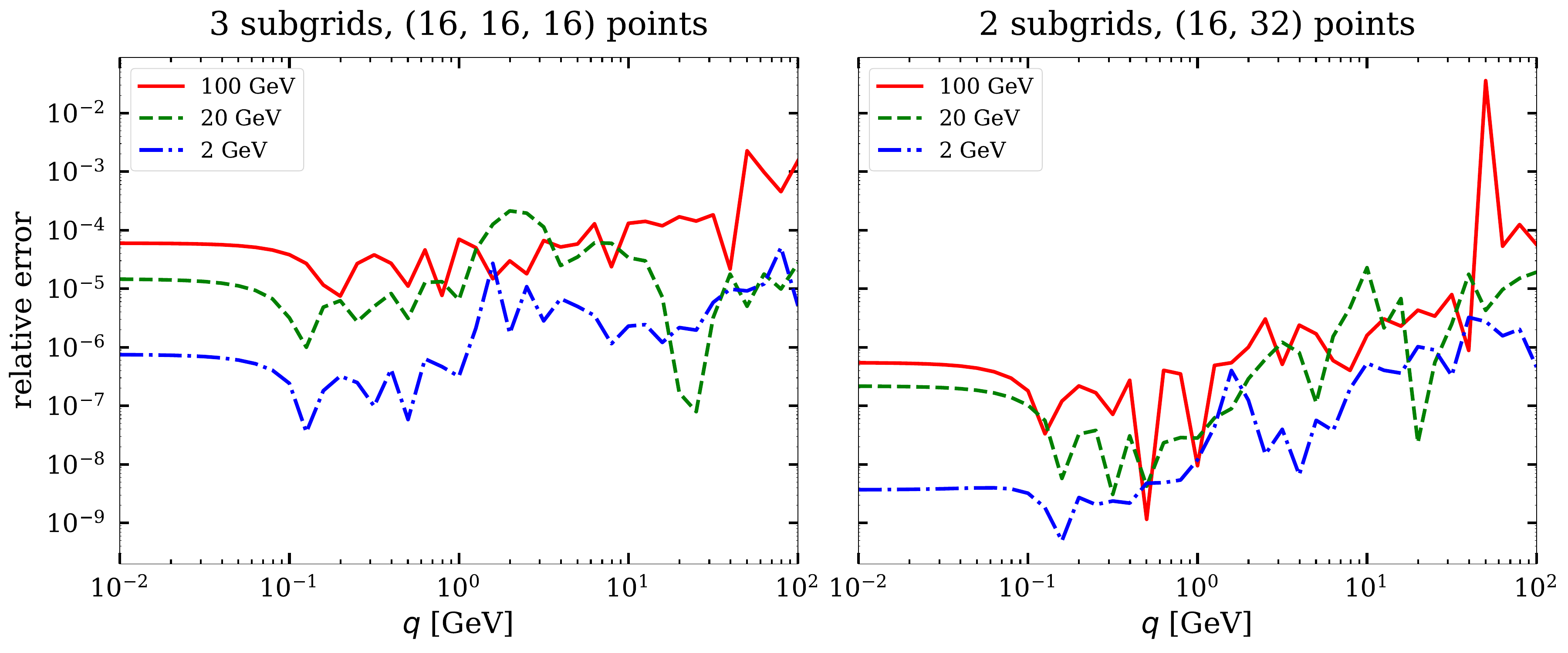}
}
\caption{\label{fig:2_vs_3_subgrids}  The same as
\fig{\protect{\ref{fig:1_vs_2_subgrids}}}, but for grids of the form $[0, z_1,
z_2, \infty]_{(16, 16, 16)}$ (left) and $[0, z_1, \infty]_{(16, 32)}$ (right).}
\end{figure}

More importantly, we find that for our integrands and $q$ values, two subgrids
are strongly preferred, as seen in \figs{\ref{fig:1_vs_2_subgrids}} and
\ref{fig:2_vs_3_subgrids}.  We hence adopt this setting for the subsequent
investigations.  We also tried settings with two subgrids and a different
partition $(n_1, n_2)$ for the number of grid points, finding no significant
improvement compared to the choice $(16, 32)$.

\subsubsection{Transformation parameters and number of grid points}
\label{sec:params}

Our method would be of limited practical use if its accuracy changed rapidly
with the choice of grid parameters, since this would require a detailed
parameter scan for each new integrand.  Fortunately, this is not the case, as we
will show now.

\begin{figure}
\centering
\subfloat[$Q = 2 \gev$, $I(q)$]{
   \includegraphics[width=0.46\textwidth]{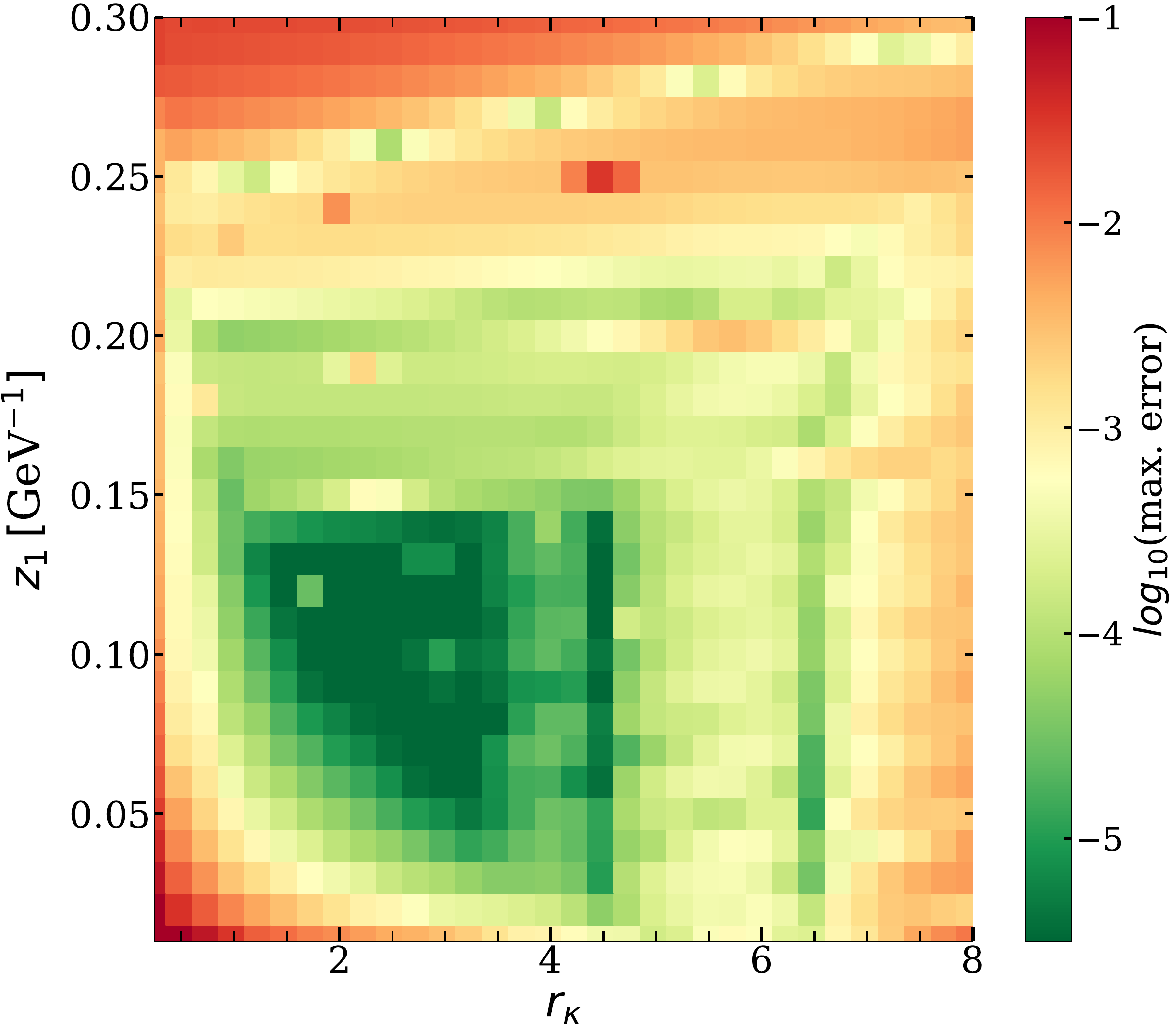}
}
\hspace{0.5em}
\subfloat[$Q = 2 \gev$, $K(q)$]{
   \includegraphics[width=0.46\textwidth]{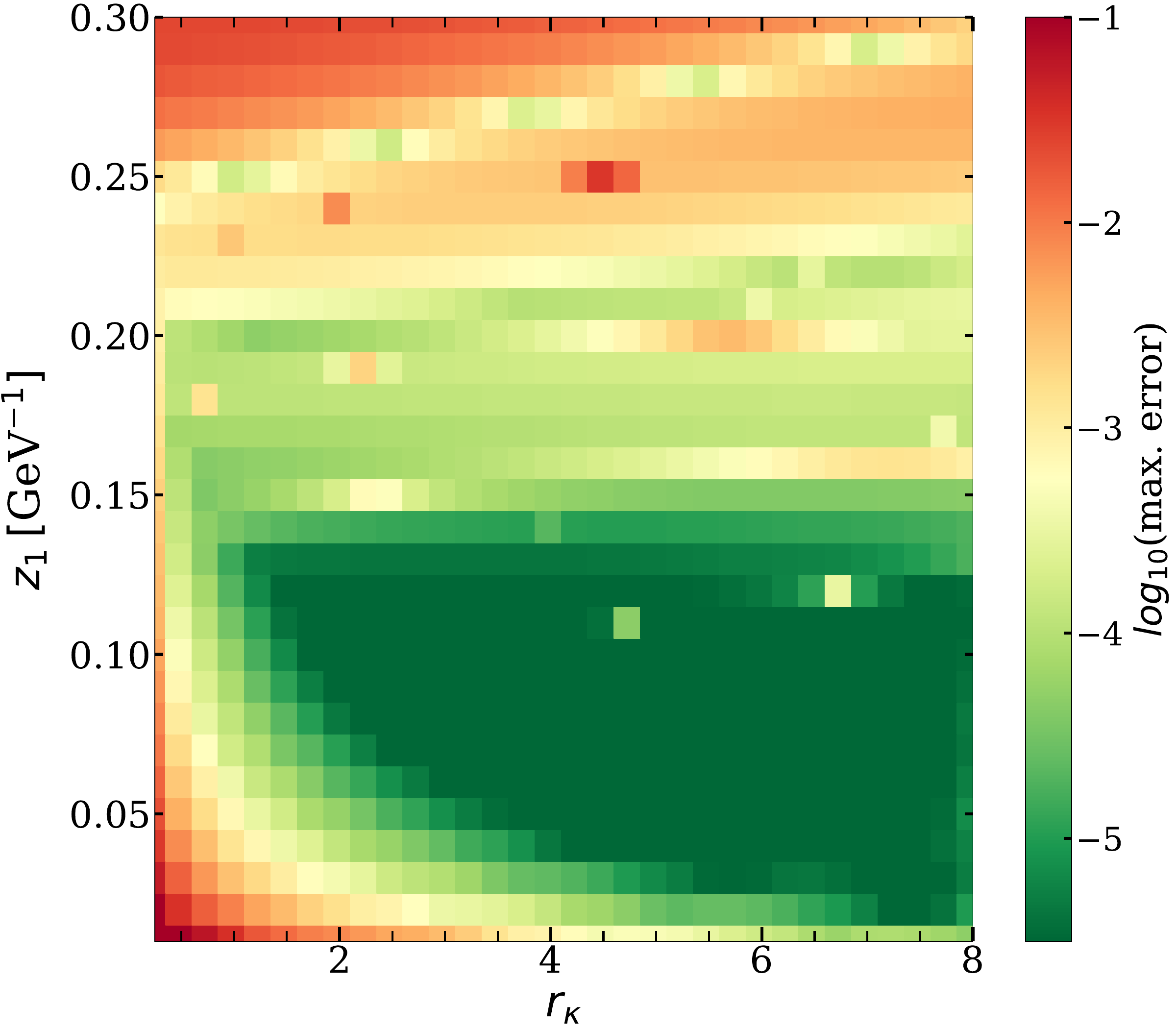}
}
\\[1.3em]
\subfloat[$Q = 20 \gev$, $I(q)$]{
   \includegraphics[width=0.46\textwidth]{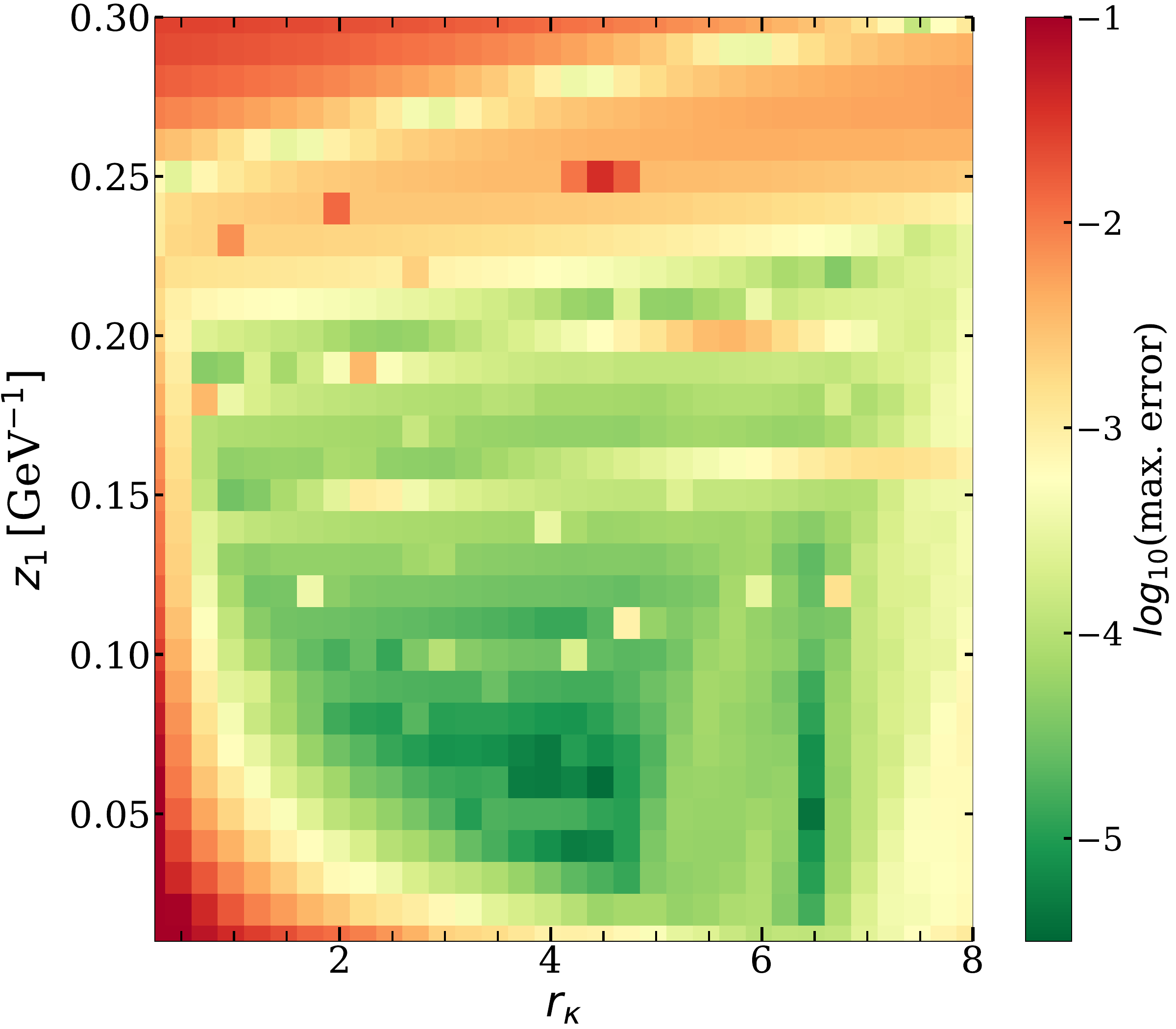}
}
\hspace{0.5em}
\subfloat[$Q = 20 \gev$, $K(q)$]{
   \includegraphics[width=0.46\textwidth]{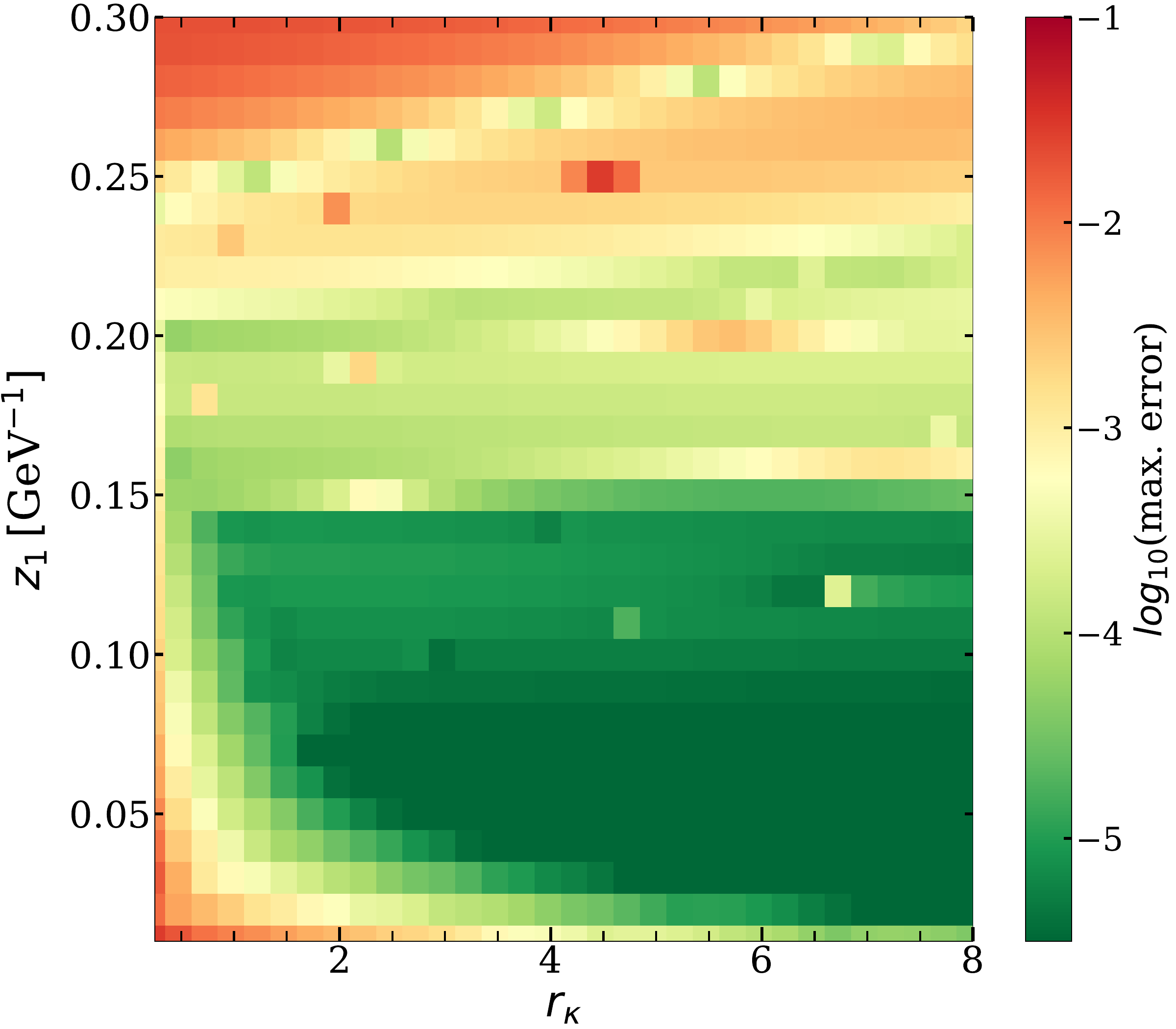}
}
\\[1.3em]
\subfloat[$Q = 100 \gev$, $I(q)$]{
   \includegraphics[width=0.46\textwidth]{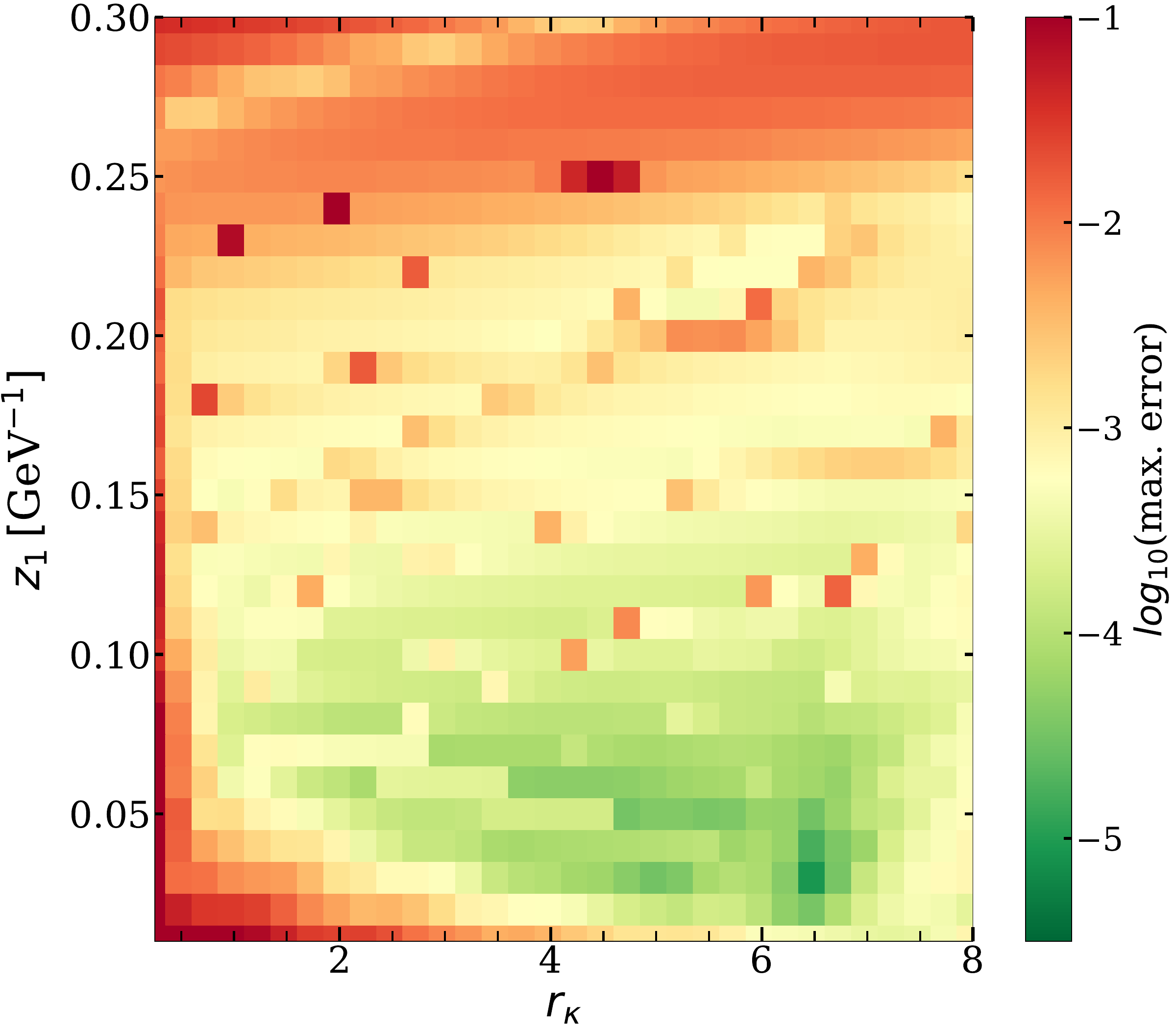}
}
\hspace{0.5em}
\subfloat[$Q = 100 \gev$, $K(q)$]{
   \includegraphics[width=0.46\textwidth]{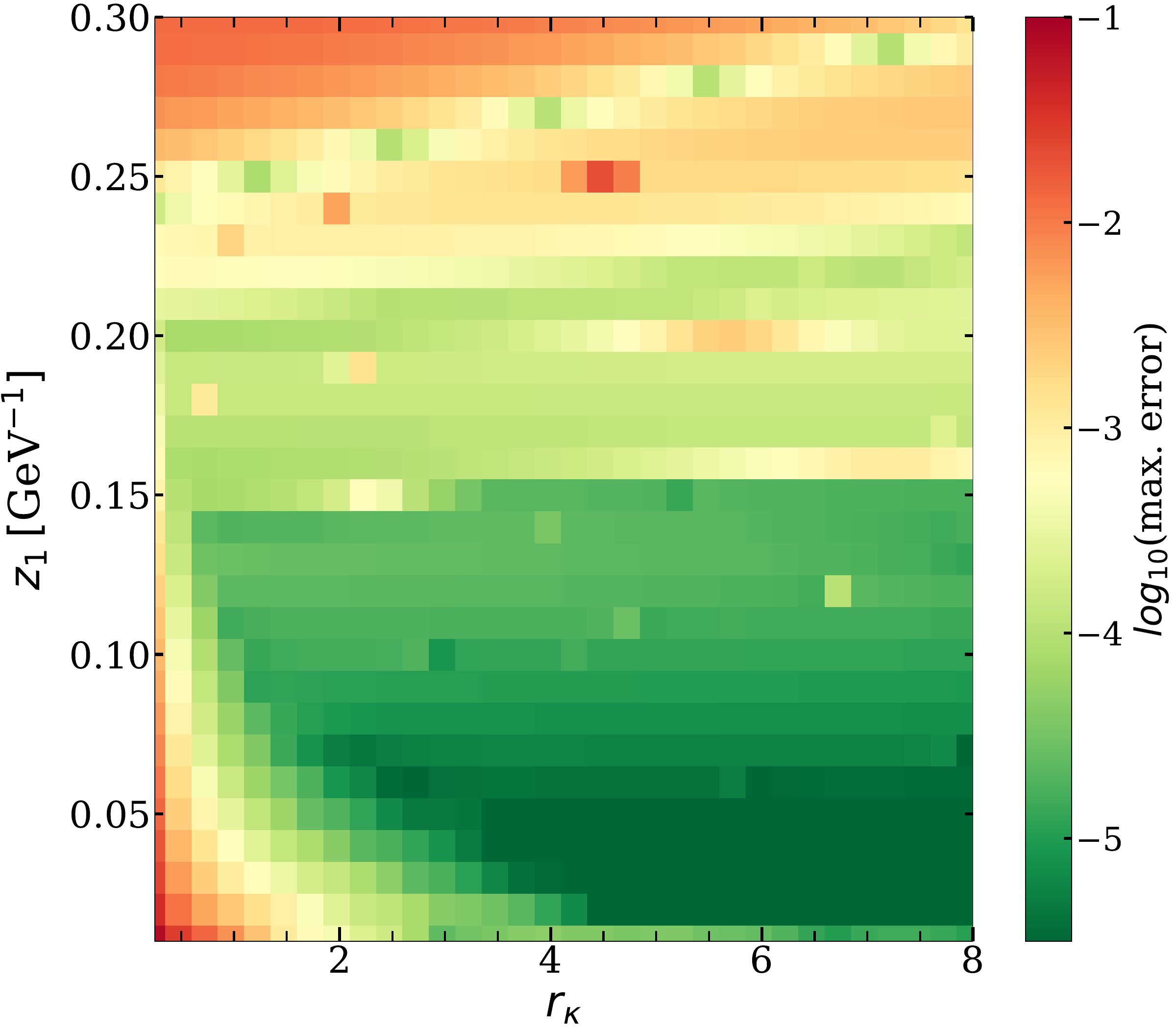}
}
\caption{\label{fig:par_Yukawa}  The relative integration error
$\varepsilon_{\text{max}}$ defined in \protect\eqref{max-err-def}, evaluated for
the Yukawa TMD as a function of the grid parameters with our default grid
structure $[0, z_1, \infty]_{(16, 32)}$.  The accuracy for $I(q)$ is shown on
the left and the one for $K(q)$ on the right.}
\end{figure}

\begin{figure}
\centering
\subfloat[$Q = 2 \gev$, $I(q)$]{
   \includegraphics[width=0.46\textwidth]{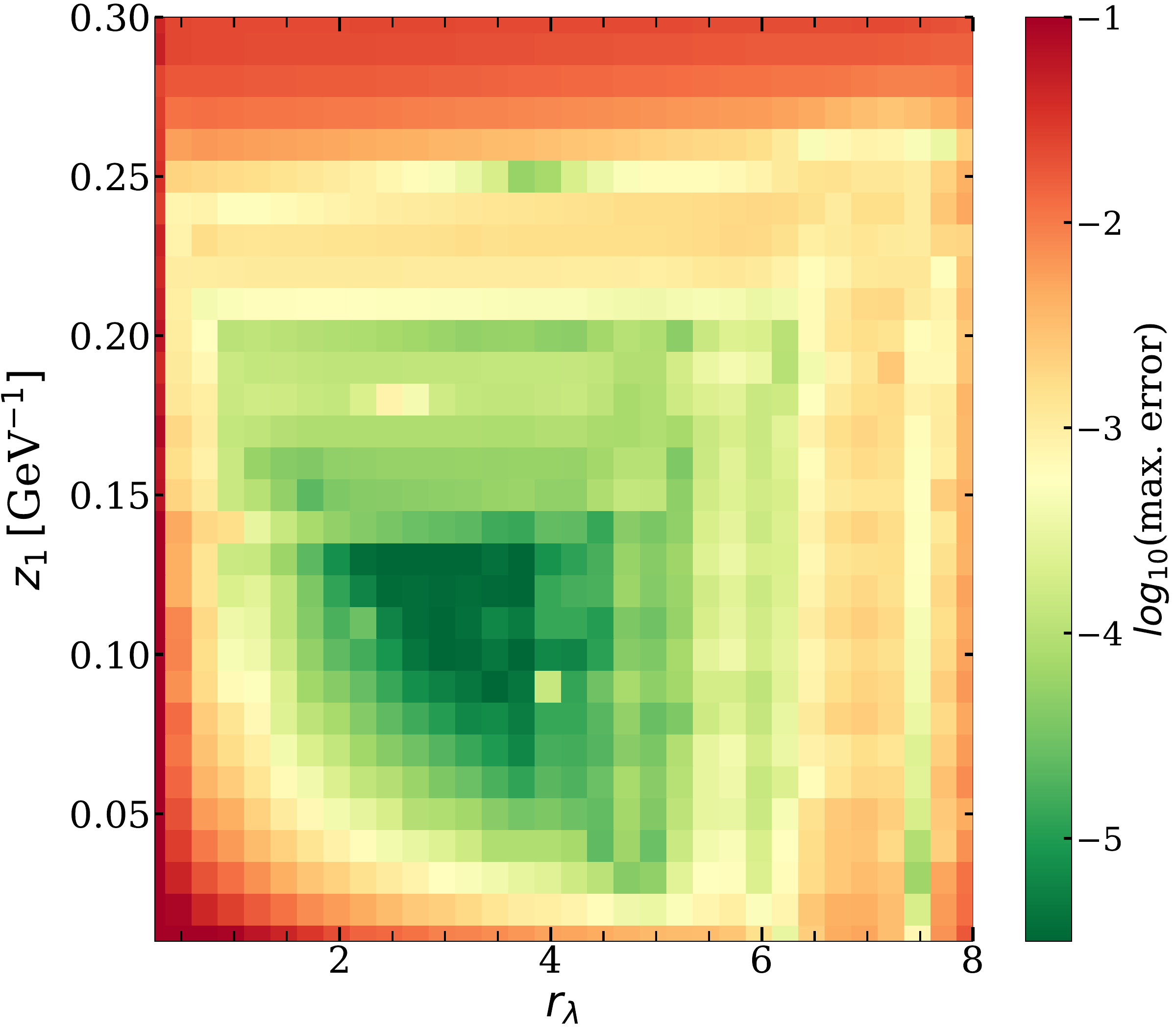}
}
\hspace{0.5em}
\subfloat[$Q = 2 \gev$, $K(q)$]{
   \includegraphics[width=0.46\textwidth]{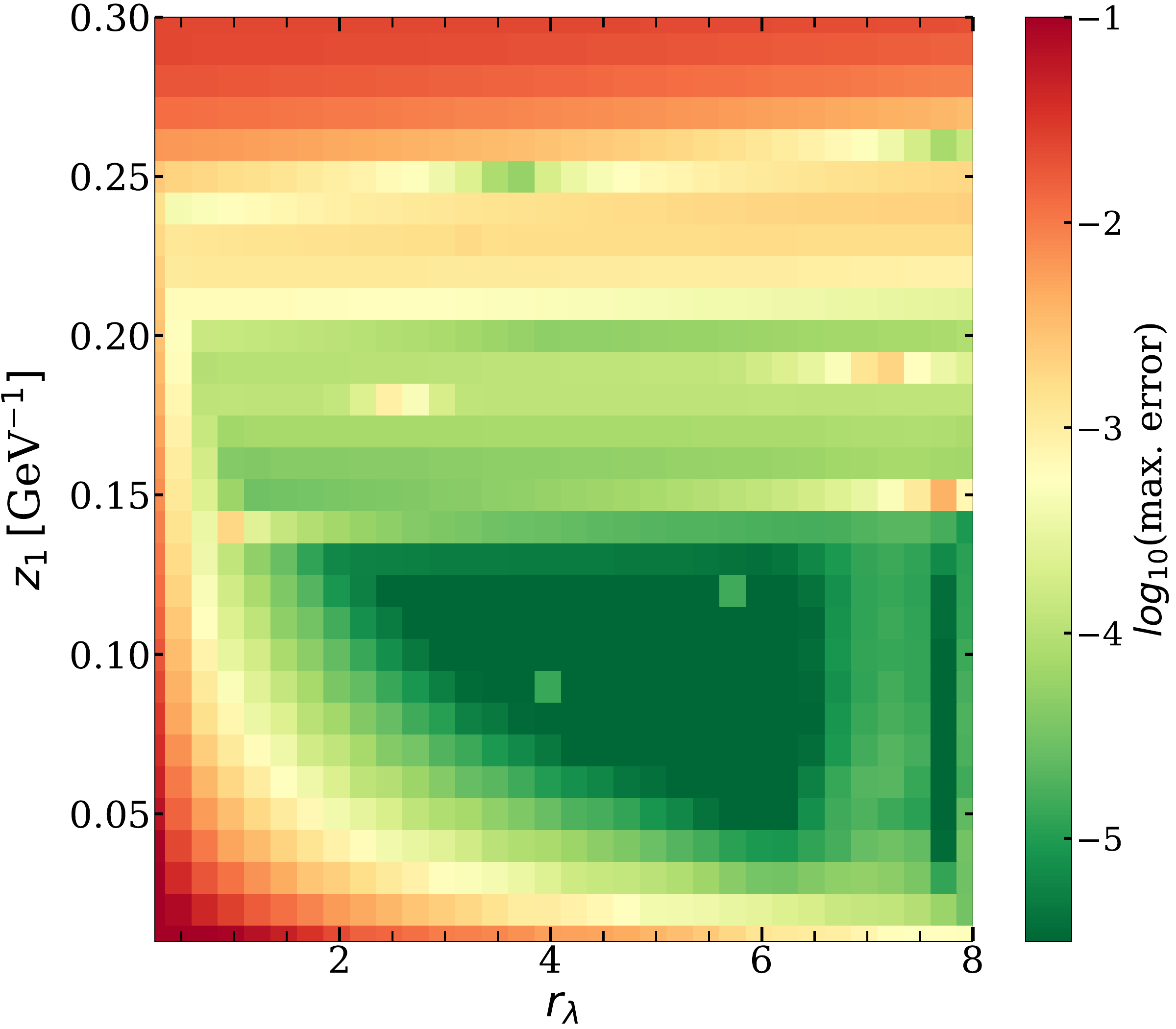}
}
\\[1.3em]
\subfloat[$Q = 20 \gev$, $I(q)$]{
   \includegraphics[width=0.46\textwidth]{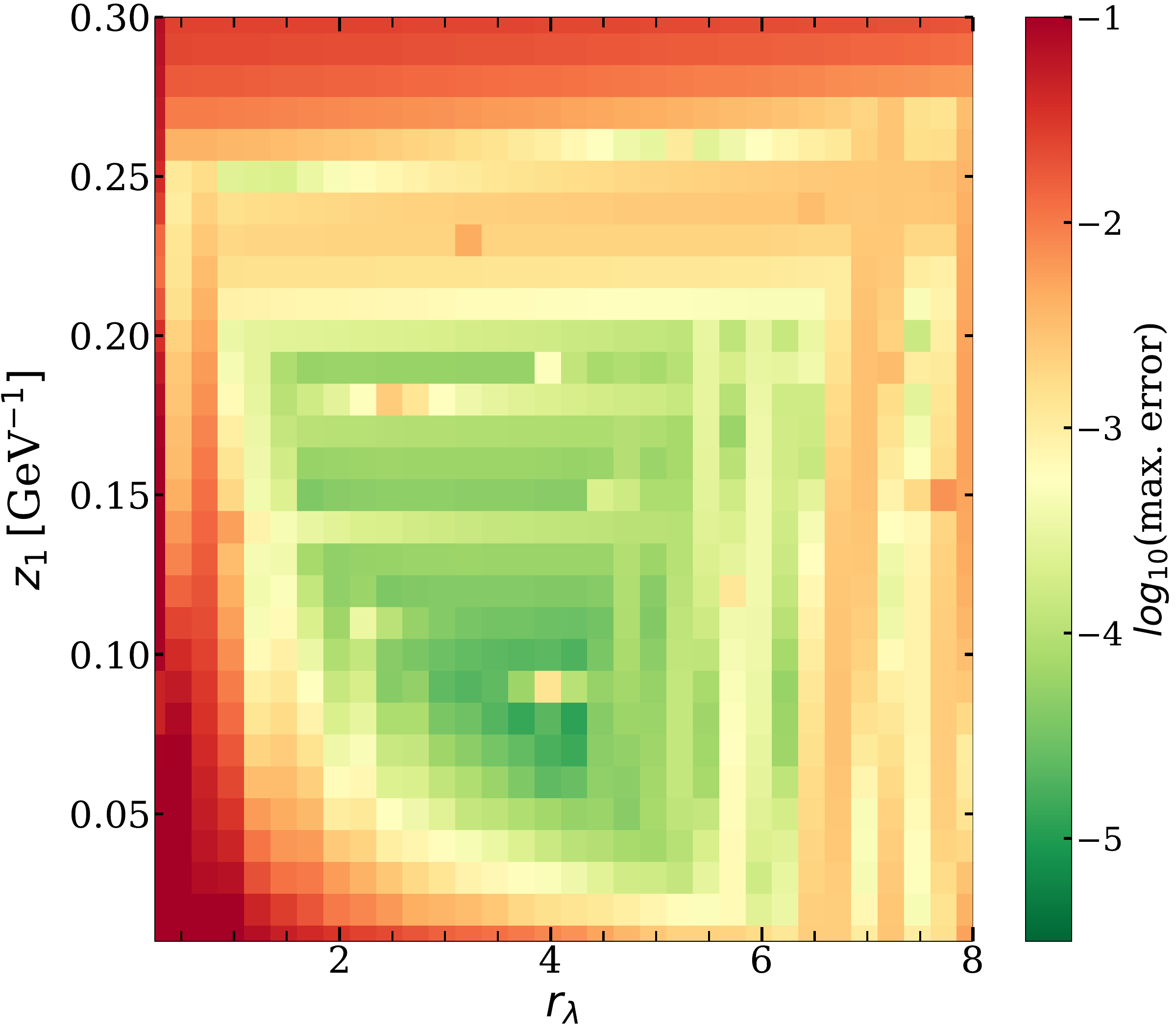}
}
\hspace{0.5em}
\subfloat[$Q = 20 \gev$, $K(q)$]{
   \includegraphics[width=0.46\textwidth]{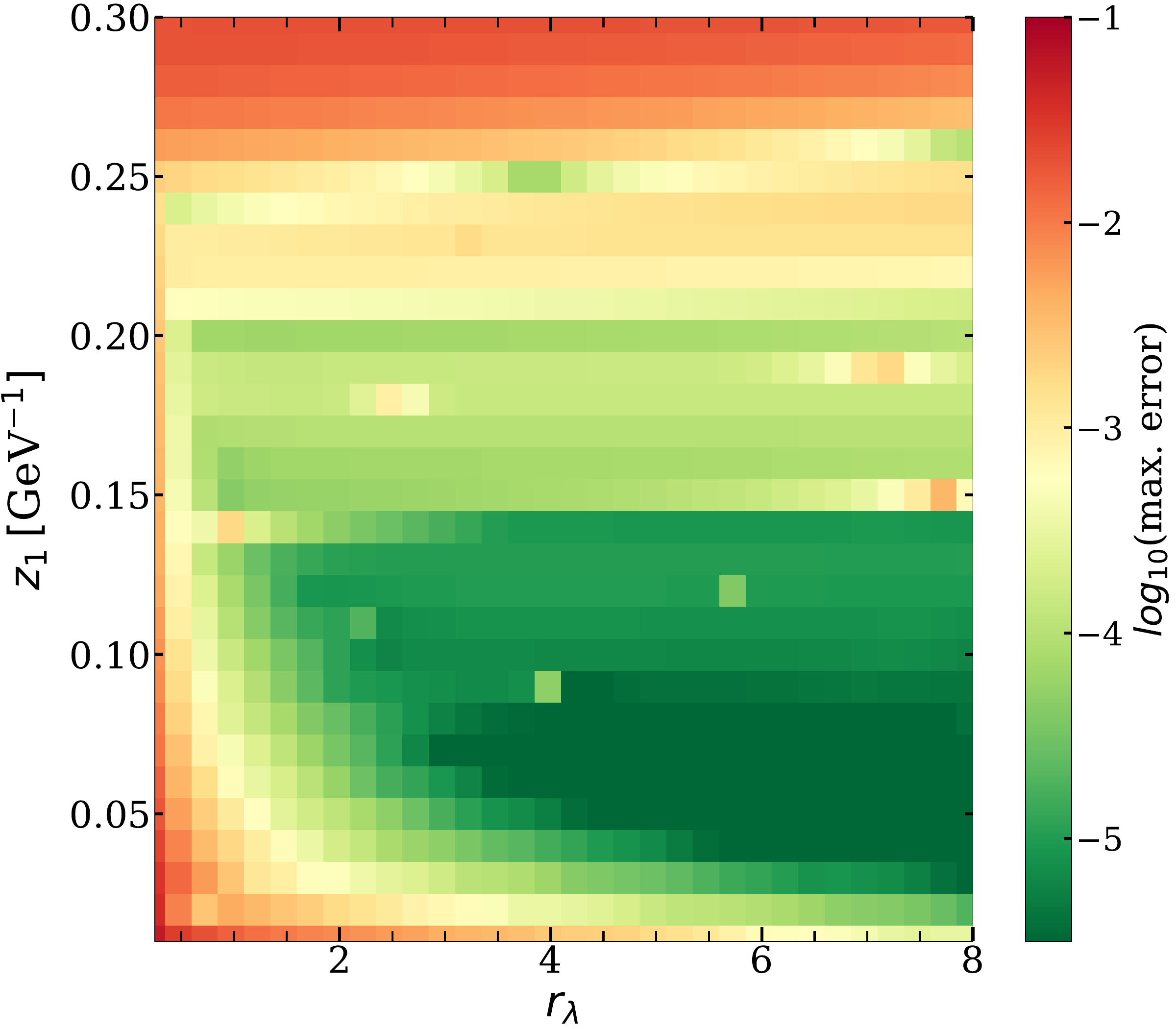}
}
\\[1.3em]
\subfloat[$Q = 100 \gev$, $I(q)$]{
   \includegraphics[width=0.46\textwidth]{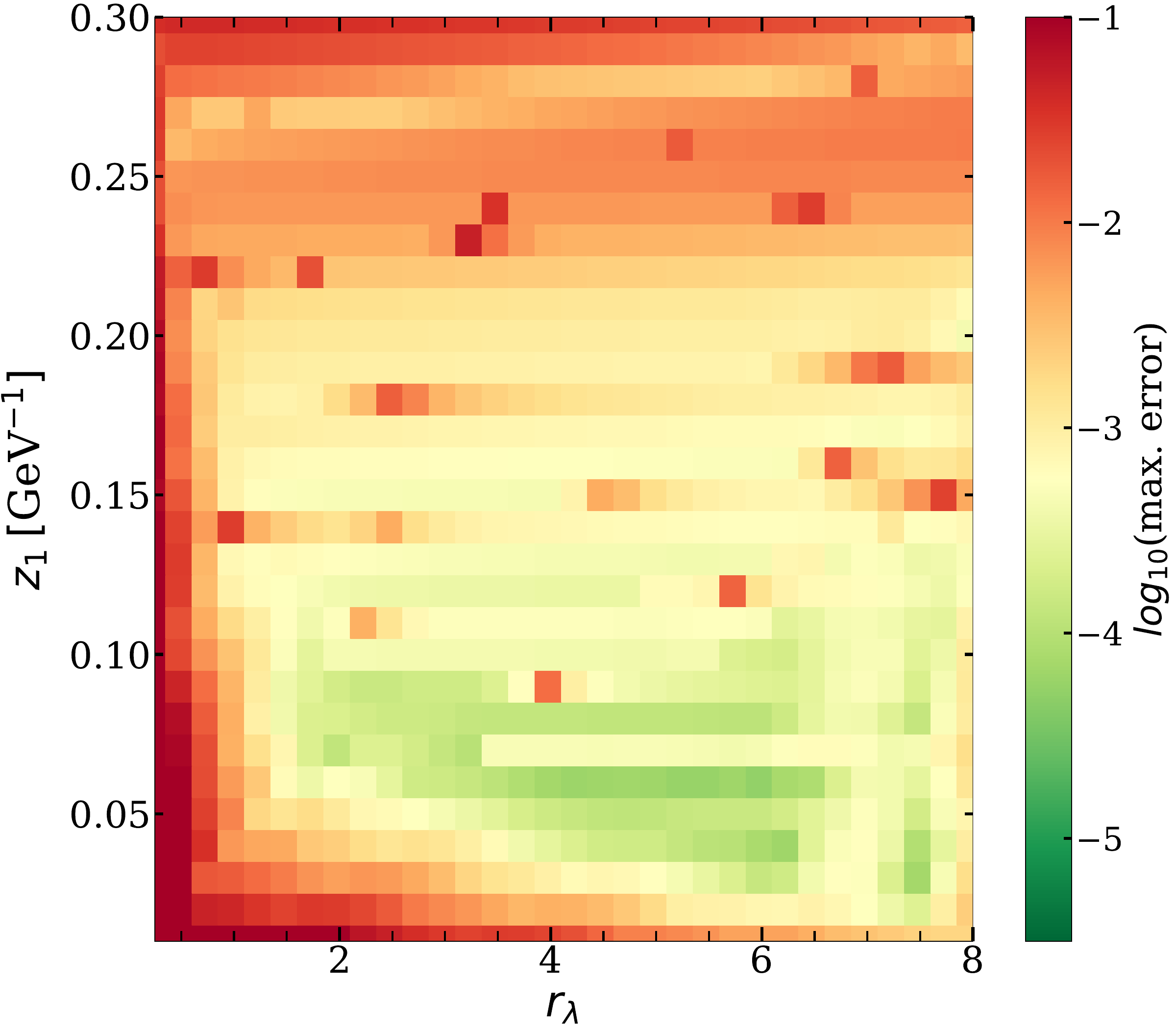}
}
\hspace{0.5em}
\subfloat[$Q = 100 \gev$, $K(q)$]{
   \includegraphics[width=0.46\textwidth]{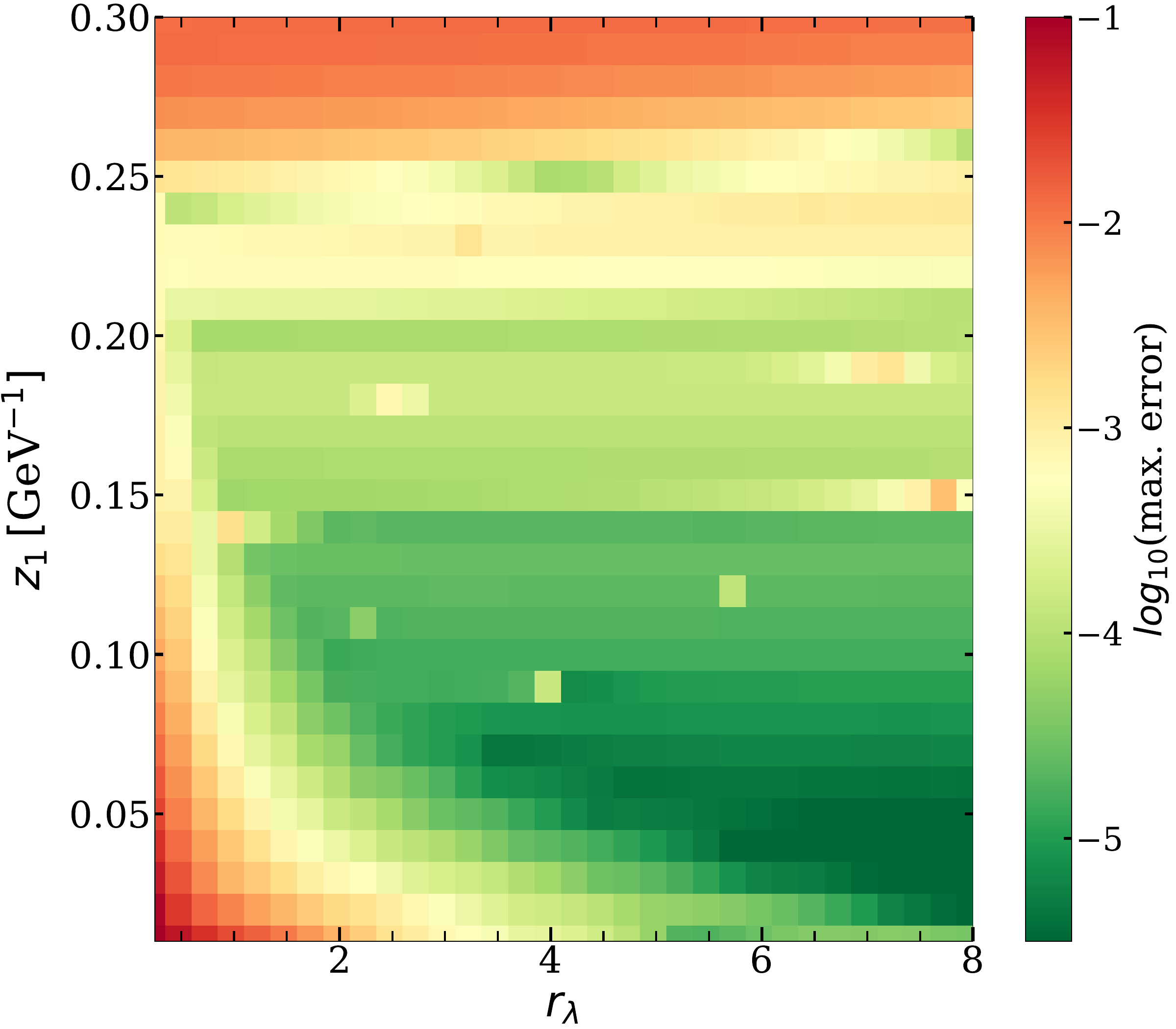}
}
\caption{\label{fig:par_Gauss}  As \fig{\protect\ref{fig:par_Yukawa}}, but for
the Gauss TMD.}
\end{figure}

\begin{figure}
\centering
\subfloat[$Q = 2 \gev$, $(12, 24)$ points]{
   \includegraphics[width=0.46\textwidth]{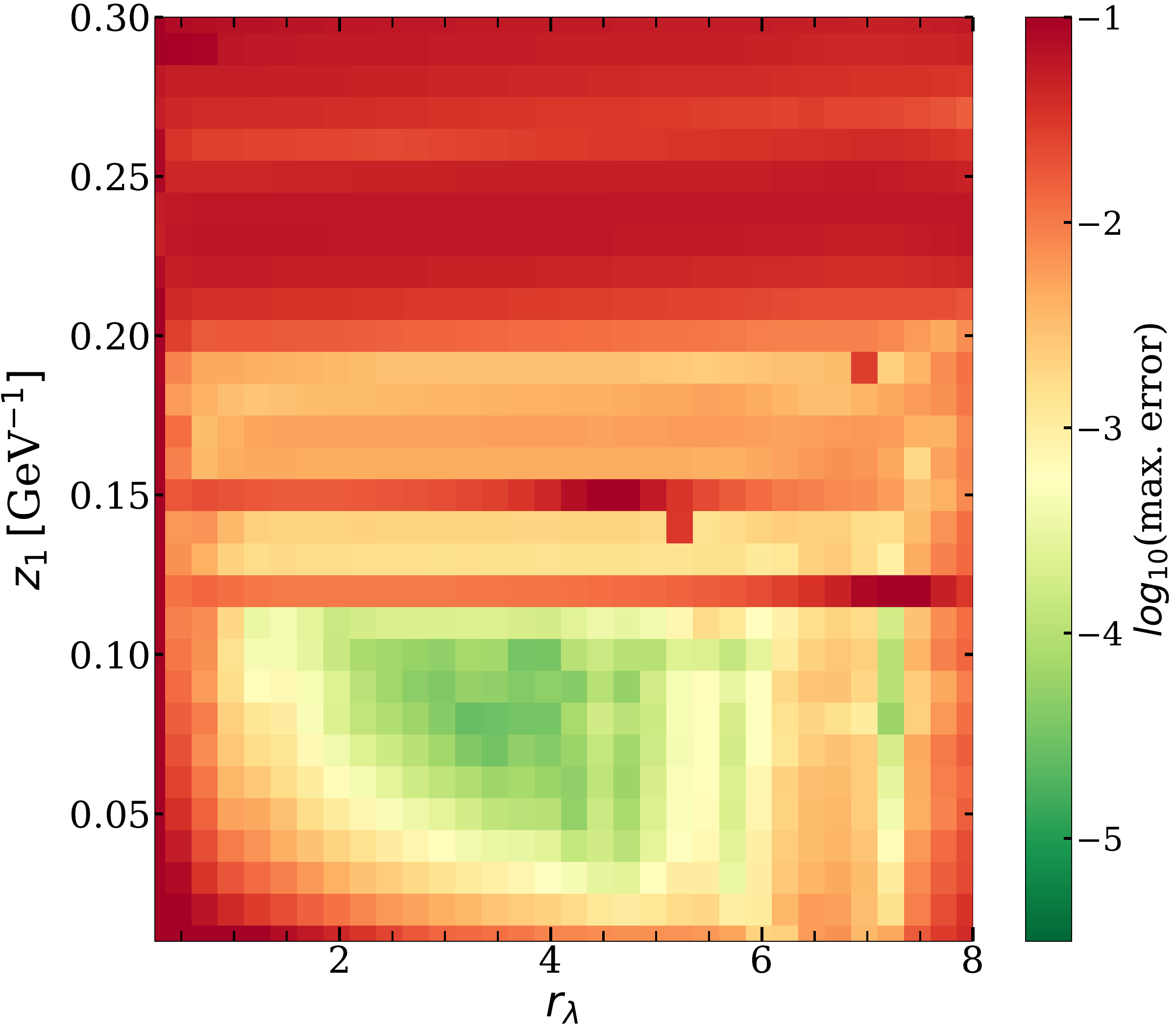}
}
\hspace{0.5em}
\subfloat[$Q = 2 \gev$, $(24, 48)$ points]{
   \includegraphics[width=0.46\textwidth]{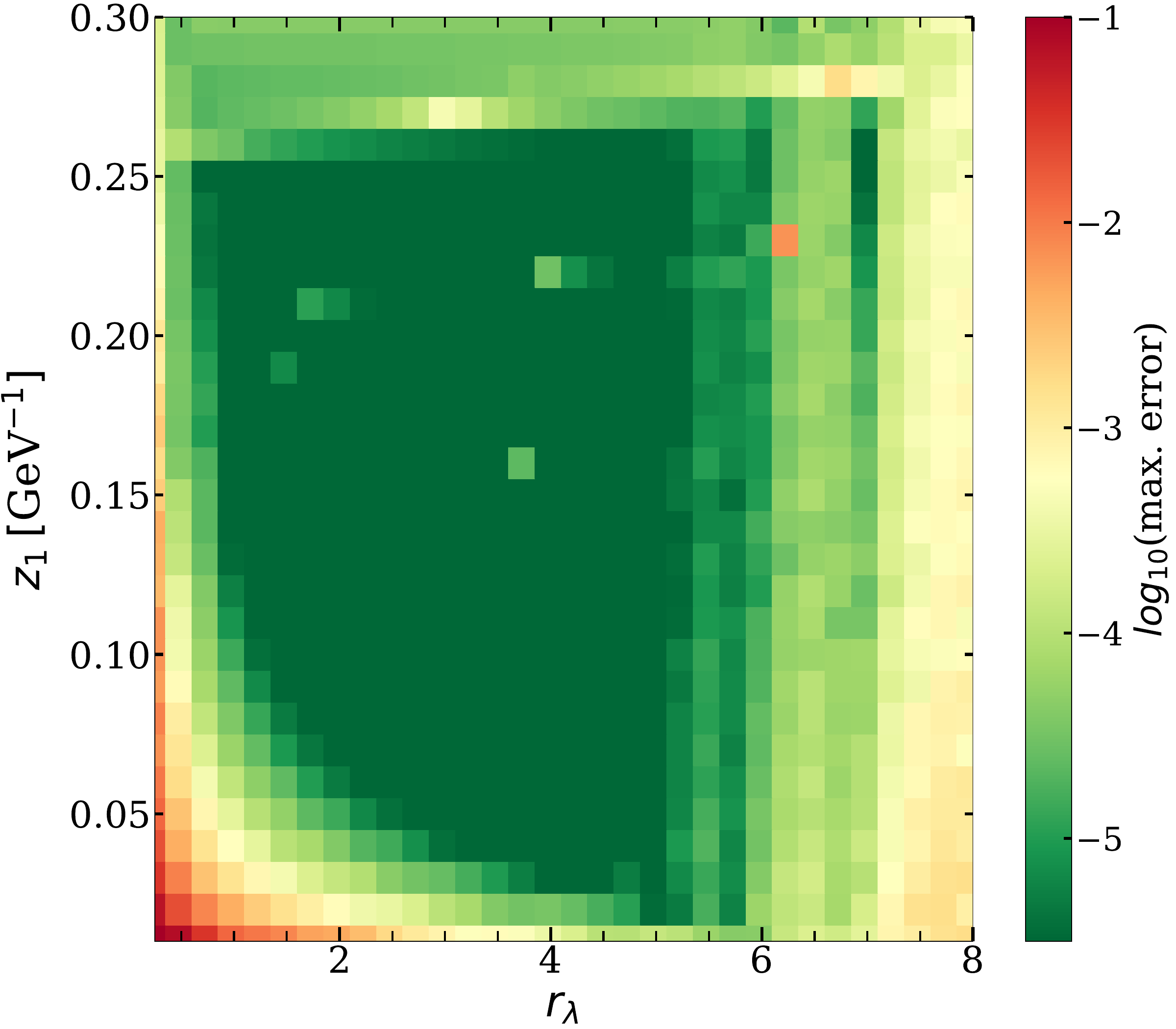}
}
\\[1.3em]
\subfloat[$Q = 20 \gev$, $(12, 24)$ points]{
   \includegraphics[width=0.46\textwidth]{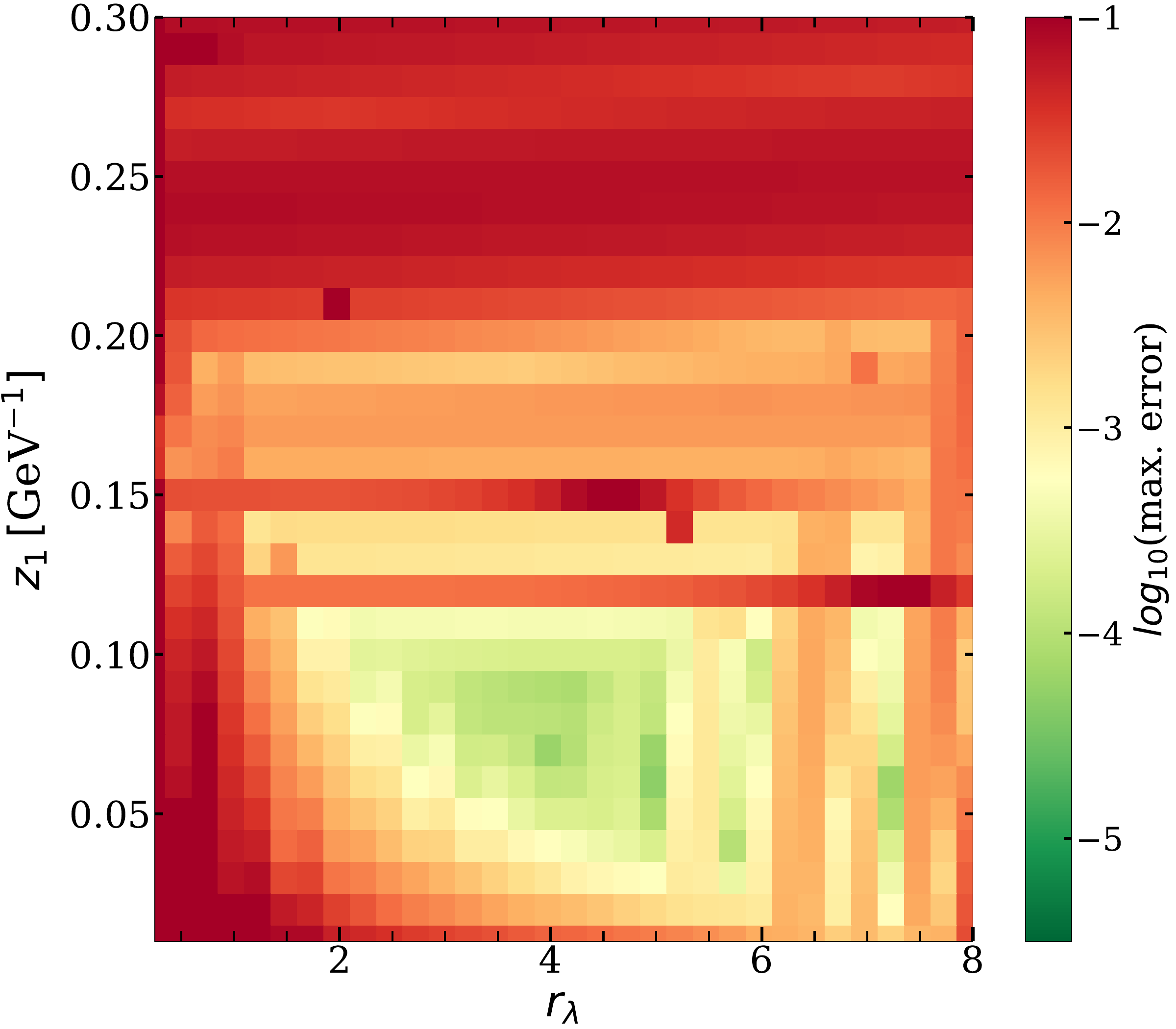}
}
\hspace{0.5em}
\subfloat[$Q = 20 \gev$, $(24, 48)$ points]{
   \includegraphics[width=0.46\textwidth]{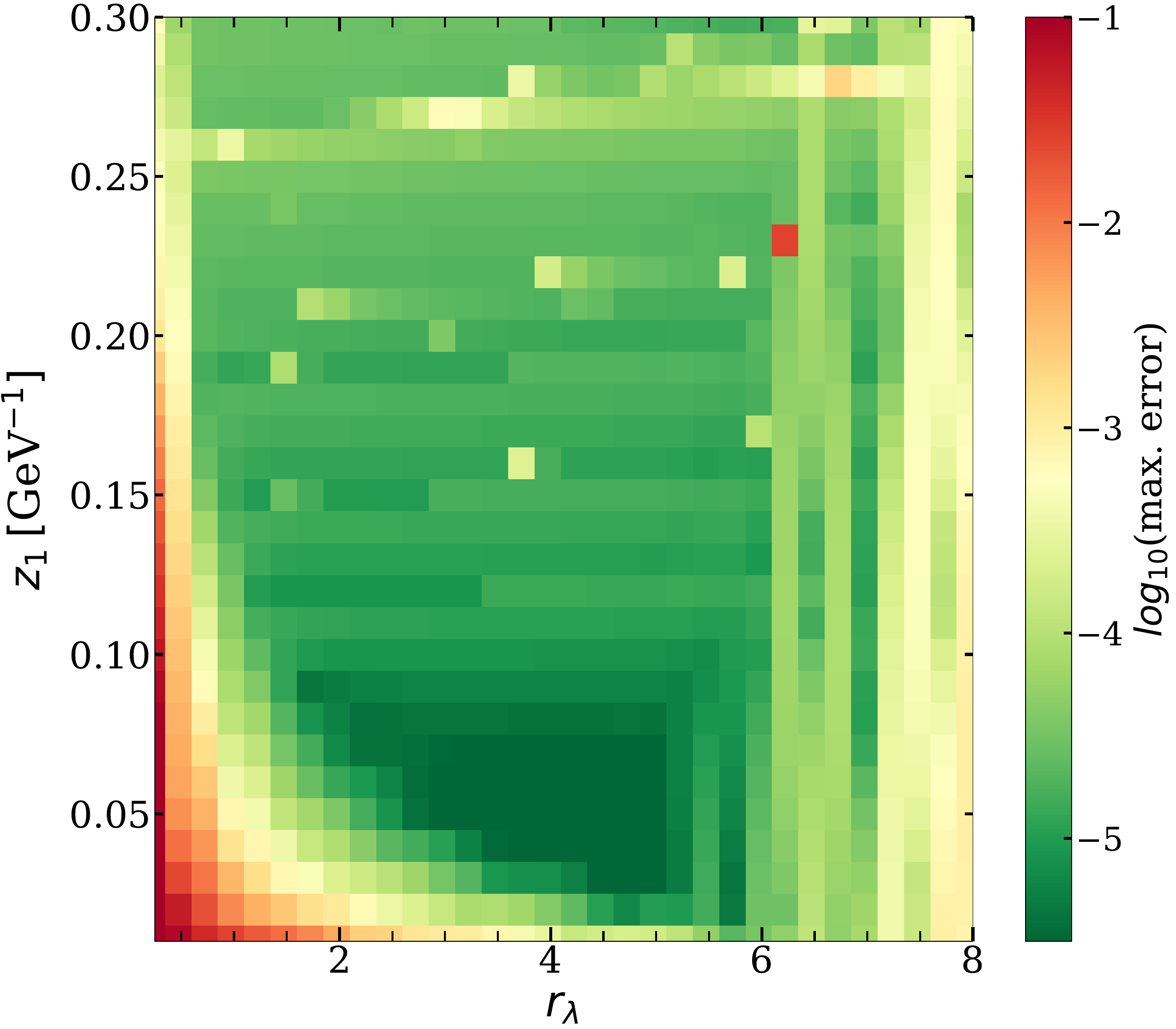}
}
\\[1.3em]
\subfloat[$Q = 100 \gev$, $(12, 24)$ points]{
   \includegraphics[width=0.46\textwidth]{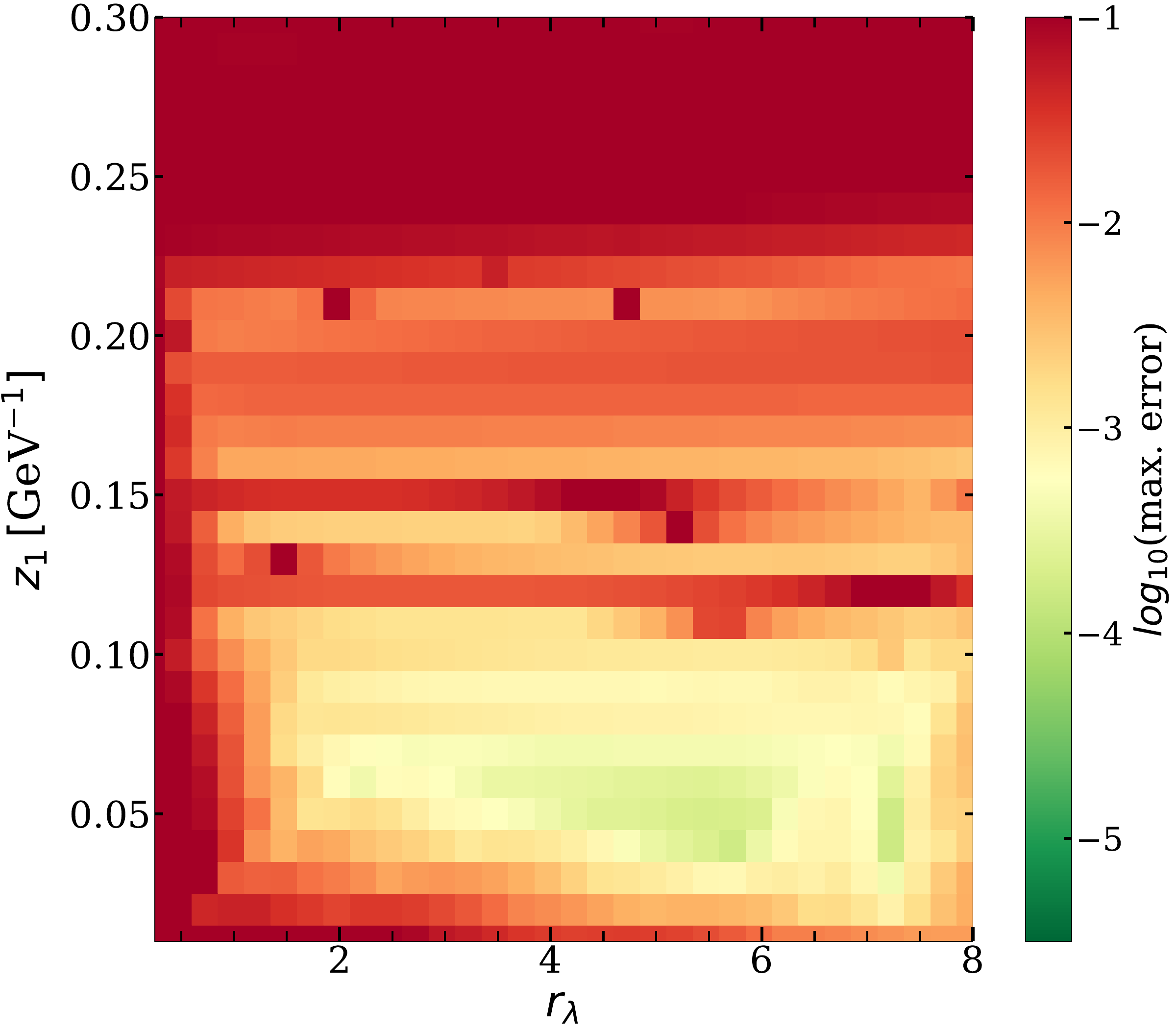}
}
\hspace{0.5em}
\subfloat[$Q = 100 \gev$, $(24, 48)$ points]{
   \includegraphics[width=0.46\textwidth]{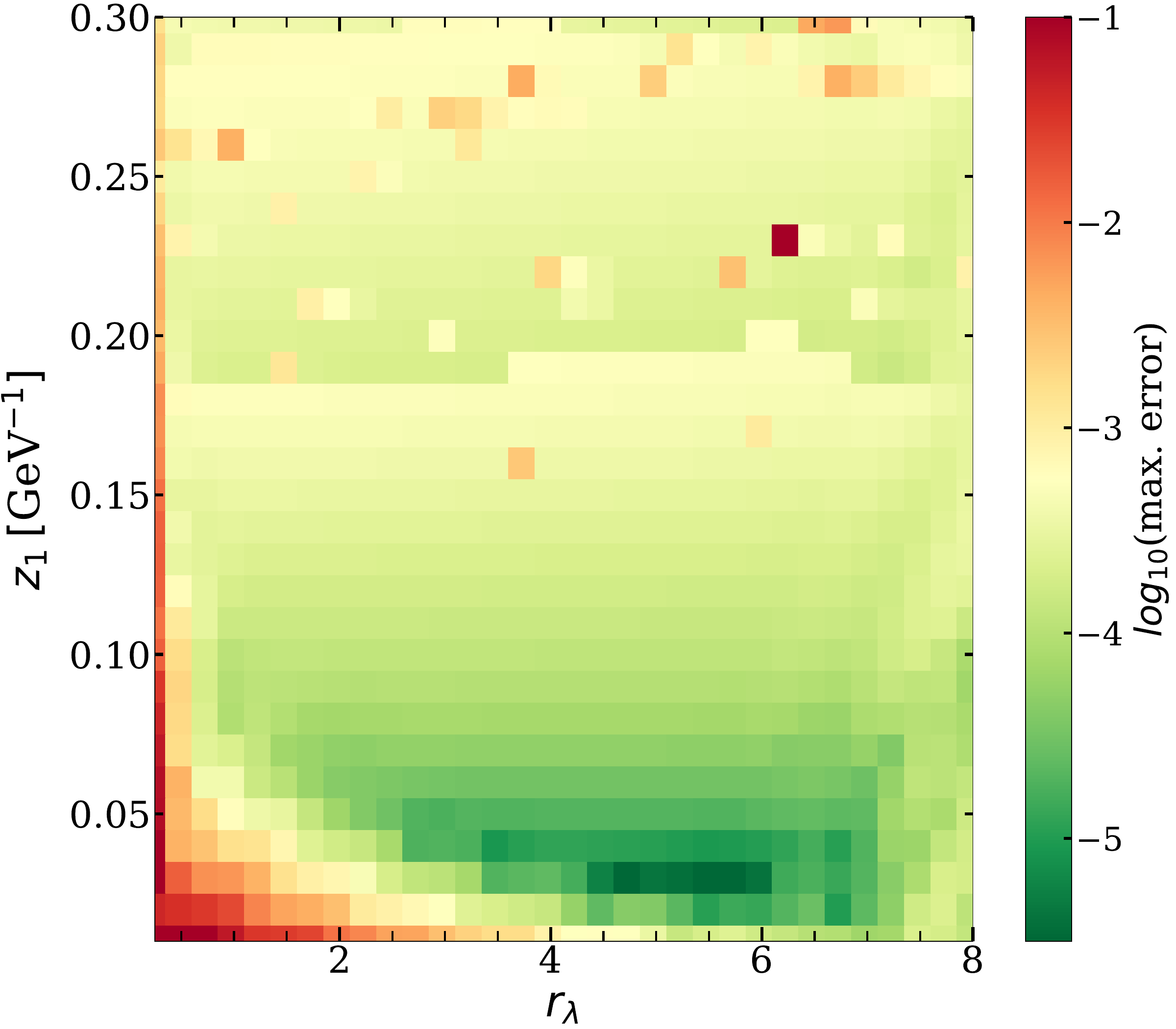}
}
\caption{\label{fig:par_grid_nos}  The relative integration error
$\varepsilon_{\text{max}}$ of $I(q)$ for the Gauss TMD, evaluated with fewer or
more grid points that in the left panels of \fig{\protect\ref{fig:par_Gauss}}.}
\end{figure}

Figures~\ref{fig:par_Yukawa} to \ref{fig:par_grid_nos} show the dependence of
the relative integration error on the parameters $z_1$ and $r_\kappa$ or
$r_\lambda$.  They give the \emph{worst} accuracy for all $q_i$ values away from
the zero crossing of $I(q)$, i.e.
\begin{align}
   \label{max-err-def}
   \varepsilon_{\text{max}}
   &=
   \max_{q_i \notin R}( \varepsilon )
   &
   \text{ with } R &= (q_0 - \Delta, q_0 + \Delta)
   \,,
\end{align}
where $\Delta$ is given in \eqref{Delta-def}.  We obtain very similar plots for
$\varepsilon_{\text{rms}}$ defined in \eqref{rms-err-def}.  Note that
$\varepsilon_{\text{max}}$ is a more stringent measure for the performance of a
parameter setting.

We see in \figs{\ref{fig:par_Yukawa}} and \ref{fig:par_Gauss} that the best
accuracy for $I(q)$ at $Q = 2$ or $20 \gev$ is obtained with $z_1$ between
$0.05$ and $0.15 \gev^{-1}$ and with $r_\kappa$ or $r_\lambda$ between $2$ and
$4$.  For $Q = 100 \gev$, the region of favourable parameters is somewhat
shifted, with $z_1$ between $0.05$ and $0.10 \gev^{-1}$ and with $r_\kappa$ or
$r_\lambda$ between $4$ and $6$.  The parameter region of good accuracy for
$K(q)$ is much wider in $r_\kappa$ or $r_\lambda$ and includes the good region
for $I(q)$. Corresponding plots for the toy TMD are similar to those for the
Yukawa TMD and not shown here.

With fewer grid points, namely $(12, 24)$ instead of $(16, 32)$, we find that
the favourable parameter regions become smaller, as is shown in the left panels
of \fig{\ref{fig:par_grid_nos}}.  Still, a relative accuracy of $10^{-3}$ or
better can be achieved in the same parameter region that is preferred for $(16,
32)$ points. In the right panels of \fig{\ref{fig:par_grid_nos}} we see that
with even more points, namely $(24, 48)$, the favourable parameter region
becomes much wider, although for $Q = 100$ there is still a clear preference for
low $z_1$.  The corresponding plots for the Yukawa and the toy TMD look quite
similar and are not shown here.

\subsection{Estimating the integration error}
\label{sec:error}

\begin{figure}
\centering
\subfloat[Yukawa TMD, $Q = 2 \gev$,
   two subgrids \mbox{$[0, z_1, \infty]_{(16, 32)}$}]{
   \includegraphics[width=0.98\textwidth,trim=10 0 0 0,clip]{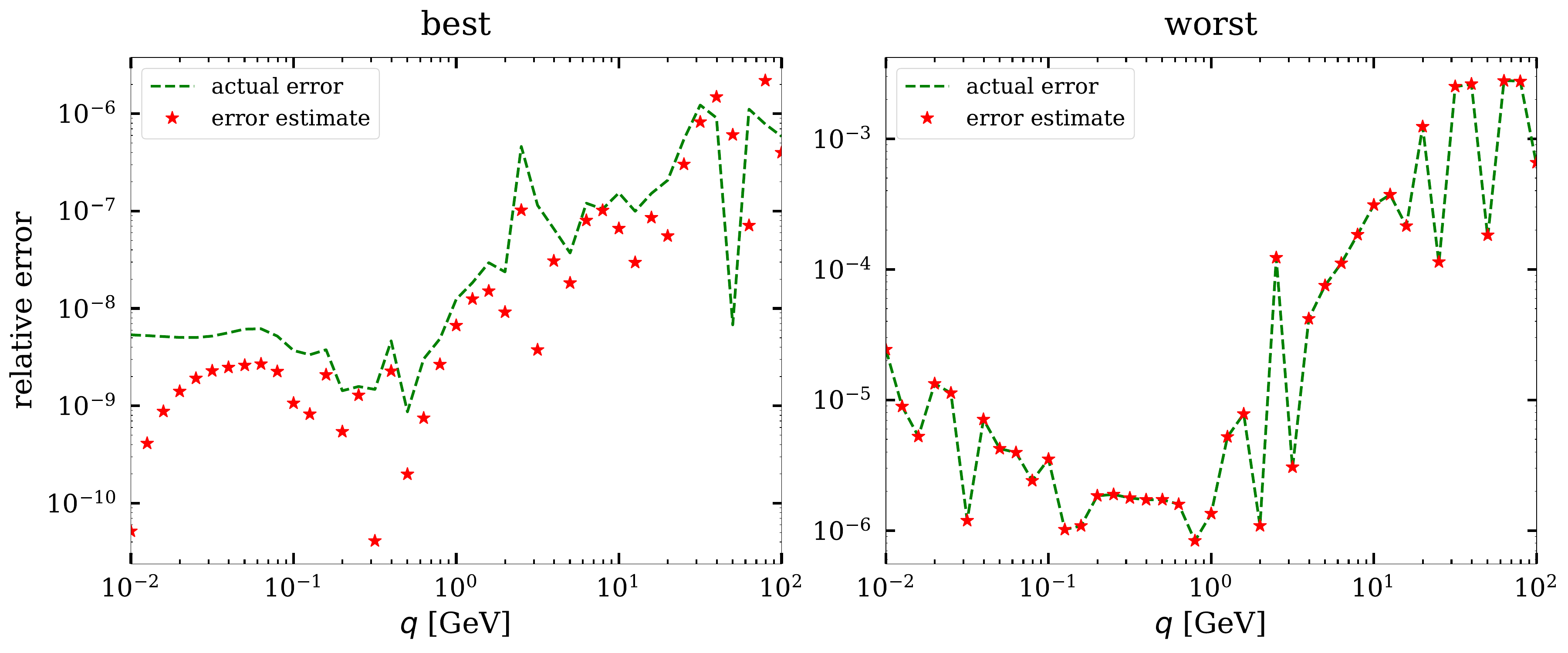}
}
\\[1.3em]
\subfloat[Toy TMD, $Q = 20 \gev$,
   one subgrid \mbox{$[0, \infty]_{(48)}$}]{
   \includegraphics[width=0.98\textwidth,trim=10 0 0 0,clip]{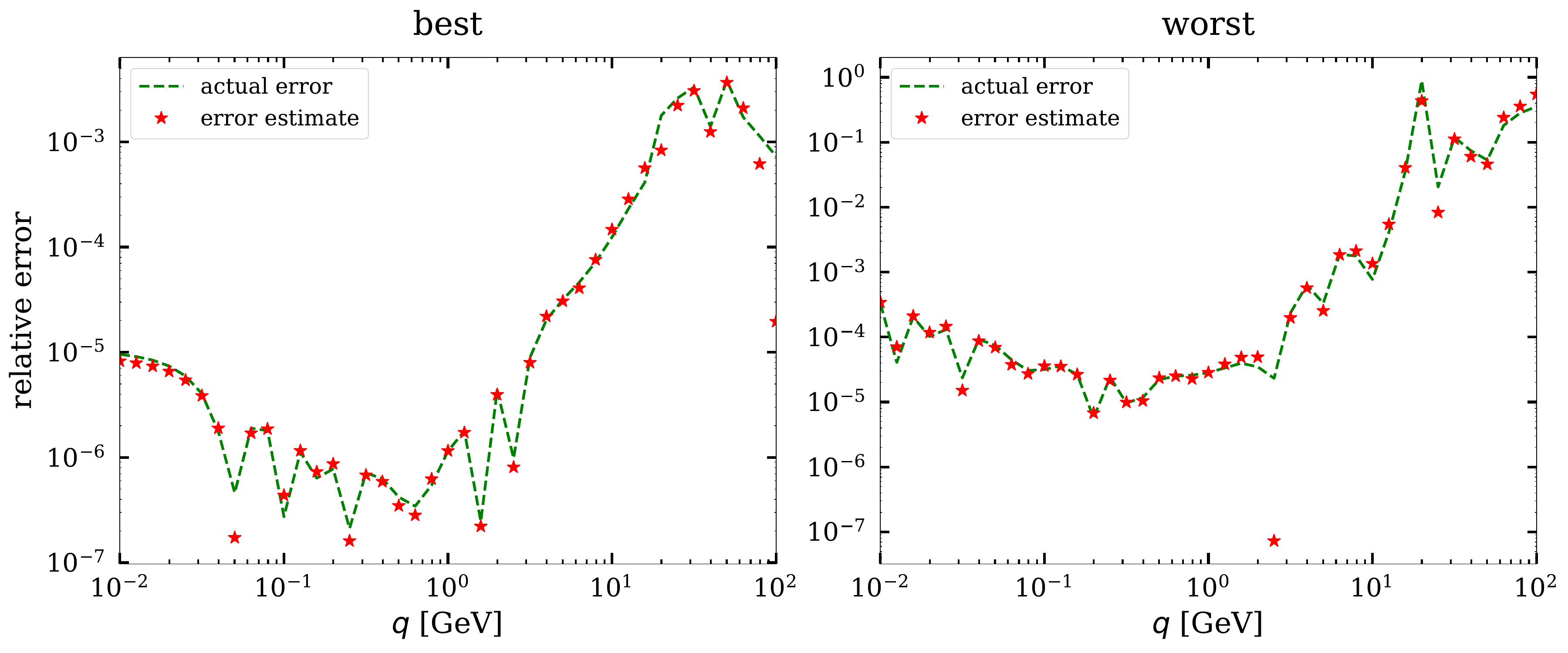}
}
\caption{\label{fig:error_estimates}  Actual and estimated relative integration
accuracy, as described in the text.  The left panels are for the best setting of
the corresponding parameter scan and the right panels for the worst one.  The
worst setting in the upper row is $(z_1, r_\kappa) = (0.03, 1.0)$, the worst
setting in the lower row is $r_\kappa=0.5$.  The best settings are given in
\tab{\protect\ref{tab:best-params}}.}
\end{figure}

To determine whether a particular grid setup in our method is satisfactory, one
needs a practical method to estimate the integration accuracy.  A straightforward
way to achieve this is to re-compute the integral with twice as many points on
each subgrid, keeping all other settings the same.  We have evaluated the
corresponding error estimates for the grids $[0, z_1, \infty]_{(16, 32)}$ and
$[0, \infty]_{(48)}$, both for the the best and for the worst parameter settings
identified in our parameter scans.  A selection of our results in shown in
\fig{\ref{fig:error_estimates}}.

In general we find that the error estimates so obtained are extremely close to
the actual integration errors.  The only notable exception to this statement is
shown in the top left panel, where we see that for most $q$ values the error is
\emph{underestimated} by doubling the number of points.  Since this problem
appears in a region where the actual relative error is well below $10^{-6}$, we
consider it to be of minor importance in practice.  We therefore do not pursue
this issue further here.

We also observe that sometimes the error estimate fails only at specific $q$
values, as seen in the bottom left panel.  This happens when the difference
between  approximate and exact integrals has a zero crossing as a function of
$q$.  As long as one monitors the error estimate also in the vicinity of such
points, this should not pose any problem.


\subsection{Comparison with the method of Kang et al.}
\label{sec:comparison-kang}

Let us finally compare our method with the algorithm of Kang et
al.~\cite{Kang:2019ctl}, which implements the Ogata method \cite{Ogata:2005}
with a particular optimisation of the parameter $h$ that controls the density of
integration nodes.  We use the C++ implementation of the algorithm that links to
the GSL library, see \url{https://github.com/UCLA-TMD/Ogata} for details. The
optimisation algorithm requires an estimate for the value of $x$ at which the
function $x^2 \ms W(x/q)$ takes its maximum; we verified that our results are
stable with respect to variations of this parameter.  The other input parameter
to the algorithm is the number $N$ of function calls to $W(z)$.

For our method, we take the preferred parameter transform for each TMD, together
with a grid $[0, z_1, \infty]_{(n_1, n_2)}$ with our default grid parameters
\eqref{tmd-param-default}.  For a total number $N$ of grid points, and thus of
calls to the function $W(z)$, we take $n_1 = \operatorname{floor}(N/3) + 1$ and
$n_2 = N - \operatorname{floor}(N/3)$ points on the respective subgrids.  We
recall that the grid parameters in \eqref{tmd-param-default} were determined for
$(n_1, n_2) = (16, 32)$, i.e.\ for $N = 47$.

When comparing the accuracy of the two algorithms at the same value of $N$, one
should bear in mind that with our method the computation of $W(z)$ at $N$ values
of $z$ allows one to evaluate the Fourier-Bessel transform for a broad range of
$q$ values, which is not the case for the Ogata method.  We nevertheless use $N$
as a measure of the computational cost in both cases.

The following plots show the accuracy achieved for the toy and Gauss TMDs.  The
corresponding accuracy for the Yukawa TMD is typically between these two cases,
both for our method and the one of \cite{Kang:2019ctl}.  We omit this in our
plots for the sake of clarity.

\begin{figure}
\centering
\subfloat[$Q = 2 \gev$, $q = 0.01 \gev$ (left) and $q = 2 \gev$ (right)]{
   \includegraphics[width=0.86\textwidth,trim=120 0 200 90,clip]{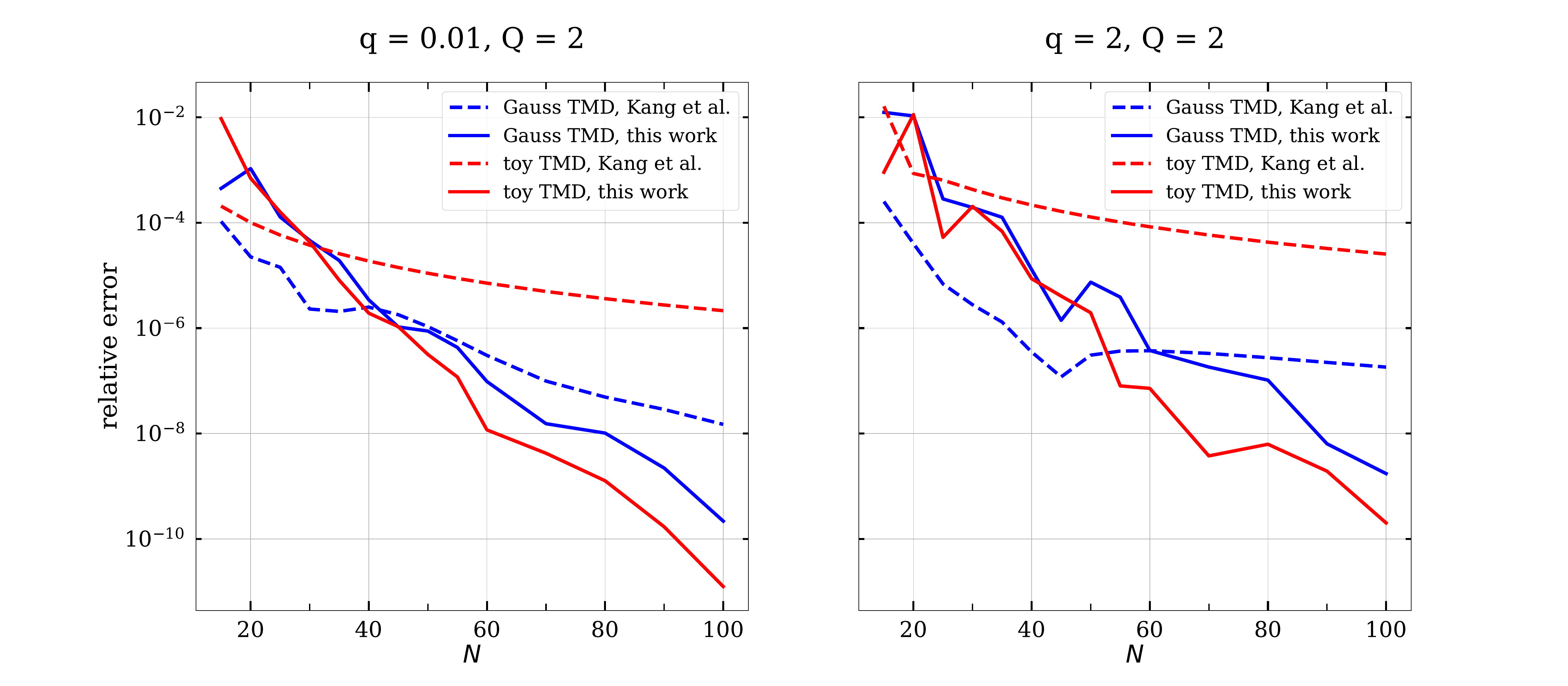}
}
\\[1.5em]
\subfloat[$Q = 20 \gev$, $q = 0.01 \gev$ (left) and $q = 2 \gev$ (right)]{
   \includegraphics[width=0.86\textwidth,trim=120 0 200 90,clip]{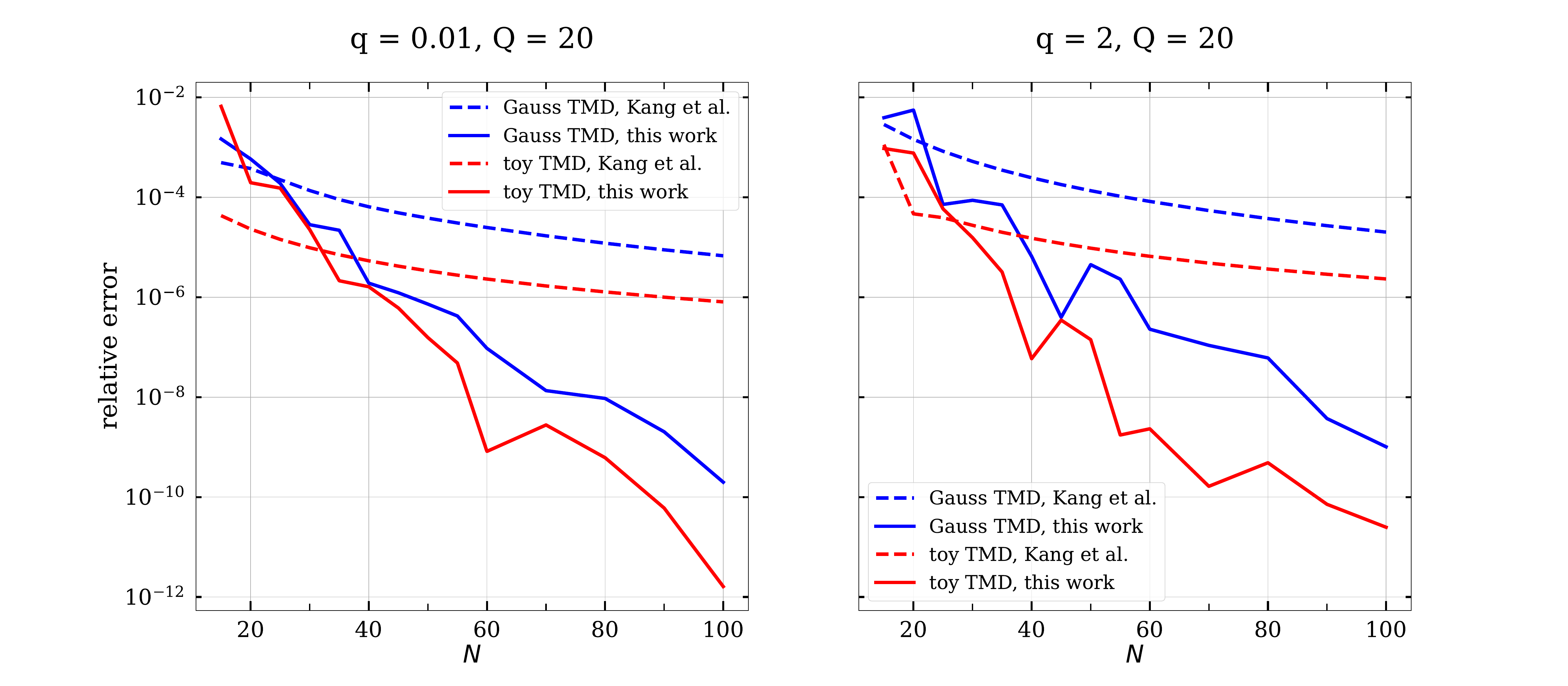}
}
\\[1.5em]
\subfloat[$Q = 100 \gev$, $q = 0.01 \gev$ (left) and $q = 2 \gev$ (right)]{
   \includegraphics[width=0.86\textwidth,trim=120 0 200 90,clip]{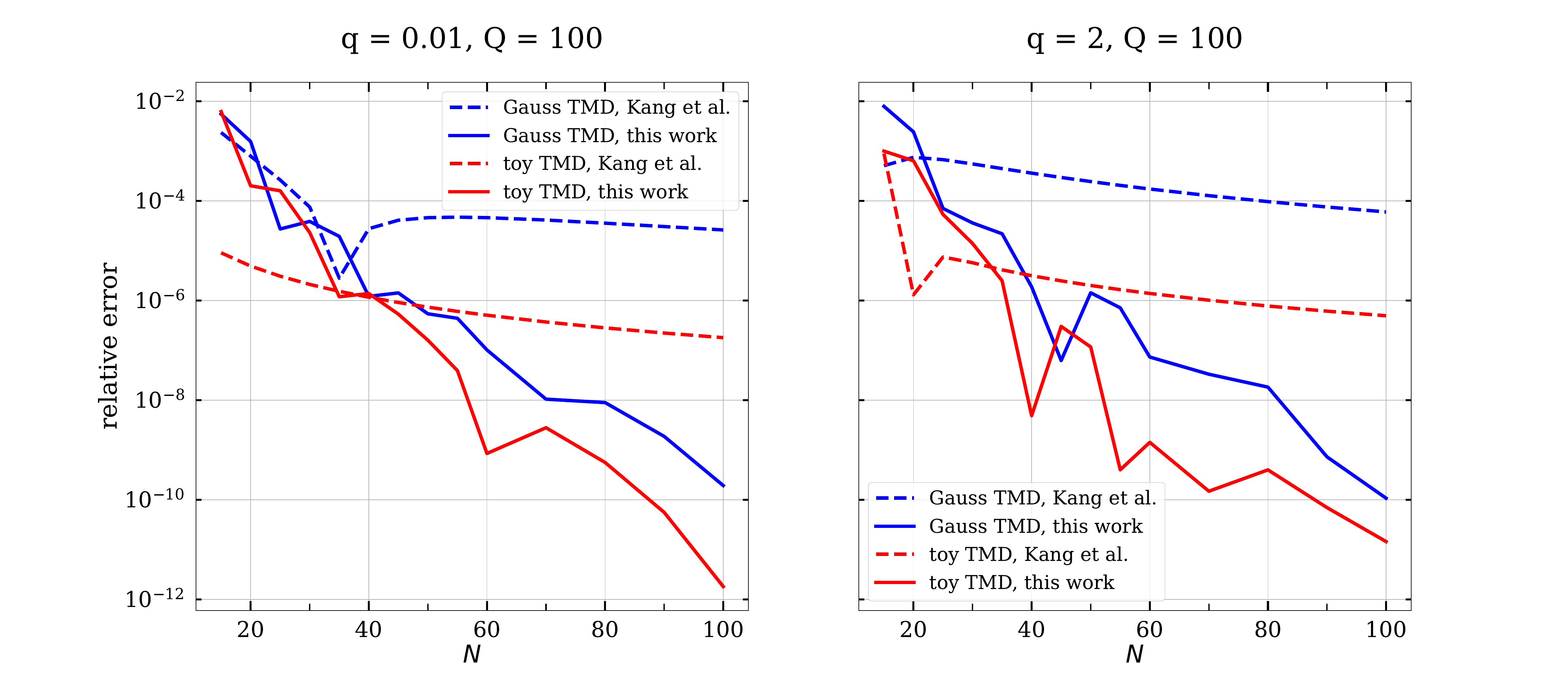}
}
\caption{\label{fig:compare_low_q}  Comparison of our method with the algorithm
of Kang et al.\ at $q = 0.01 \gev$ (left) and $q = 2 \gev$ (right) for different
values of $Q$.  The relative integration error is evaluated for $N$ calls to
the function $W(z)$, with $N$ varying in steps of $5$ or $10$.  Continuous
curves are shown as a guide for the eye.}
\end{figure}

\begin{figure}
\centering
\subfloat[$Q = 2 \gev$, $q = 20 \gev$ (left) and $q = 100 \gev$ (right)]{
   \includegraphics[width=0.86\textwidth,trim=120 0 200 90,clip]{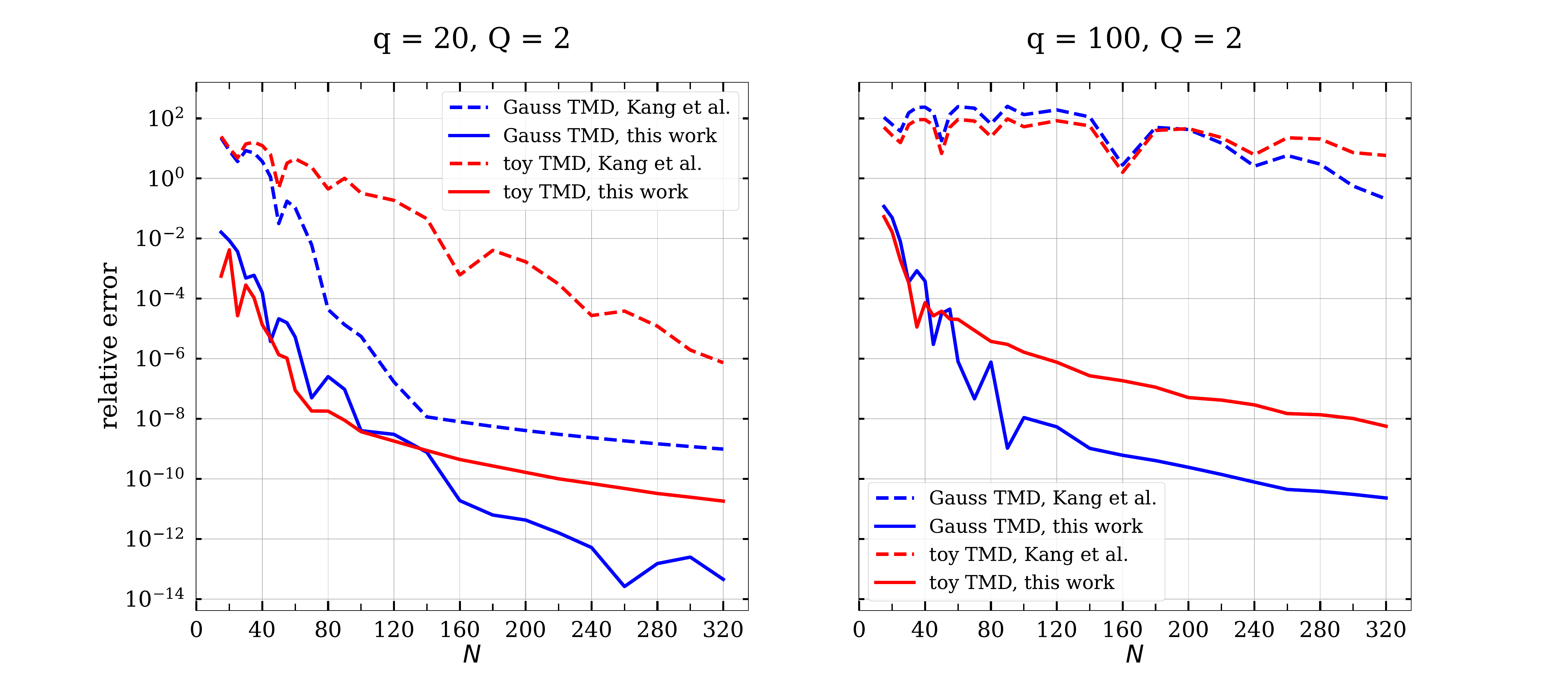}
}
\\[1.5em]
\subfloat[$Q = 20 \gev$, $q = 20 \gev$ (left) and $q = 100 \gev$ (right)]{
   \includegraphics[width=0.86\textwidth,trim=120 0 200 90,clip]{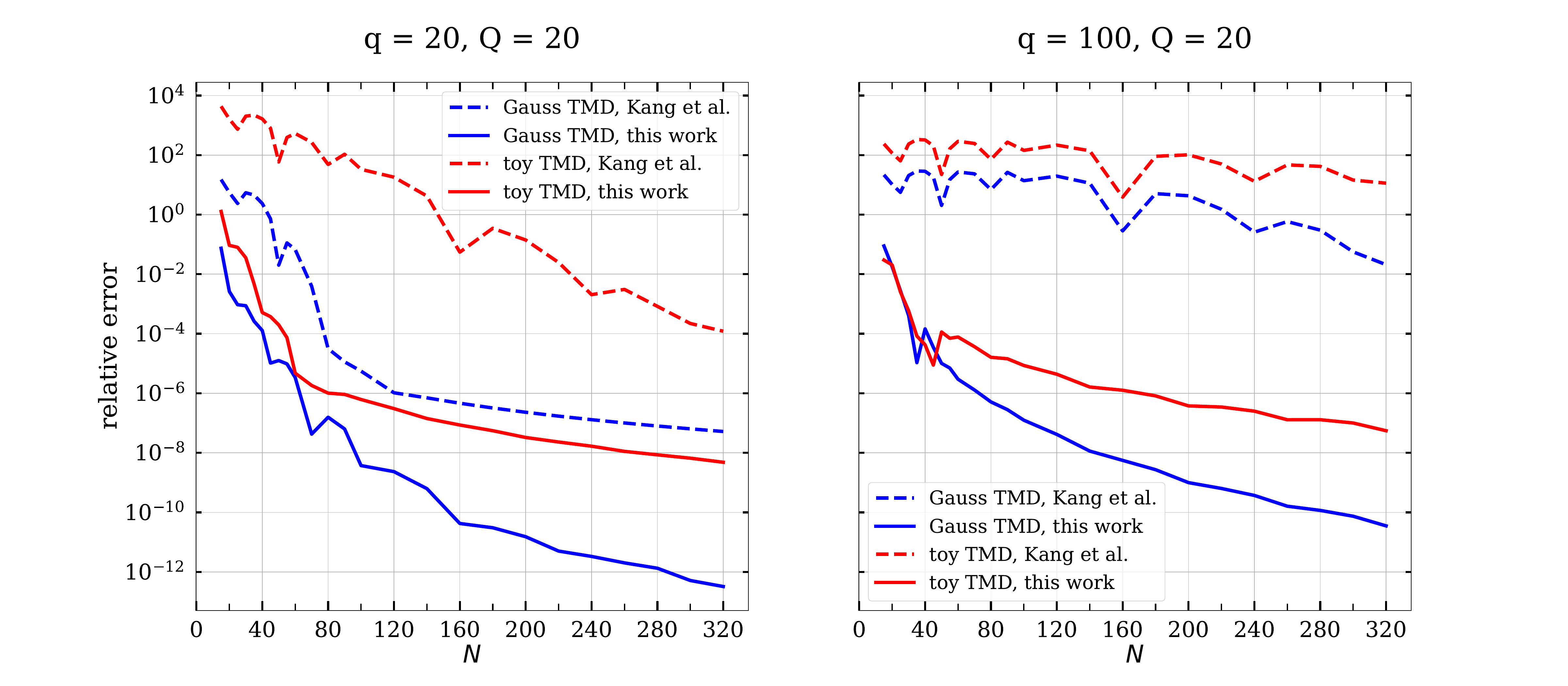}
}
\\[1.5em]
\subfloat[$Q = 100 \gev$, $q = 20 \gev$ (left) and $q = 100 \gev$ (right)]{
   \includegraphics[width=0.86\textwidth,trim=120 0 200 90,clip]{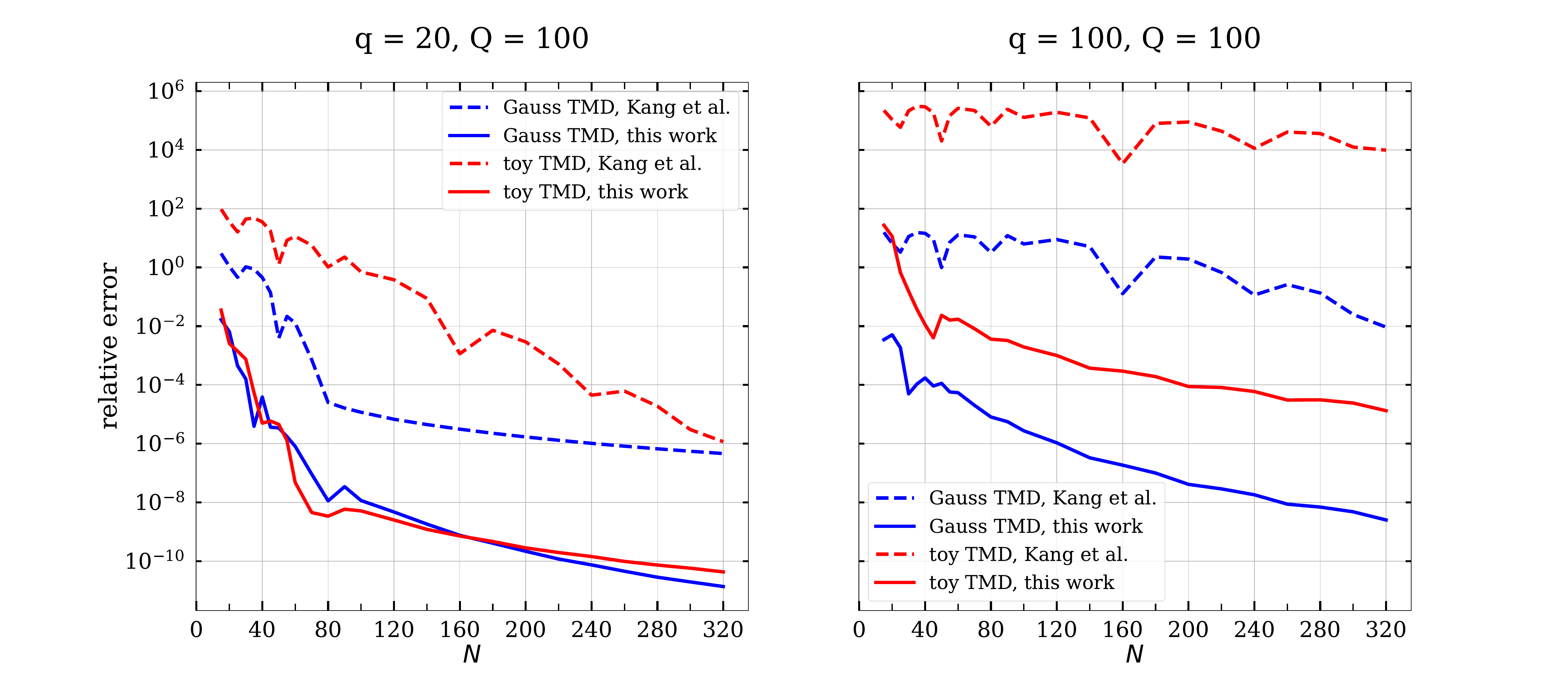}
}
\caption{\label{fig:compare_high_q}  As \fig{\protect\ref{fig:compare_low_q}} but
for $q = 20 \gev$ (left) and $q = 100 \gev$ (right).  $N$ varies in steps of
$20$ for $N \ge 100$.}
\end{figure}

We see in \fig{\ref{fig:compare_low_q}} that for low $q$ the algorithm of
\cite{Kang:2019ctl} works very well.  It often performs better than
our method for small $N$.
The situation is very different for $q = 20 \gev$ and $q = 100 \gev$, as shown
in \fig{\ref{fig:compare_high_q}}.  With $N=60$ points our method achieves
relative errors below $10^{-4}$, except for the toy TMD at $q = Q = 100 \gev$,
where $I(q)$ is close to its zero crossing.
By contrast, the algorithm of \cite{Kang:2019ctl} requires a much higher number
of nodes for $q = 20 \gev$ (well above $100$ for the toy TMD), and at $q = 100
\gev$, no satisfactory result is obtained for the toy TMD even with $320$ nodes.

We observe that the intermediate parameter $h_u$ in the optimisation procedure
of Kang et al.\ grows linearly with $q$ until it reaches its maximum allowed
value $h_{\text{max}} = 2$.  As discussed in \sect{2} of \cite{Kang:2019ctl},
the value of $h_u$ used in the algorithm is increasingly far away from its
optimum as $q$ increases, and a reliable result can then only be achieved by
increasing the number $N$ of function calls.

In summary, we find that both methods require more grid points as $q$ increases,
but that these requirements are much less severe for our method.

\section{Formally divergent integrals}
\label{sec:special-int}

In this section, we take a closer look at Fourier-Bessel integrals from $0$ to
$\infty$ with integrands that do not decrease at large $z$.  The integrals
we have considered so far are either absolutely convergent, or they can be
defined as
\begin{align}
   \label{int-limit-def}
   \int_{0}^{\infty} dz \, J_\nu(z) \, \ftil(z)
   &=
   \lim_{z_b\to \infty} \int_{0}^{z_b} dz \, J_\nu(z) \, \ftil(z)
   \,.
\end{align}
The situation is more complicated if $\ftil(z)$ grows like $\sqrt{z}$ or faster
at large $z$.  An example is the case $\nu=0$ and $\ftil(z) = z$, where
\begin{align}
   \int_{0}^{z_b} dz \, J_0(q z) \, z
   &=
   \frac{z_b}{q} \, J_1(q z_b)
\end{align}
diverges at $z_b \to \infty$ for any $q > 0$, whereas the integral up to
$\infty$ can be defined as a distribution
\begin{align}
   \label{delta-fct}
   \int_0^\infty dz \, J_0(q z) \, z
   &=
   \frac{\delta(q)}{q}
   =
   2 \ms \delta(q^2)
   \,.
\end{align}
The latter relation can be obtained from the representation of a two-dimensional
$\delta$ distribution as the Fourier transform of a constant.

In the context of TMD studies, one encounters Fourier-Bessel transforms with
functions $\ftil(z)$ increasing even faster than $z$.  Writing the fixed-order
perturbative cross section up to order $\alpha_s^n$ as a Fourier-Bessel
integral, one gets an integrand of the form $J_0(q z)$ times
\begin{align}
   \label{ftil-perturb}
   \ftil(z)
   &=
   z \sum_{k=0}^{2 n} c_k \ln^k(z Q)
\end{align}
with coefficients $c_k$ that are independent of $z$.
The corresponding integral can be defined in the sense of distributions, see
\cite[\app{C}]{Ebert:2016gcn} for a systematic investigation and
\cite[\app{B}]{Nadolsky:2001sf} for an earlier reference.

As shown in the papers just quoted, the Fourier-Bessel transform of
\eqref{ftil-perturb} can be carried out analytically, but this is no longer true
for more complicated cases.  For instance, the method of \cite{Cal:2023mib}
involves a smooth interpolation between the form \eqref{ftil-perturb} at $q \sim
Q$ and an expression including all-order resummation at $q \ll Q$.  In this
scheme, the function $\ftil$ depends on both $z$ and $q$.  It generically has a
rather complicated form, and numerical integration is required to compute its
Fourier-Bessel transform.

These considerations motivate us to extend our method to integrands that
increase with $z$.  We do this for general $\nu \ge 0$, which presents only a
modest overhead compared with the case $\nu = 0$ required for the physics
application just sketched.


\subsection{Defining the integral}

In the following we consider functions that satisfy
\begin{align}
   \label{f-cond}
   z^{\nu} \, \Bigl( z\ms \frac{d}{dz} \Bigr)^k \ms \ftil(z, q)
   & \;\underset{z \to 0}{\rightarrow}\;
   \mathit{const}
   \,,
   &
   z^{-3/2} \, \Bigl( z\ms \frac{d}{dz} \Bigr)^k \, \ftil(z, q)
   & \;\underset{z \to \infty}{\rightarrow}\;
   0
   & \qquad
   \text{for } k = 0,1,2,
\end{align}
where it is understood that $\mathit{const}$ may be finite or zero.  If $\ftil$
depends on $q$, we also require that
\begin{align}
   \label{dfdq-cond}
   z^{\nu} \ms \frac{d}{dq} \ms \ftil(z, q)
   & \;\underset{z \to 0}{\rightarrow}\;
   \mathit{const}
   \,,
   &
   z^{-3/2} \, \frac{d}{dq} \ms \ftil(z, q)
   & \;\underset{z \to \infty}{\rightarrow}\;
   0
\end{align}
and
\begin{align}
   \label{deriv-commute}
   \frac{d}{d z} \ms \frac{d}{d q} \, \ftil(z, q)
   &=
   \frac{d}{d q} \ms \frac{d}{d z} \, \ftil(z, q)
   \,.
\end{align}
The integral of interest is
\begin{align}
   \label{sing-int}
   I(q)
   &=
   \int_{0}^\infty dz \, J_{\nu}(q z) \, \ftil(z, q)
\end{align}
with $\nu \ge 0$.  The relation \eqref{Bessel-diff} for Bessel functions of the
first kind implies
\begin{align}
   J_{\nu}(q z)
   &=
   \frac{1}{z} \, \frac{1}{q^{\nu+1}} \, \frac{d}{dq} \,
   \bigl[ q^{\nu+1} \, J_{\nu+1}(q z) \bigr]
   \,,
\end{align}
which gives
\begin{align}
   I(q)
   &=
   \frac{1}{q^{\nu+1}}
   \int_0^\infty dz\, \frac{d}{dq} \,
   \Biggl[ \ms q^{\nu+1} \, J_{\nu+1}(q z) \, \frac{\ftil(z, q)}{z}
   \Biggr]
   - \int_0^\infty dz\, J_{\nu+1}(q z) \;
      \frac{d}{dq} \, \frac{\ftil(z, q)}{z}
\end{align}
when inserted into \eqref{sing-int}.  Interchanging differentiation and
integration, we get
\begin{align}
   \label{sing-int-def}
   I(q)
   &=
   \frac{1}{q^{\nu+1}} \, \frac{d}{dq}
   \Biggl[ \ms q^{\nu+1} \!
   \int_0^\infty dz\, J_{\nu+1}(q z) \, \frac{\ftil(z, q)}{z}
   \Biggr]
   - \int_0^\infty dz\, J_{\nu+1}(q z) \;
      \frac{d}{dq} \, \frac{\ftil(z, q)}{z}
   \,.
\end{align}
The conditions \eqref{f-cond} and \eqref{dfdq-cond} guarantee the convergence of
\eqref{sing-int-def}, but not the one of \eqref{sing-int}.  If $\ftil$ grows so
fast that the integral in \eqref{sing-int} is divergent, then we \emph{define}
that integral by \eqref{sing-int-def}.

Obviously, \eqref{sing-int-def} cannot be used at $q=0$.  In
\app{\ref{sec:distribution}} we show how \eqref{sing-int-def} can be extended to
a distribution in $q$, with a plus-prescription and a $\delta$ distribution at
$q=0$.  In the remainder of this section we consider the region $q>0$.

Let us rewrite \eqref{sing-int-def} such that explicit differentiation w.r.t.\
$q$ is not necessary.  We start by inserting
\begin{align}
   \label{replace-J-nu-p1}
   J_{\nu+1}(q z)
   &=
   \frac{1}{q} \, \frac{1}{z^{\nu+2}} \, \frac{d}{dz} \,
   \bigl[ z^{\nu+2} \, J_{\nu+2}(q z) \bigr]
\end{align}
into \eqref{sing-int-def} and use integration by parts in both terms.  Together
with the limiting behaviour of the Bessel functions at small and large $z$, the
conditions \eqref{f-cond} and \eqref{dfdq-cond} ensure that the boundary terms
vanish, so that we get
\begin{align}
   I(q)
   &=
   - \frac{1}{q^{\nu+1}} \, \frac{d}{dq}
   \Biggl[ \, q^{\nu} \! \int_0^\infty dz\, J_{\nu+2}(q z) \,
   z^{\nu+2} \, \frac{d}{dz} \, \frac{\ftil(z, q)}{z^{\nu+3}}
   \Biggr]
   \notag \\[0.2em]
   & \quad
   + \frac{1}{q}
      \int_0^\infty dz\, J_{\nu+2}(q z) \;
      z^{\nu+2} \frac{d}{dz} \, \frac{d}{dq} \, \frac{\ftil(z, q)}{z^{\nu+3}}
   \,.
\end{align}
Using the product rule for the $q$ derivative in the first term, we get
\begin{align}
   \label{int-sing-med}
   I(q)
   &=
   - \frac{1}{q^{\nu+1}}
      \int_0^\infty dz\; \frac{d}{dq}
      \biggl[ q^{\nu} J_{\nu+2}(q z) \biggr] \,
   z^{\nu+2} \, \frac{d}{dz} \, \frac{\ftil(z, q)}{z^{\nu+3}}
   \,,
\end{align}
where we have used the condition \eqref{deriv-commute}. Remarkably, terms that
involve the $q$ dependence of $\ftil$ have disappeared.  In this sense, our
procedure can be regarded as local in $q$.
The two relations in \eqref{Bessel-basic} imply
\begin{align}
   \frac{d}{d q} \, J_{\nu+2}(q z)
   &=
   z J_{\nu+1}(q z) - \frac{\nu+2}{q} \, J_{\nu+2}(q z)
   \,,
\end{align}
which allows us to replace the remaining $q$ derivative in \eqref{int-sing-med}.
The result is
\begin{align}
   I(q)
   &=
   \frac{1}{q^2}
   \int_0^\infty dz\, \bigl[ 2 J_{\nu+2}(q z) - q z \ms J_{\nu+1}(q z) \bigr] \,
   z^{\nu+2} \, \frac{d}{dz} \, \frac{\ftil(z, q)}{z^{\nu+3}}
   \,.
\end{align}
We now use \eqref{replace-J-nu-p1} and integrate by parts once more.  The
boundary terms vanish again by virtue of \eqref{f-cond} and \eqref{dfdq-cond},
so that
\begin{align}
   I(q)
   &=
   \frac{1}{q^2}
   \int_0^\infty dz\, J_{\nu+2}(q z) \, z^{\nu+2} \;
   \frac{d}{dz}  \biggl[
      \frac{2 \ms \ftil(z, q)}{z^{\nu+3}}
      + z\, \frac{d}{dz} \, \frac{\ftil(z, q)}{z^{\nu+3}}
   \biggr]
   \,.
\end{align}
This can be brought into the standard form \eqref{int-final-def},
\begin{align}
   \label{int-sing-master}
   I(q)
   &=
   \frac{1}{q^2}
   \int_0^\infty dz\, J_{\nu+2}(q z) \, \biggl( \frac{1+z}{z} \biggr)^{\nu+2}
   f_1(z)
\end{align}
with
\begin{align}
   \label{f1-from-ftilde}
   f_1(z)
   &=
   \frac{z^{\nu}}{(1+z)^{\nu + 2}} \,
   \biggl[
      z^2 \frac{d^2}{d z^2}\, \ftil(z)
   - (2\nu + 3) \, z\ms \frac{d}{dz}\, \ftil(z)
   + (\nu + 3) (\nu + 1) \, \ftil(z)
   \biggr]
   \,,
\end{align}
where from now on we drop the $q$ dependence of $\ftil$ again, as we do in the
rest of this paper.  With the conditions in \eqref{f-cond}, the function
$f_1$ is finite or zero at the integration boundaries, so that the integral
\eqref{int-sing-master} exists in the usual sense and can be evaluated using our
adaptation of Levin's method in \eqref{int-final-def} and
\eqref{Levin-int-final}.


\paragraph{Discretisation.}
For discretising the integrand, we introduce the scaled function
\begin{align}
   \label{g-def}
   g(z)
   &=
   \biggl( \frac{z}{1+z} \biggr)^{\nu} \, \; \frac{\ftil(z)}{(1+z)^2}
   \,,
\end{align}
which according to \eqref{f-cond} is finite or zero at the integration
boundaries.  Inserting this into \eqref{f1-from-ftilde}, we get
\begin{align}
   \label{f1-from-g}
   f_1(z)
   &=
   \Bigl( z\ms \frac{d}{dz} \Bigr)^2 \; g(z)
   - 2\, \biggl[ \nu + \frac{\nu+2}{1+z} \biggr] \,
      z\ms \frac{d}{dz}\, g(z)
   \notag \\[0.3em]
   & \quad
   + \biggl[ \nu^2-1 + \frac{(2\nu+1) (\nu+2)}{1+z}
   + \frac{(\nu+1) (\nu+2)}{(1+z)^2} \biggr] \,
      \, g(z)
   \,.
\end{align}
For the variable transform from $z$ to $u$ used for discretisation, we require
that
\begin{align}
   z \, \frac{d u}{d z}
   &
   \,\underset{z \to \infty}{\longrightarrow}\,
   0
   \,,
\end{align}
which is fulfilled for all cases in \tab{\ref{tab:var-trf}}.  This allows us to
approximate the derivatives in \eqref{f1-from-g} using Chebyshev interpolation:
\begin{align}
   z\, \frac{d}{dz} \, g(z) \,\biggr|_{z = z_j}
   & \approx \sum_{k} D^{(1)}_{j k} \; g(z_k)
   \,,
   &
   \biggl( z\ms \frac{d}{dz} \biggr)^2 \; g(z) \,\biggr|_{z = z_j}
   & \approx \sum_{k} D^{(2)}_{j k} \; g(z_k)
   \,,
\end{align}
with
\begin{align}
   D^{(1)}_{j k}
   &=
   z_j \, \frac{du}{dz} (z_j) \, D^u_{j k}
   \,,
   &
   D^{(2)}_{j k}
   &=
   \sum_{l} D^{(1)}_{j l} \, D^{(1)}_{l k}
   \,.
\end{align}
In close analogy with \eqref{IBP-f1-f0-discrete}, we can thus use
\eqref{f1-from-g} to compute $f_1(z_j)$ by multiplying $g(z_k)$ with a matrix.
This matrix depends only on $\nu$ and the interpolation grid and can hence be
precomputed.


\paragraph{Generalisation.}
We note that using the relation
\begin{align}
   J_\nu(q z)
   &=
   \frac{1}{z^j} \, \frac{1}{q^\nu}
   \biggl( \frac{1}{q}\, \frac{d}{dq} \biggr)^j
      \bigl[ q^{\nu + j} \, J_{\nu+j}(q z) \bigr]
\end{align}
with a suitable value of $j$, it is straightforward to generalise the previous
method to functions $\ftil$ growing like $z^{3/2}$ or faster as $z \to \infty$.
We checked explicitly that with $j=2$ and two steps of integration by parts, one
obtains a representation of \eqref{sing-int} as an integral over $J_{\nu+3}(q
z)$ times a function depending on $(d/dz)^{k} \, \ftil$ with $k=0,1,2,3$.


\subsection{Numerical accuracy}
The evaluation of the integral \eqref{int-sing-master} using the discretised
version of \eqref{g-def} and \eqref{f1-from-g} is implemented as a separate
class for ``special integrals'' in \bestlime. The integration is performed using
the $J_\nu$ call of the standard Bessel integrator of the library.

The timing of integration calls for the special integrator agrees within $10\%$
with the timing of $J_{\nu-1}$ calls for the standard integrator.  This is
plausible, because in both cases the discretised integrand is multiplied by a
matrix for differentiation, before being passed to the algorithm that implements
Levin's method.

Let us investigate the accuracy of our method for the functions $\ftil(z) = z
\ln^k (z^2 \ms Q^2 / b_0^2)$, which form a basis of the fixed-order perturbative
expression \eqref{ftil-perturb}.  The constant $b_0$ is defined in
\eqref{b0-def} and gives a simpler form of the Fourier-Bessel transform.  The
exact result for $I(q)$ at $q > 0$ is readily obtained by evaluating
\eqref{sing-int-def} analytically, which can be done using
\begin{align}
   (\ln z)^k
   &=
   (d / dz)^k \, z^\mu \,\bigr|_{\mu=0}
\end{align}
and
\begin{align}
   \int_0^\infty dz\; J_{\nu+1}(q z) \, z^\mu
   &=
   \frac{\Gamma\bigl( \frac{\nu}{2} + \frac{\mu}{2} + 1 \bigr)}{
         \Gamma\bigl( \frac{\nu}{2} - \frac{\mu}{2} + 1 \bigr)} \,
   \frac{2^\mu}{q^{\mu+1}}
   \,.
\end{align}
Explicit results can also be found in \cite[\app{C}]{Ebert:2016gcn}.  In the
following, we set the hard scale to $Q = 100 \gev$.

\begin{figure*}
\centering
\subfloat[\label{fig:log-trf-nu0-k4}
   $\int dz \, J_0(q z) \, z \, \ln^k w(z)$, \quad $k = 1,2,3,4$]{
   \includegraphics[width=0.45\textwidth,trim=0 0 0 55,clip]{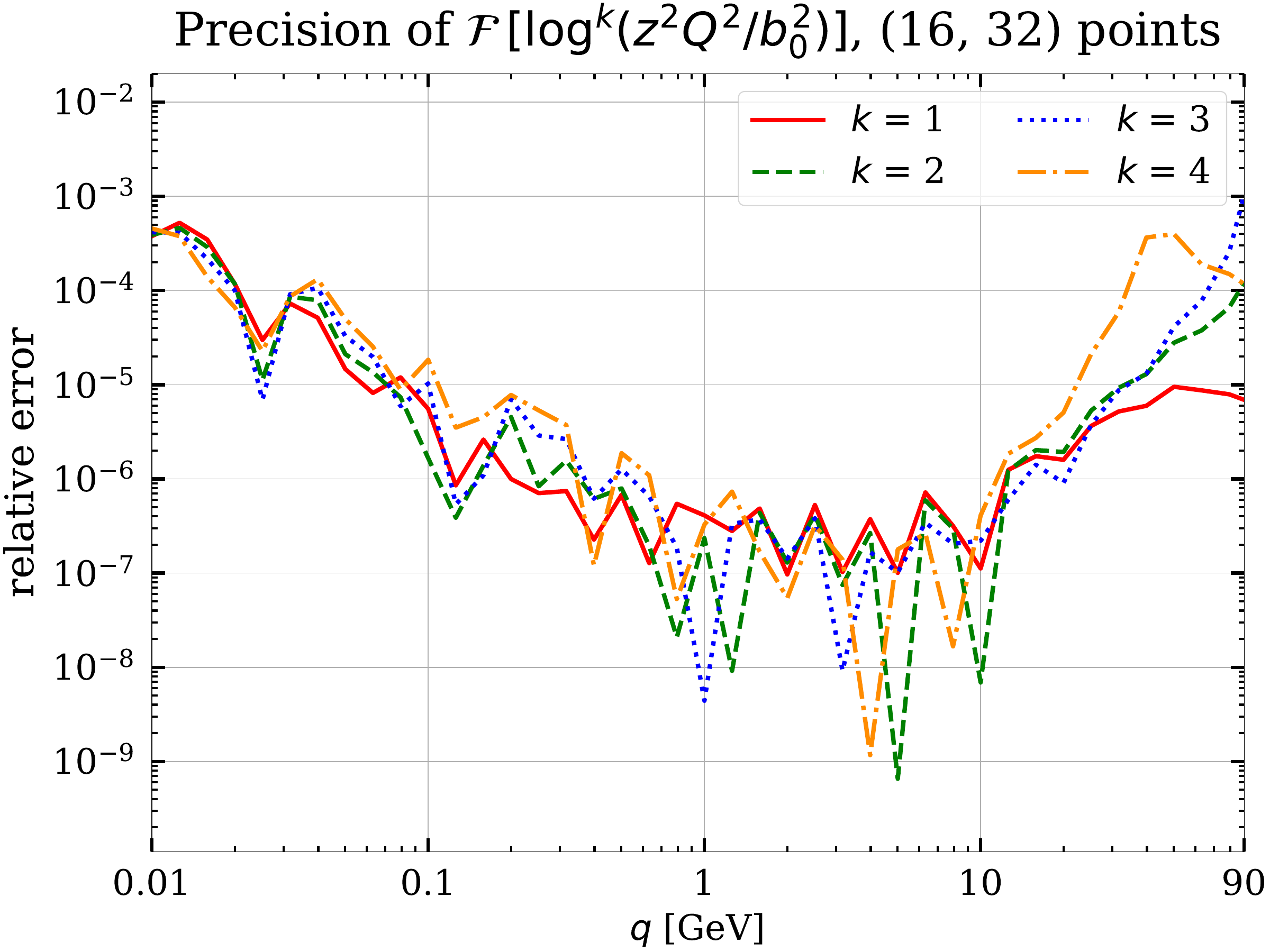}
}
\hspace{0.5em}
\subfloat[\label{fig:log-trf-nu0-k8}
   $\int dz \, J_0(q z) \, z \, \ln^k w(z)$, \quad $k = 5,6,7,8$]{
   \includegraphics[width=0.45\textwidth,trim=0 0 0 55,clip]{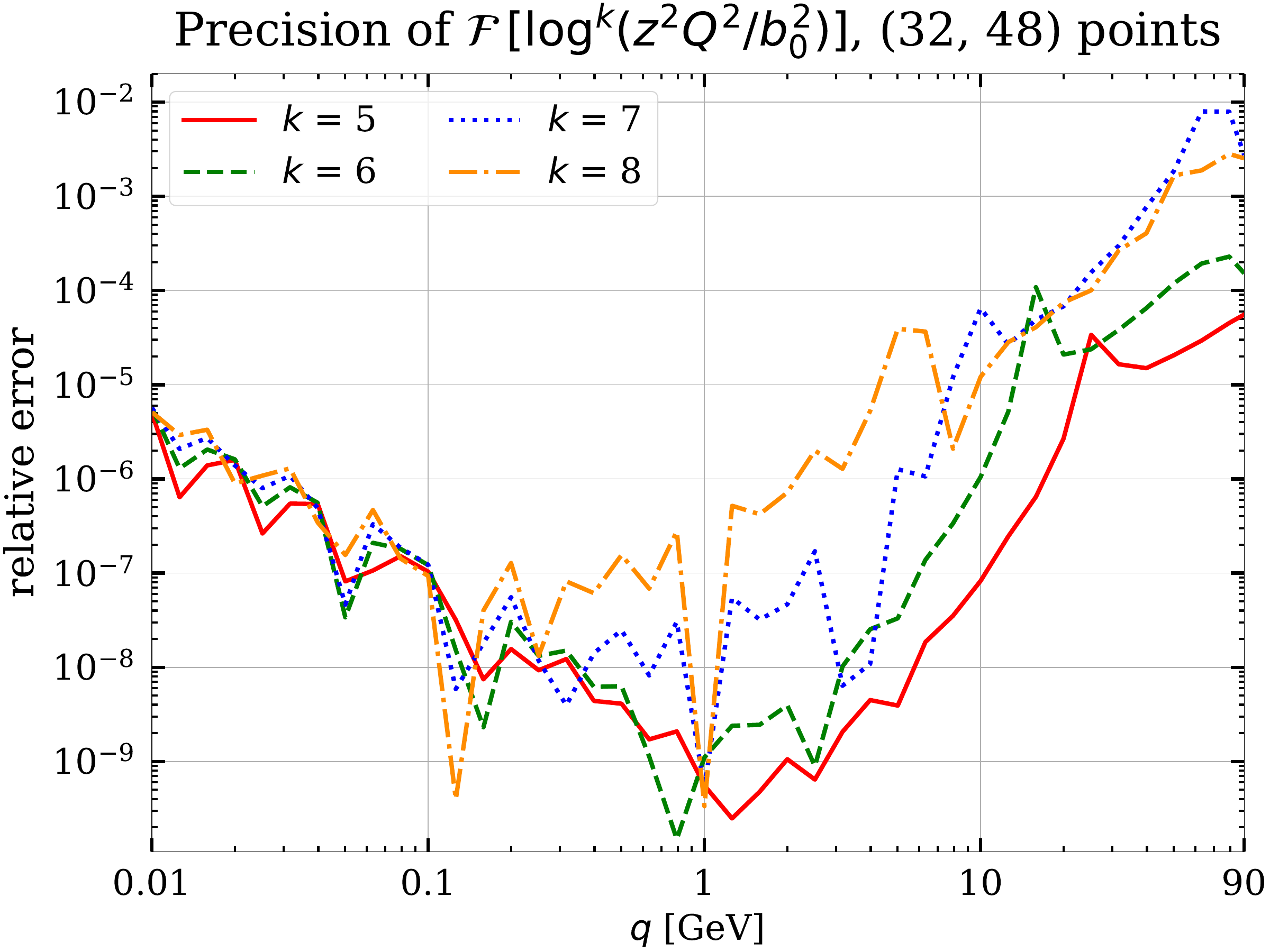}
}
\\[0.5em]
\subfloat[\label{fig:log-trf-nu012-k1}
   $\int dz \, J_\nu(q z) \, z \, \ln w(z)$, \quad $\nu = 0,1,2$]{
   \includegraphics[width=0.45\textwidth,trim=0 0 0 65,clip]{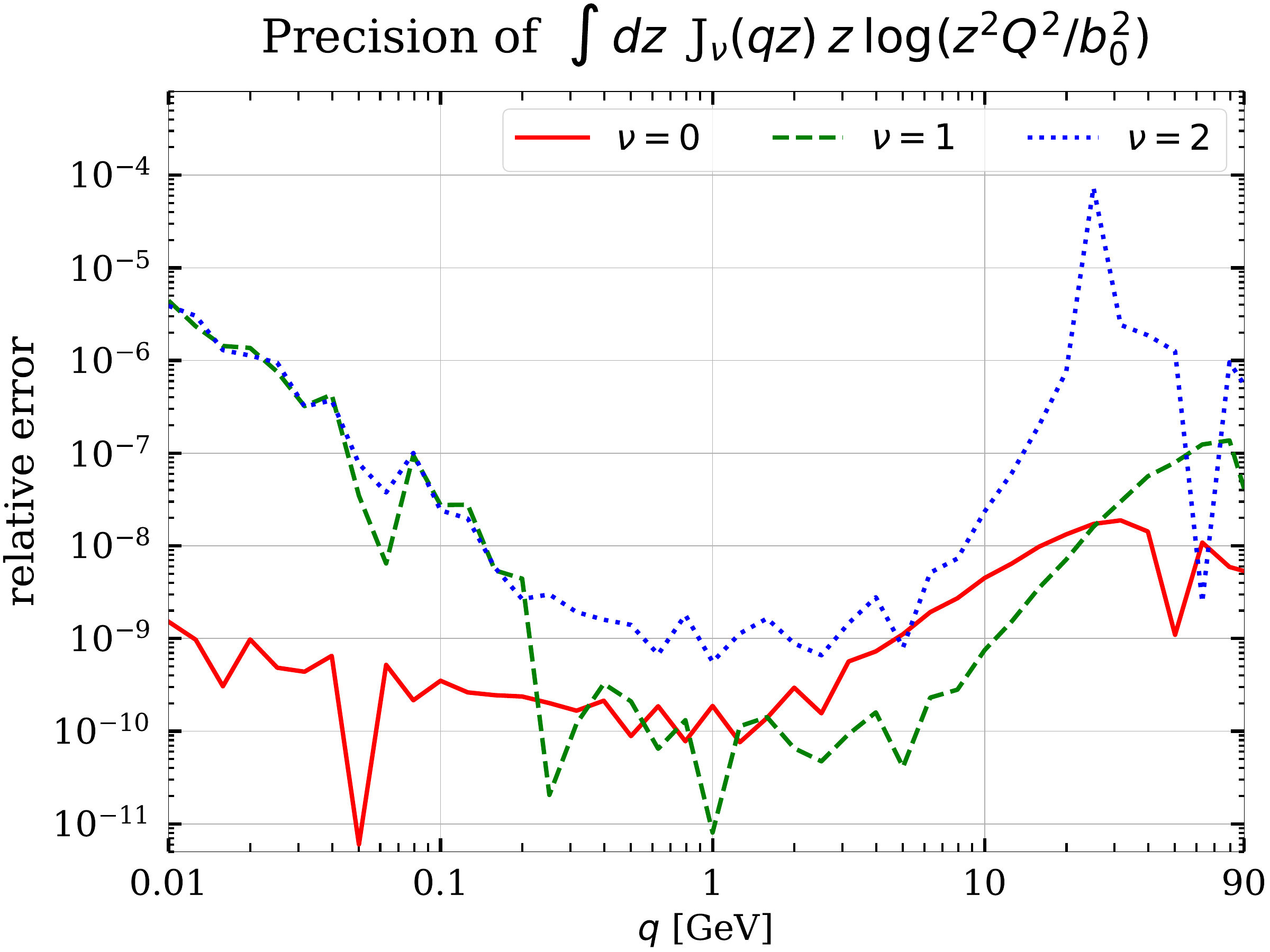}
}
\hspace{0.5em}
\subfloat[\label{fig:log-trf-nu012-k2}
   $\int dz \, J_\nu(q z) \, z \, \ln^2 w(z)$, \quad $\nu = 0,1,2$]{
   \includegraphics[width=0.45\textwidth,trim=0 0 0 65,clip]{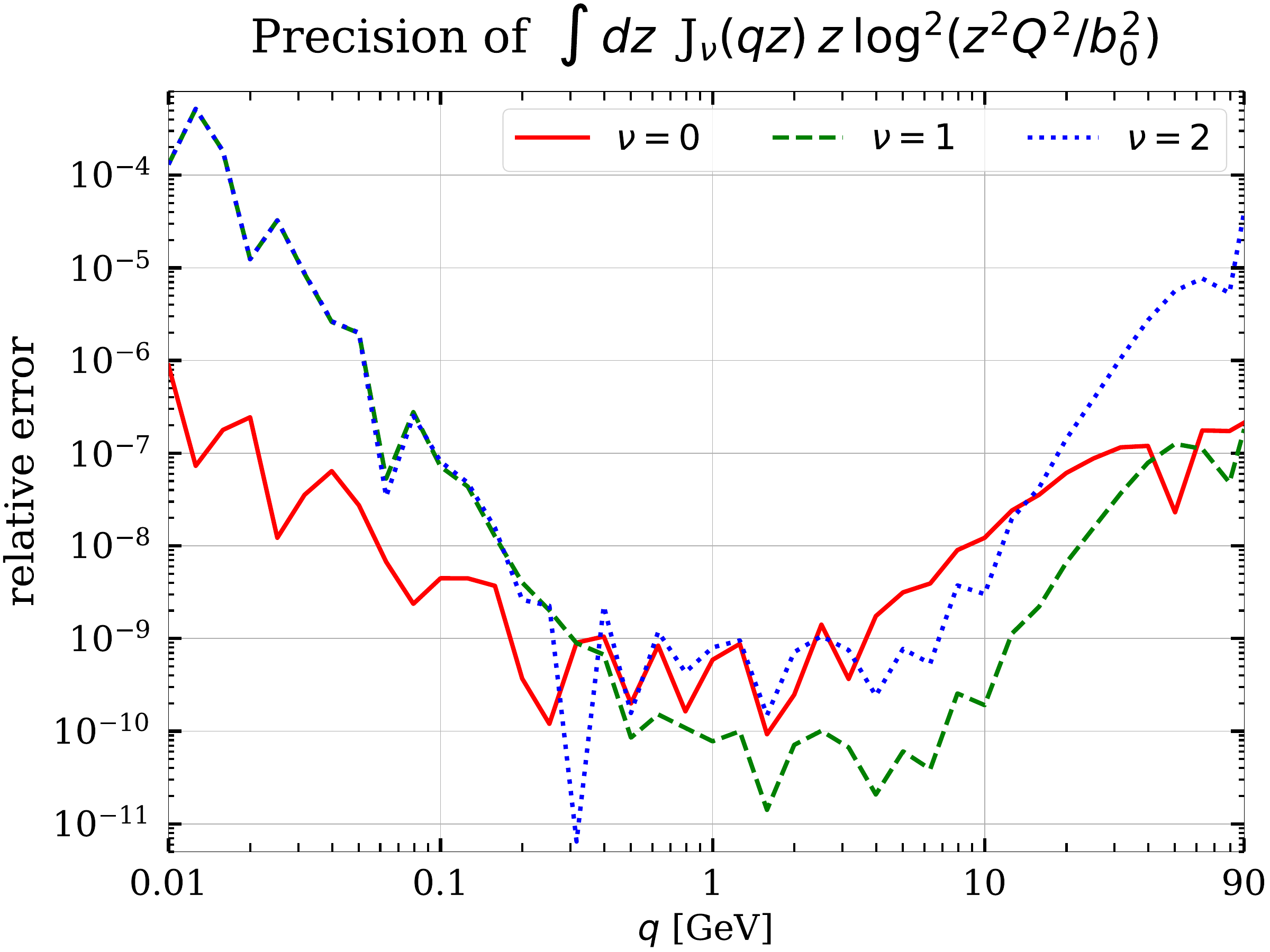}
}
\caption{\label{fig:log-trf} Relative accuracy of the integral
\protect\eqref{sing-int} evaluated via \protect\eqref{int-sing-master} and
\protect\eqref{f1-from-ftilde} for different functions.  The integrands are
specified in the legends, where we abbreviate $w(z) = (z \ms Q / b_0)^2$.  The
grid settings are specified in the text.}
\end{figure*}

We find that for $k \le 4$, quite good integration accuracy can be achieved
with the \texttt{inv pow} transformation that we already used for the benchmark
integrals in \sect{\ref{sec:benchmarks}}.  For higher powers of the logarithm,
the results become however increasingly inaccurate unless one takes a very large
number of grid points.  A much better performance for all $k$ is obtained with
the \texttt{log pow} transformation in \tab{\ref{tab:var-trf}}.
For $\nu = 0$ and $k$ up to $8$ (corresponding to a perturbative calculation up
to $\alpha_s^4$ in the TMD context), we take a
\begin{align}
   \label{log-trf-grids}
   &
   \text{\texttt{log pow} grid with } [0, 0.05, \infty]_{(n_1, n_2)}
   \,,
   &
   \alpha = 0.2 ,
   \quad
   z_{\text{lo}} = 10^{-8} ,
   \quad
   z_{\text{hi}} = 0.1
\end{align}
and
\begin{align}
   (n_1, n_2)
   &=
   \begin{cases}
      (16, 32) & \text{ for } k = 1,2,3,4, \\
      (32, 48) & \text{ for } k = 5,6,7,8,
   \end{cases}
\end{align}
where the transformation parameters and the subinterval division have been
determined by a parameter scan similar to the one presented in
\sect{\ref{sec:params}}.  The resulting integration accuracy is shown in
\fig{\ref{fig:log-trf-nu0-k4}} and \ref{fig:log-trf-nu0-k8}.  We plot the
accuracy only up to $q = 90 \gev$, because for some values of $k$ the integrals
have a zero crossing near $q = 100 \gev$.

In \fig{\ref{fig:log-trf-nu012-k1}} and \fig{\ref{fig:log-trf-nu012-k2}} we show
that our method also works well for $\nu > 0$.  The grid used for these cases is
a
\begin{align}
   \label{log-trf-grids-012}
   &
   \text{\texttt{log pow} grid with } [0, 0.1, \infty]_{(24, 48)}
   \,,
   &
   \alpha = 0.25 ,
   \quad
   z_{\text{lo}} = 5 \times 10^{-7} ,
   \quad
   z_{\text{hi}} = 0.1
   \,.
\end{align}
We note that with increasing order $\nu$ of $J_\nu(q z)$, the region close to
$z=0$ is increasingly suppressed in the integral, so that a single grid setting
for different $\nu$ is a compromise to some extent.

\begin{figure*}
\centering
\subfloat[$\int dz \, J_0(q z) \, z \, W_{\text{toy}}(z)$]{
   \includegraphics[width=0.45\textwidth,trim=0 0 0 55,clip]{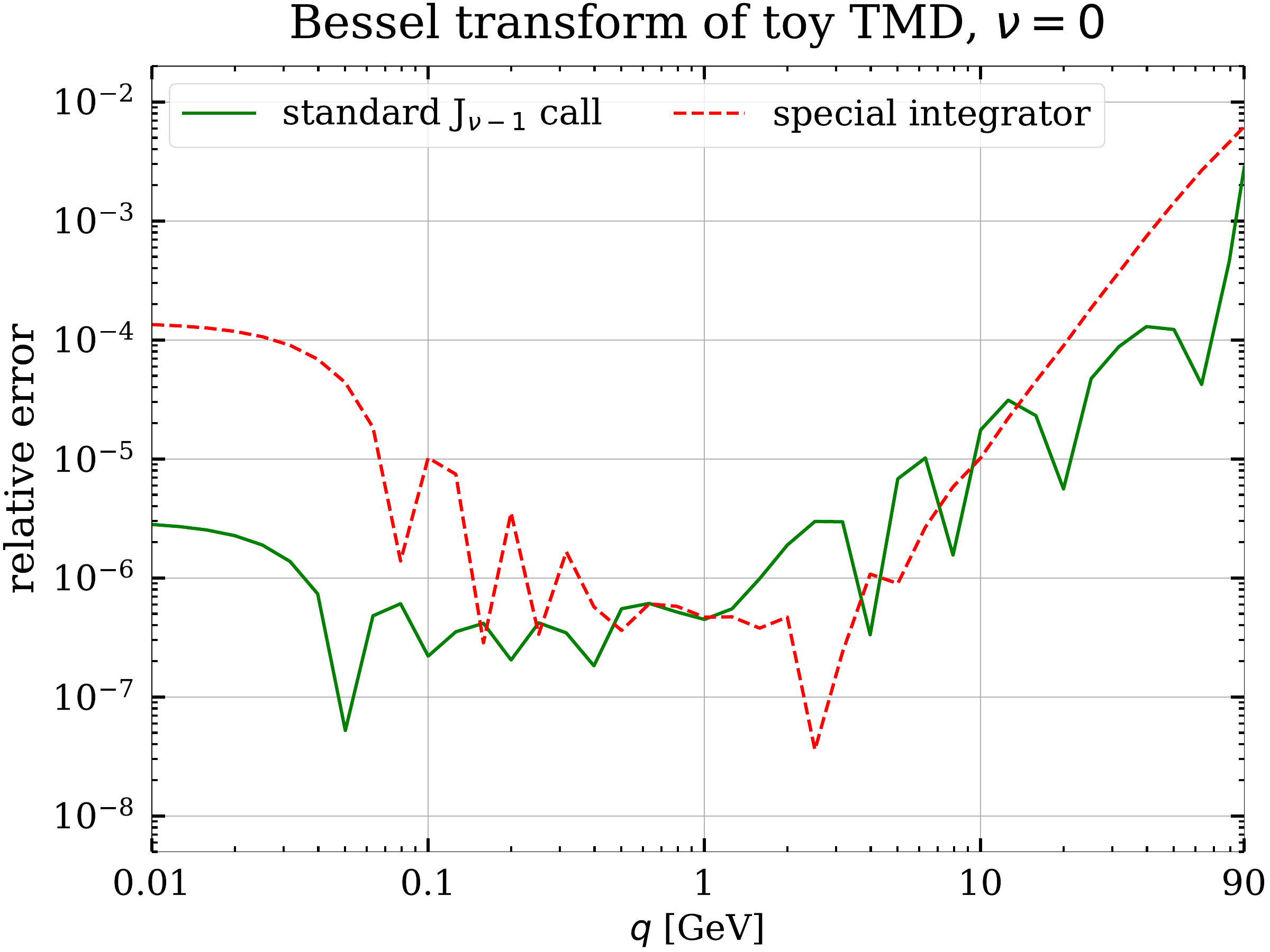}
}
\hspace{0.5em}
\subfloat[\label{fig:method-comp-cumul}
   $q \ms \int dz \, J_1(q z) \, W_{\text{toy}}(z)$]{
   \includegraphics[width=0.45\textwidth,trim=0 0 0 55,clip]{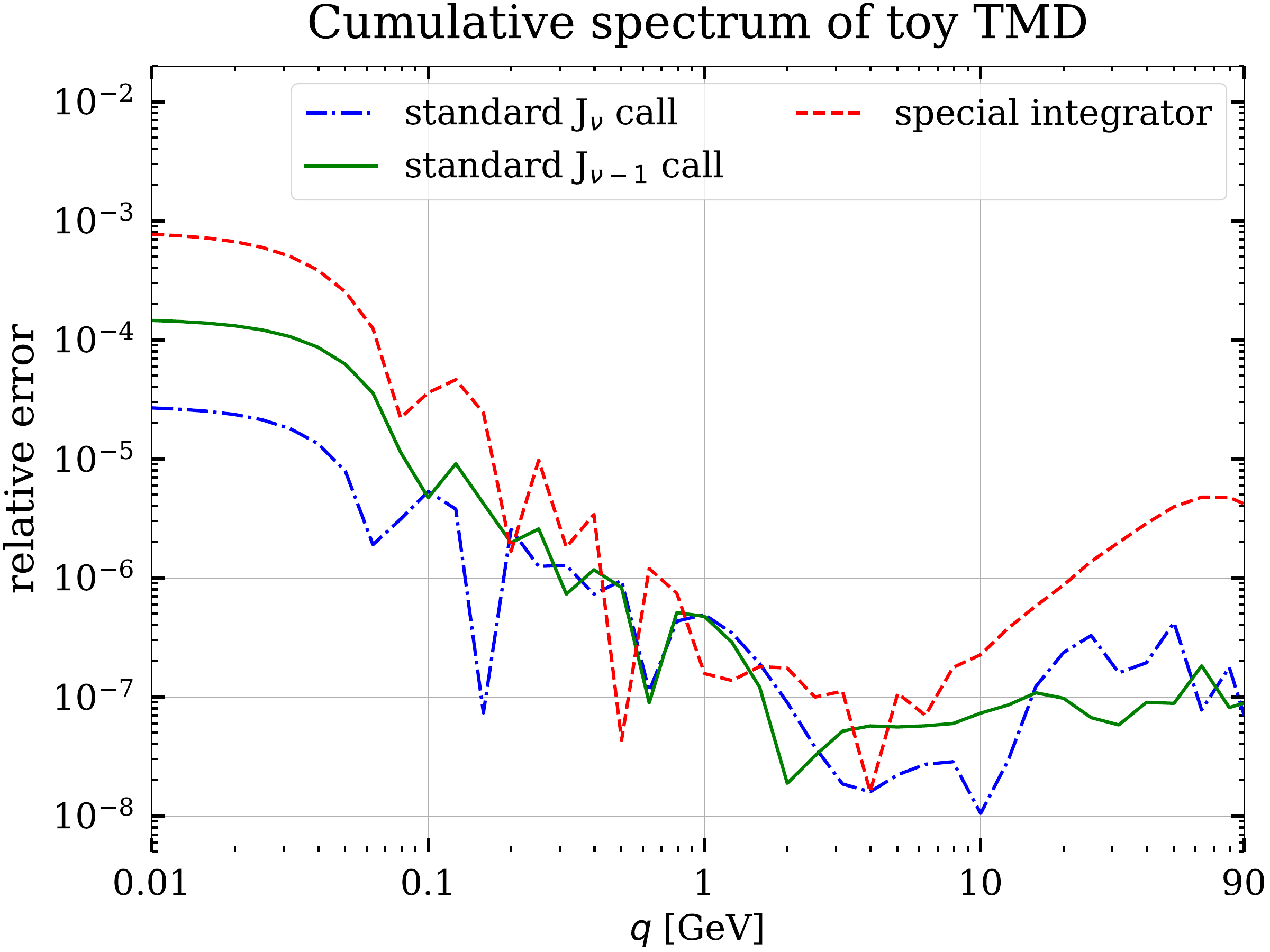}
}
\caption{\label{fig:method-comp} Comparison of the accuracy obtained with
different integration methods for the $q$ spectrum (left) and the cumulative $q$
spectrum (right) of the cross section computed with the toy TMD in
\protect\eqref{toy-TMD-W}.  More detail is given in the text.}
\end{figure*}

The derivation of our master formulae \eqref{int-sing-master} and
\eqref{f1-from-ftilde} is of course also valid if the original integral
\eqref{sing-int} is convergent.  In this case, we can compare the respective
performance of our standard method from \sect{\ref{sec:method-final}} and of
the method developed here.  This is done in \fig{\ref{fig:method-comp}} for the
toy TMD specified in \eqref{toy-TMD-W} to \eqref{toy-TMD-params}. Shown is the
accuracy for the ``special integrator'' described in the present section and for
the $J_{\nu-1}$ call of our standard integrator.  The curves in
\fig{\ref{fig:method-comp-cumul}} are for an integral involving $J_1(q z)$, so
that we can also compare with the $J_\nu$ call of our standard integrator.
Whilst we see no pronounced difference in performance at intermediate $q$
values, there is some trend for the special integrator to be slightly less
precise at very small or very large $q$.  This may be due to the fact that two
numerical derivatives are taken in that case.
The results in \fig{\ref{fig:method-comp}} are for an
\begin{align}
   \label{special-exp-sqrt-grid}
   &
   \text{\texttt{exp sqrt} grid with } [0, 0.05, \infty]_{(16, 32)}
   \,,
   &
   m = r_\kappa \, \kappa ,
   \qquad
   r_\kappa = 4 ,
\end{align}
which coincides with our default settings in \eqref{tmd-grid-default} to
\eqref{tmd-param-default} except for the value of $r_\kappa$.  With our default
choice $r_\kappa = 3$, the discrepancy between the special and standard
integrators becomes even larger at small $q$.  A more thorough comparison
between the methods should cover a broader range of integrands and allow
different grid settings for the different methods.  We do not pursue this here,
because the main purpose of the special integrator is to evaluate integrals that
cannot be handled with the standard one.

\section{Summary}
\label{sec:summary}

We have presented a method to compute Fourier-Bessel transforms, $\int dz \,
J_\nu(q z) \, \ftil(z)$, with $z$ ranging from $0$ to $\infty$ or being
restricted to a smaller interval.  The method works for general index $\nu \ge
0$ of the Bessel function.  The function $\ftil(z)$ may be non-analytic at
$z=0$ (details are given in \sect{\ref{sec:final_scaling_z_behaviour}}), and for
$z\to \infty$ it should decrease or tend to a constant.

Our method is a modification of an algorithm by Levin \cite{Levin:1996}.  For
$\nu \ge 1$ the integration problem is transformed to the system of ODEs given
in \eqref{ODE-final}, and integration by parts provides access to Bessel
functions of order $\nu < 1$ as specified in \eqref{IBP}.  The system of ODEs is
solved by collocation, with Chebyshev polynomials in a transformed variable
$u(z)$ as basis functions.  The resulting matrix equation is handled with
standard linear algebra methods.  \rev{The derivative required for integration
by parts is evaluated by approximating $\ftil(z)$ with the same Chebyshev
polynomials.}  As a complement, we use Clenshaw-Curtis
quadrature in $z$ intervals where $q z$ does not exceed the first zero of
$J_\nu(x)$.  This avoids most of the cases in which this matrix equation is
poorly conditioned, so that its solution would require a numerically expensive
singular value decomposition.
The details of our algorithm are laid out in \sect{\ref{sec:algorithm}} and
implemented in the C++ library \bestlime, which can be downloaded from
\cite{BestLime:release}.

The method requires a number of presets to be decided by the user.  The first
one is a variable transform $u(z)$ that maps the integration interval in $z$
onto a finite interval in $u$ and should be adapted to the large-$z$ behaviour
of $\ftil(z)$.  The $z$-grid on which $\ftil(z)$ must be evaluated is then
specified by the number of points, with the possibility of dividing the total
integration range into several subintervals.  As we have shown, such a
subdivision is beneficial for both the accuracy and the computational cost of
integration.  This set of choices is admittedly more involved than for other
methods (the Ogata method \cite{Ogata:2005} for instance requires a step
parameter $h$ and the total number $N$ of function evaluations).  On the other
hand, we found that with about $50$ to $80$ grid points, a high integration
accuracy can be reached for a wide range of grid parameters and over a wide
range of $q$, so that a detailed tuning of the grid settings should in general
not be necessary.  Settings that we found to be satisfactory are given in
\sect{\ref{sec:benchmarks}} for a wide class of functions, and in
\eqs{\eqref{tmd-grid-default}} to \eqref{tmd-param-default} for integrands
appearing in TMD cross sections.

\rev{In fixed-order perturbative TMD calculations one encounters functions
$\ftil(z)$ that grow with $z$, such that the corresponding Fourier-Bessel
integrals do not converge.  As shown in \sect{\ref{sec:special-int}} and
\app{\ref{sec:distribution}}, these integrals can be defined in the
distributional sense, and their regular parts at $q>0$ can be rewritten such
that they can be evaluated numerically with our method.  We obtain good accuracy
for $\ftil(z) = z \, \ln^k \! z$ with $k$ up to $8$, which is the highest power
appearing at order~$\alpha_s^4$.}

\rev{As explained in \sect{\ref{sec:algorithm}},} our method is most efficient
for computing Bessel-weighted integrals of several
functions $\ftil(z)$ at the same value of $q$.  Changing the value of $q$
requires somewhat more computation time for linear algebra manipulations, but
does \emph{not} necessitate a new evaluation of the function(s) $\ftil(z)$,
since the grid in $z$ does not depend on $q$.  This presents an important
advantage if the function $\ftil(z)$ is expensive to compute.  In an ongoing
separate project, we found that computing two-fold Fourier-Bessel transforms
$\int d z_1 \int d z_2\, J_{\nu_1}(q_1 z_1) \, J_{\nu_2}(q_2 \ms z_2) \,
\ftil(z_1, z_2)$ can be done accurately and efficiently with the implementation
of our method in \bestlime.

\section*{Acknowledgements}

It is a pleasure to thank Frank Tackmann, Alexey Vladimirov, and Pavel Nadolsky
for useful discussions.
The implementation of our method in \bestlime\ is heavily drawing on the
\chilipdf\ library \cite{Diehl:2021gvs, Diehl:2023cth}, which is under
development.  We gratefully acknowledge the contributions of our collaborators
Florian Fabry, Peter Pl\"o{\ss}l, Riccardo Nagar, and Frank Tackmann to that
project.  Special thanks go to Riccardo Nagar and Frank Tackmann for their
consent to releasing the present code before a full release of \chilipdf.

This work is in part supported by the Deutsche Forschungsgemeinschaft (DFG,
German Research Foundation) -- grant number 409651613 (Research Unit FOR 2926)
and grant number 491245950.
O.G.\ is supported by the German Academic Scholarship Foundation.

\appendix

\section{Useful relations for Bessel and Lommel functions}
\label{sec:functions}

In this appendix, we collect results about Bessel and Lommel functions that are
used in the main body of our work.

\subsection{Bessel functions}

The properties of Bessel functions are well documented in the literature.  The
following relations can for instance be found in
\cite[\href{https://dlmf.nist.gov/10}{Chapter 10}]{NIST:DLMF}.

\paragraph{Relations.}

Functions of different order are related by
\begin{align}
   \label{Bessel-basic}
   Z_{\nu}(x)
   &=
   \frac{x}{2 \nu} \, \Bigl[ Z_{\nu-1}(x) + Z_{\nu+1}(x) \Bigr]
   \,,
   \notag \\[0.5em]
   \frac{d}{dx} \, Z_{\nu}(x)
   &=
   \frac{1}{2} \, \Bigl[ Z_{\nu-1}(x) - Z_{\nu+1}(x) \Bigr]
   \,,
\end{align}
which implies
\begin{align}
   \label{Bessel-diff}
   \frac{d}{dx} \ms \Bigl[ x^\nu Z_{\nu}(x) \Bigr]
   &=
   x^\nu \, Z_{\nu-1}(x)
\end{align}
and
\rev{
\begin{align}
   \label{Bessel-diff-2}
   \frac{d}{dx} \ms \Bigl[ x^{-\nu} Z_{\nu}(x) \Bigr]
   &=
   - x^{-\nu} \, Z_{\nu+1}(x)
   \,,
\end{align}
where} $Z_{\nu}$ stands for $J_{\nu}$ or $Y_{\nu}$, i.e.\ for the Bessel
functions of the first or the second kind.  The two sets of functions are
related by
\begin{align}
   Y_{\nu}(x)
   &=
   \frac{\cos(\nu \pi) \, J_{\nu}(x) - J_{-\nu}(x)}{\sin(\nu\pi)}
\end{align}
for non-integer $\nu$.  For $n \in \mathbb{Z}$, one can take the limit $\nu \to
n$ of the r.h.s.\ to define $Y_{n}$.


\paragraph{Behaviour at large arguments.}

The leading asymptotic behaviour for $x\to \infty$ of $J_\nu(x)$ and $Y_\nu(x)$
consists of oscillations and a global decrease like $x^{-1/2}$, namely
\begin{align}
   \label{Bessel-J-Y-large-x}
   J_{\nu}(x)
   &
   \sim
   \sqrt{\frac{2}{\pi x}} \,
   \cos\Bigl( x - \nu \ms \frac{\pi}{2} - \frac{\pi}{4} \Bigr)
   \,,
   &
   Y_{\nu}(x)
   &
   \sim
   \sqrt{\frac{2}{\pi x}} \,
   \sin\Bigl( x - \nu \ms \frac{\pi}{2} - \frac{\pi}{4} \Bigr)
   \,.
\end{align}


\paragraph{Series representations.}

For $\nu \ge 0$ one has
\begin{align}
   \label{Bessel-J-Y-series}
   x^\nu \ms J_{\nu}(x)
   &=
   x^{2 \nu} \, \ser(x^2)
   \,,
   \notag \\
   x^\nu \, Y_{\nu}(x)
   &=
   \begin{cases}
      \ser(x^2) + x^{2 \nu} \ms \ln(x) \, \ser(x^2)
         & \text{ for } \nu \in \mathbb{Z}
      \,,
      \\
      \ser(x^2) + x^{2 \nu} \, \ser(x^2)
         & \text{ for } \nu \notin \mathbb{Z}
      \,,
   \end{cases}
\end{align}
where we use our generic notation $\ser(x^2)$ for a power series in $x^2$,
explained after \eqn{\eqref{gen-series-def}}.


\paragraph{Fourier-Bessel transform.}

\rev{The Fourier-Bessel transform of order $\nu$
\begin{align}
   \label{FB-trf}
   I(q) &= \int_0^\infty dz\ms z\, J_{\nu}(q z) \, W(z)
\intertext{is its own inverse:}
   \label{FB-inverse}
   W(z) &= \int_0^\infty dq\ms q\, J_{\nu}(q z) \, I(q)
   \,.
\end{align}
This can be expressed in terms of the closure relation
\begin{align}
   \label{Bessel-closure}
   \int_0^\infty dq\ms q\, J_\nu(q y) \, J_\nu(q z)
   &=
   \frac{\delta(y - z)}{z}
   \,.
\end{align}
Equations~\eqref{FB-trf} to \eqref{Bessel-closure} require that $\nu \ge -1/2$.}


\subsection{Lommel functions}

The literature about Lommel functions is much more sparse than for Bessel
functions.  This holds especially for the Lommel functions of the second kind,
which are relevant for the discussion in \sect{\ref{sec:rescale_lin}}.  We
therefore provide some more detail here and in particular derive a
representation in terms of power series.
Unless explicitly mentioned, the following relations can be found in
\cite[\href{https://dlmf.nist.gov/11.9}{Chapter 11.9}]{NIST:DLMF} or in
\cite[Chapter 8.57]{Gradshteyn}.

The Lommel functions of the first and second kind, $s_{\mu, \nu}(x)$ and
$S_{\mu, \nu}(x)$, solve the second-order differential equation
\begin{align}
   \label{Lommel-ODE}
   x^2 \, w''(x) + x \, w'(x) + \bigl[ x^2 - \nu^2 \bigr] \, w(x)
   &=
   x^{\mu + 1}
   \,.
\end{align}
The Lommel functions of first kind are given in terms or a power series as
\begin{align}
   \label{1st-Lommel-def}
   s_{\mu, \nu}(x)
   &=
   x^{\mu + 1} \, \ser(x^2)
\end{align}
and oscillate like the Bessel functions for large $x$.  For the Lommel functions
of the second kind one has
\begin{align}
   \label{2nd-Lommel-def}
   S_{\mu, \nu}(x)
   &=
   s_{\mu, \nu}(x) + 2^{\mu - 1} \,
      \Gamma\biggl( \frac{\mu+\nu+1}{2} \biggr) \,
      \Gamma\biggl( \frac{\mu-\nu+1}{2} \biggr)
   \nonumber \\
   & \quad \times
      \biggl[ \sin \Bigl(\bs (\mu-\nu) \frac{\pi}{2} \,\Bigr) \, J_{\nu}(x)
         - \cos \Bigl(\bs (\mu-\nu) \frac{\pi}{2} \,\Bigr) \, Y_{\nu}(x)
      \biggr]
   \,.
\end{align}
If $\mu + \nu$ or $\mu - \nu$ is a negative odd integer, several coefficients of
the power series in \eqref{1st-Lommel-def} become infinite, as do the $\Gamma$
functions in \eqref{2nd-Lommel-def}.  The function $S_{\mu, \nu}$ can be defined
by a limiting procedure in these cases, as shown in \cite[Chapter
3.4]{Luke:1962} or \cite[Chapter 10.73]{Watson:1944}.


\paragraph{Relations.}

The Lommel functions of the second kind satisfy
\begin{align}
   \label{Lommel-symm}
   S_{\mu, \nu}(x) &= S_{\mu, -\nu}(x)
\end{align}
and
\begin{align}
   \label{Lommel-basic}
   S_{\mu, \nu}(x)
   &=
   \frac{x}{2 \nu} \, \Bigl[ (\mu+\nu-1) \, S_{\mu-1, \nu-1}(x)
      - (\mu-\nu-1) \, S_{\mu-1, \nu+1}(x) \Bigr]
   \,,
   \notag \\
   \frac{d}{dx} \, S_{\mu, \nu}(x)
   &=
   \frac{1}{2} \, \Bigl[ (\mu+\nu-1) \, S_{\mu-1, \nu-1}(x)
      + (\mu-\nu-1) \, S_{\mu-1, \nu+1}(x) \Bigr]
   \,,
\end{align}
which implies
\begin{align}
   \label{Lommel-diff}
   \frac{d}{dx} \ms \Bigl[ x^\nu S_{\mu, \nu}(x) \Bigr]
   &=
   (\mu+\nu-1) \, x^{\nu} \, S_{\mu-1, \nu-1}(x)
   \,.
\end{align}
Notice the close analogy of these relations with \eqref{Bessel-basic} and
\eqref{Bessel-diff}.
The identity
\begin{align}
   \label{Lommel-recursion}
   S_{\mu + 2, \nu}(x)
   &=
   x^{\mu + 1}
   - (\mu + \nu + 1) (\mu - \nu + 1) \, S_{\mu, \nu}(x)
\end{align}
implies that
\begin{align}
   \label{Lommel-simple}
   S_{\nu+1, \nu}(x) &= x^{\nu}
   \,,
   &
   S_{-\nu+1 , \nu}(x) &= x^{-\nu}
   \,,
\end{align}
and by recursion
\begin{align}
   \label{Lommel-polynom}
   x^{-\nu} \, S_{\nu+1+2 k, \nu}(x)
   &=
   \sum_{j=0}^{k} c_{\nu,\ms k,\ms j} \, x^{2j}
   \,,
   &
   x^{\nu} \, S_{-\nu+1+2 k, \nu}(x)
   &=
   \sum_{j=0}^{k} c_{-\nu,\ms k,\ms j} \, x^{2j}
\end{align}
for $k \in \mathbb{N}$, with coefficients $c_{\nu,\ms k,\ms j}$ whose explicit
form we do not need.


\paragraph{Behaviour at large arguments.}
For $x\to \infty$, the Lommel functions of the second kind have an asymptotic
expansion
\begin{align}
   \label{Lommel-large-x}
   x^\nu \ms S_{\mu, \nu}(x)
   & \sim
   x^{\mu + \nu - 1} \, \ser(x^{-2})
   \,,
\end{align}
where the first term in the series $\sigma$ is equal to $1$.  This expansion is
also valid for the special cases mentioned below \eqn{\eqref{2nd-Lommel-def}}.


\paragraph{Series representation.}

As in the case of Bessel functions, we assume $\nu \ge 0$.  Combining
\eqref{Bessel-J-Y-series}, \eqref{1st-Lommel-def}, and \eqref{2nd-Lommel-def},
one obtains a series representation
\begin{align}
   x^\nu \ms S_{\mu, \nu}(x)
   &=
   \begin{cases}
      \ser(x^2) + x^{\mu + \nu + 1} \, \ser(x^2)
      + x^{2 \nu} \ms \ln(x) \, \ser(x^2)
         & \text{ for } \nu \in \mathbb{Z}
      \,,
      \\
      \ser(x^2) + x^{\mu + \nu + 1} \, \ser(x^2)
      + x^{2 \nu} \, \ser(x^2)
         & \text{ for } \nu \notin \mathbb{Z}
      \,,
   \end{cases}
\end{align}
provided that neither $\mu + \nu$ nor $\mu - \nu$ is a negative odd integer.
For these exceptional values of $\mu$ and $\nu$, we need separate derivations.
We limit ourselves to values satisfying
\begin{align}
   \label{pos-condition}
   \mu + \nu > -1
   \,,
\end{align}
which is sufficient for the discussion in \sect{\ref{sec:rescale_lin}}.
This leaves us with the possibility that $\mu - \nu$ is a negative odd integer.
We then can distinguish two cases:
\begin{itemize}
\item $\nu \in \mathbb{Z}$.  Then $\mu + \nu$ is a positive odd integer, because
we impose  \eqref{pos-condition}.  We can therefore use the second
representation in \eqref{Lommel-polynom} and find that $x^\nu \, S_{\mu,
\nu}(x)$ is a polynomial in $x^2$.
\item $\nu \notin \mathbb{Z}$.  In this case we can use a representation derived
in \cite[Chapter 3.4]{Luke:1962} and also in \cite[Chapter 10.73]{Watson:1944}.
It has the form
\begin{align}
   \label{Lommel-special-1}
   x^\nu \ms S_{\nu - 1, \nu}(x)
   &=
   x^{2 \nu} \ms \ser(x^2)
   + 2^{\nu-1} \ms \Gamma(\nu)
      \biggl[ x^\nu J_\nu(x) \, \ln\bs\frac{x}{2}
         - \frac{\pi}{2} \, x^\nu \ms Y_\nu(x)
      \biggr]
   \notag \\[0.1em]
   &=
   \ser(x^2) + x^{2 \nu} \, \ser(x^2) + x^{2 \nu} \ms \ln(x) \, \ser(x^2)
\end{align}
and is valid for all $\nu > 0$.  Starting from this, we can repeatedly apply the
recursion relation \eqref{Lommel-recursion} to obtain
\begin{align}
   \label{Lommel-special-2}
   x^\nu \ms S_{\mu = \nu - 1 - 2 k,\ms \nu}(x)
   &=
   d_{\mu, \nu} \, x^\nu \, S_{\nu - 1, \nu}(x)
   + x^{\mu + \nu + 1} \ms \ser(x^2)
   &
   \text{for }
   k = 1, 2, 3, \ldots ,
\end{align}
where $d_{\mu, \nu}$ is some number and the series multiplying $x^{\mu + \nu +
1}$ is in fact a polynomial.
\end{itemize}

Putting everything together, we find that one can always write
\begin{align}
   \label{Lommel-gen-series}
   x^\nu \ms S_{\mu, \nu}(x)
   &=
   \ser(x^2) + x^{\mu + \nu + 1} \, \ser(x^2) + x^{2 \nu} \, \ser(x^2)
      + x^{2 \nu} \ms \ln(x) \, \ser(x^2)
\end{align}
for $\mu + \nu > -1$ and $\nu \ge 0$. If $\mu + \nu$ is a positive odd integer,
then the r.h.s.\ reduces to a polynomial in $x^2$.  Otherwise, the term with
$\ln(x)$ is present if $\nu$ is an integer or if $\nu - \mu$ is a negative odd
integer.

\section{Integrands with a Gaussian decrease in \texorpdfstring{$z$}{z}}
\label{sec:gaussian}

In \sect{\ref{sec:rescale_lin_large_z}} we solved the system
\eqref{ODE-h3-lin-f1}, \eqref{ODE-h3-lin-f2} of differential equation for functions $f_1$ or $f_2$ with
an exponential decrease at large arguments.  Let us do the same for the case of
functions with a Gaussian decrease.  We treat the functions $f_1$ and $f_2$ in
turn, since the corresponding solutions are slightly different.

We start with $f_2 = 0$ and
\begin{align}
   f_1(z)
   &=
   c_1 \ms z^{\mu + \nu} \,
   \exp\bigl( - \lambda^2 z^2 - \kappa z \ms\bigr)
   \,,
\end{align}
where $\lambda > 0$, whilst $\kappa$ may be positive, negative, or zero.
With the ansatz
\begin{align}
   \label{gauss-h-hat-ansatz}
   h_i(z, q)
   &=
   \hat{h}_i(z, q) \, \exp( - \lambda^2 z^2 - \kappa z )
   &
   (i=1, 3)
\end{align}
the system \eqref{ODE-h3-lin-f1}, \eqref{ODE-h3-lin-f2} of ODEs becomes
\begin{align}
\label{ODE-gauss-f1}
   \begin{pmatrix} c_1 \ms z^{\mu + \nu} \\ 0 \end{pmatrix}
   &=
   - \begin{pmatrix}
        2 \lambda^2 z \hat{h}_1 \\
        q \ms \hat{h}_1 + 2 \lambda^2 z^2 \ms \hat{h}_3
     \end{pmatrix}
   - \kappa \begin{pmatrix} \hat{h}_1 \\ z \hat{h}_3 \end{pmatrix}
   + \begin{pmatrix}
      d\ms \hat{h}_1 / dz + q \ms z \hat{h}_3 \\
      z\ms d\ms \hat{h}_3 / dz - 2\nu \ms \hat{h}_3
     \end{pmatrix}
   \,,
\end{align}
which can in turn be solved with the ansatz
\begin{align}
   \hat{h}_1(z, q)
   &=
   z^{\mu + \nu - 1} \, \sum_{k=0}^\infty b_{1,k}(q) \ms z^{-k}
   \,,
   \notag \\
   \hat{h}_3(z, q)
   &=
   z^{\mu + \nu - 3} \, \sum_{k=0}^\infty b_{3,k}(q) \ms z^{-k}
   \,.
\end{align}
We find that for each $k$ in the sums, the first term on r.h.s. of
\eqref{ODE-gauss-f1} is leading at large $z$, whilst the second term is down by
$z^{-1}$ and the third term is down by $z^{-2}$.   One can therefore solve for
the coefficients recursively, starting with
\begin{align}
   b_{1,0} &= - c_1 \ms/ (2 \lambda^2) \,,
   &
   b_{3,0} &= - q \ms b_{1,0} \ms/ (2 \lambda^2)
   \,.
\end{align}

We now turn to the case where $f_1 = 0$ and
\begin{align}
   f_2(z)
   &=
   c_2 \ms z^{\mu + \nu} \,
   \exp\bigl( - \lambda^2 z^2 - \kappa z \ms\bigr)
   \,.
\end{align}
If we make again the ansatz \eqref{gauss-h-hat-ansatz}, the system
\eqref{ODE-h3-lin-f1}, \eqref{ODE-h3-lin-f2} of ODEs takes the form
\begin{align}
\label{ODE-gauss-f2}
   \begin{pmatrix} 0 \\ c_2 \ms z^{\mu + \nu} \end{pmatrix}
   &=
   - \begin{pmatrix}
        2 \lambda^2 z \hat{h}_1 - q \ms z \hat{h}_3 \\
        2 \lambda^2 z^2 \ms \hat{h}_3
     \end{pmatrix}
   - \kappa \begin{pmatrix} \hat{h}_1 \\ z \hat{h}_3 \end{pmatrix}
   + \begin{pmatrix}
      d\ms \hat{h}_1 / dz \\
      z\ms d\ms \hat{h}_3 / dz - 2\nu \ms \hat{h}_3 - q \ms \hat{h}_1
     \end{pmatrix}
     \,.
\end{align}
We now set
\begin{align}
   \hat{h}_1(z, q)
   &=
   z^{\mu + \nu - 2} \, \sum_{k=0}^\infty b_{1,k}(q) \ms z^{-k}
   \,,
   \notag \\
   \hat{h}_3(z, q)
   &=
   z^{\mu + \nu - 2} \, \sum_{k=0}^\infty b_{3,k}(q) \ms z^{-k}
   \,,
\end{align}
such that the three terms on the r.h.s.\ of \eqref{ODE-gauss-f2} are leading,
subleading in $z^{-1}$, and subleading in $z^{-2}$, respectively.  One can then
solve for the coefficients iteratively, starting with
\begin{align}
   b_{3,0} &= - c_2 \ms/ (2 \lambda^2) \,,
   &
   b_{1,0} &= q \ms b_{3,0} \ms/ (2 \lambda^2)
   \,.
\end{align}

\section{Chebyshev interpolation}
\label{sec:cheb-details}

In this appendix, we collect formulae for Chebyshev interpolation that are
relevant in our method.  We use the same notation as in \cite{Diehl:2021gvs,
Diehl:2023cth}, where references or derivations for the following results can be
found.

We start with functions defined on the interval $[-1, 1]$, as is customary in
the discussion of Chebyshev polynomials.  The generalisation to any other finite
interval is trivial and given at the end of this appendix.


\paragraph{Functions with support $t \in [-1, 1]$.}
The Chebyshev polynomials of the first kind, $T_k(t)$, can be defined by
\begin{align}
   \label{T-def}
   T_k(\cos \theta) &=  \cos (k \ms \theta) \,,
\end{align}
where $k \ge 0$ is an integer.  Using the multiple-angle formula for the cosine,
one readily verifies that $T_k(t)$ is a polynomial in $t$ of order $k$.  The set
of all $T_k(t)$ is orthonormal with respect to the integration measure
$1 / \sqrt{1 - t^2}$.

For a given integer $N$, the \emph{Chebyshev points} are given by
\begin{align}
   \label{cheb_points}
   t_j
   &= \cos \theta_j
   \,,
   &
   \theta_j
   &= \frac{j \pi}{N}
   &
   \text{ with }
   j = 0, \ldots, N
   \,.
\end{align}
They form a descending series from $t_0 = 1$ to $t_N = -1$ and obey the
reflection property $t_{N-j} = - t_j$.  The polynomial $T_N(t)$ assumes its
maxima $+1$ and minima $-1$ at the Chebyshev points, as is easily seen from
\eqref{T-def}.  We note that the density of Chebyshev points increases from the
centre toward the end points of the interval $[-1, 1]$. This feature is crucial
to avoid Runge's phenomenon for equispaced interpolation grids.

The \emph{Chebyshev interpolant} $p_N(t)$ of a function $f(t)$ is the unique
polynomial that satisfies
\begin{align}
   p_N(t_j) &= f(t_j)
   &
   \text{ for }
   j = 0, \ldots, N
\end{align}
at the Chebyshev points $t_j$.  It can be written in the form
\begin{align}
   \label{cheb_interp}
   p_N(t) &= \sum_{k=0}^{N}\, \beta_k\ms c_k\ms T_{k}(t)
\end{align}
with coefficients
\begin{align}
   \label{cheb_interp_coeff}
   c_k &= \frac{2}{N} \sum_{j=0}^{N}\, \beta_j \ms T_k(t_j) \, f(t_j)
   \,,
\end{align}
where $\beta_0 = \beta_N = 1/2$ and $\beta_j = 1$ otherwise.

For functions $f(t)$ that are sufficiently regular in the interval $[-1, 1]$
(including the end points), one can show that the difference $|f(t) - p_N(t)|$
uniformly tends to zero on the full interval in the limit $N\to \infty$.  A
precise statement of this convergence theorem can be found in
\cite[section~2]{Diehl:2021gvs}.  We note that a function is \emph{not}
sufficiently regular for the purpose of this theorem if its first derivative
diverges at an endpoint of the interval.  This explains our insistence on a
finite first derivative of $h_1$ and $h_3$ in the discussion of
\sect{\ref{sec:rescale_lin}}.


\paragraph{Differentiation.}
Given the Chebyshev interpolant $p_N(t)$ for a function $f(t)$, one can
approximate the derivative $f'(t) = d f/ d t$ of the function by the derivative
$d p_N /d t$ of its interpolant.  Note that in general $d f/ d t$ is not equal
to $d p_N /d t$ at the Chebyshev points; this is obvious because one cannot
compute the exact values of $f'(t_j)$ from the discrete set of function values
$f(t_j)$.
At the Chebyshev points, the derivative of the interpolating polynomial can be
computed as
\begin{align}
\label{cheb_deriv}
\frac{d}{d t} \ms p_N(t_j) &= \sum_{k=0}^{N}\, D_{j k}\, f(t_k)
\end{align}
with a differentiation matrix given by
\begin{align}
   \label{basic-diff-matrix}
   D_{0 0}
   &= - D_{N N} = \frac{2 N^2 + 1}{6}
   \,,
   \notag \\
   D_{j j}
   &=
   - \dfrac{\cos\theta_j}{2 \sin^2\theta_j}
   &&
   \text{for}\quad j \neq 0, N,
   \notag \\
   D_{j k} &= \dfrac{\beta_k}{\beta_j}\,
            \dfrac{(-1)^{j+k}}{t_j - t_k}
   &&
   \text{for}\quad j \neq k
   \,.
\end{align}
Note that the matrix multiplication \eqref{cheb_deriv} maps a vector $f(t_k)
= \text{\it const}$ onto the zero vector, as it must be because the derivative
of a constant function is zero.


\paragraph{Integration.}
Approximating a function by its Chebyshev interpolant, one can derive an
integration rule
\begin{align}
   \label{CC-rule}
   \int_{-1}^1 d t \, f(t)
   &\approx
   \sum_{\genfrac{}{}{0pt}{}{k=0}{\text{even}}}^{N}\,
      \frac{2 \ms \beta_k \ms c_k}{1-k^2}
   = \sum_{j=0}^{N}\, w_j\ms f(t_j)
\end{align}
with weights
\begin{align}
   \label{CC-weights}
   w_j
   &=
   \frac{4 \beta_j}{N}\,
   \sum_{\genfrac{}{}{0pt}{}{k=0}{\text{even}}}^{N}\,
   \beta_k\, \frac{\cos (k \ms \theta_j)}{1-k^2}
   \,.
\end{align}
This is known as \emph{Clenshaw-Curtis quadrature}.  A detailed discussion of
its accuracy, as well as a comparison with Gauss quadrature, is given in
\cite[chapter 19]{Trefethen:2012} and in \cite{Trefethen:2008sia}.


\paragraph{Functions with support $u \in [u_a, u_b]$.}
It is easy to generalise the preceding results to an arbitrary finite interval
in the variable $u$ by introducing the linear transformation
\begin{align}
   \label{u-t-transform}
   u(t) &= \frac{u_a - u_b}{2} \, t + \frac{u_a + u_b}{2}
   \,.
\end{align}
With the Chebyshev points
\begin{align}
   \label{u-Cheb-points}
   u_j
   &= u(t_j)
   =
   \frac{u_a - u_b}{2} \, \cos\frac{j \pi}{N}  + \frac{u_a + u_b}{2}
\end{align}
in the new variable, we obtain the following rules for differentiation and for integration:
\begin{align}
   \label{diff-mat-u}
   \frac{d}{d u} \ms f(u_j)
   & \approx
   \sum_{k=0}^{N}\, D^u_{j k}\, f(u_k)
   \,,
   &
   D^u_{j k} &= \frac{2}{u_a - u_b} \, D_{j k}
\end{align}
with $D_{j k}$ from \eqref{basic-diff-matrix} and
\begin{align}
   \label{CC-quadrature-u}
   \int_{u_a}^{u_b} d u \, f(u)
   & \approx
   \sum_{j=0}^{N}\, w^u_j \ms f(u_j)
   \,,
   &
   w^u_j &= \frac{u_b - u_a}{2} \, w_j
\end{align}
with $w_j$ from \eqref{CC-weights}.

\section{Fourier-Bessel integrals as distributions}
\label{sec:distribution}

In this appendix, we show how the formally divergent integrals discussed in
\sect{\ref{sec:special-int}} can be defined in the distributional sense.

For a given $\nu$ and a function $g(q)$, we define the plus-distribution $[ g(q)
]_{+}^{q_0}$ by its action on a test function, which we require to be of the
form $q^{\nu+1} \ms \varphi(q)$:
\begin{align}
   \label{plus-def}
   &
   \int_0^\infty dq  \, q^{\nu+1} \ms
      \varphi(q) \, \bigl[ g(q) \bigr]_{+}^{q_0}
   =
   \int_0^\infty dq  \, q^{\nu+1} \ms
      \bigl[ \varphi(q) - \theta(q_0 - q) \ms \varphi(0) \bigr]\ms  \, g(q)
   \,.
\end{align}
This can also be written as
\begin{align}
   \label{plus-limit}
   \bigl[ g(q) \bigr]_{+}^{q_0}
   &=
   \lim_{\epsilon \to 0} \,
   \biggl[
      \theta(q - \epsilon) \, g(q)
      - \frac{\delta(q)}{q^{\nu+1}} \,
      \int_{\epsilon}^{q_0} d\tilde{q} \; \tilde{q}^{\nu+1} \ms g(\tilde{q})
   \biggr]
   \,,
\end{align}
where it is understood that the limit $\epsilon \to 0$ should be taken
\emph{after} integrating over $q$ with a test function.
Note that the factor $1 / q^{\nu+1}$ in \eqref{plus-limit} does not give rise to
singularities because we consider test functions proportional to $q^{\nu+1}$.
Alternatively, one can write $\delta(q) / q^{\nu+1} = (\nu+2) \,
\delta(q^{\nu+2})$ and integrate over $q^{\nu+2}$ in \eqref{plus-def}.

If $g(q)$ decreases faster than $1 / q^{\nu+2}$ at large $q$, then the upper
integration limit $q_0$ in the subtraction term of \eqref{plus-def} and
\eqref{plus-limit} can be set to $\infty$.  Otherwise, $q_0$ must be finite.

For $\nu = 0$ our definition is equivalent to the one in
\cite[\app{C}]{Ebert:2016gcn} if the latter is applied to two-dimensional test
functions $\varphi(\mvec{q})$ with azimuthal symmetry.

We now extend our previous definition \eqref{sing-int-def} to the distributional
form
\begin{align}
   \label{BT-distr}
   I(q)
   &=
   \int_{0}^\infty dz \, J_{\nu}(q z) \, \ftil(z, q)
   \notag \\[0.2em]
   &=
   \biggl[ \frac{1}{q^{\nu+1}} \, \frac{d}{dq} \, K(q)
   \biggr]_{+}^{q_0}
   + \frac{\delta(q)}{q^{\nu+1}} \, K(q_0)
   - \int_0^\infty dz\, J_{\nu+1}(q z) \; \frac{d}{dq} \, \frac{\ftil(z, q)}{z}
   \,,
\end{align}
where
\begin{align}
   \label{K-def}
   K(q)
   &=
   q^{\nu+1} \! \int_0^\infty dz\, J_{\nu+1}(q z) \; \frac{\ftil(z, q)}{z}
   \,.
\end{align}
Using \eqref{plus-limit}, we can rewrite \eqref{BT-distr} as
\begin{align}
   I(q)
   &=
   \lim_{\epsilon \to 0} \,
   \Biggl[
      \frac{\theta(q - \epsilon)}{q^{\nu+1}} \,
      \frac{d}{dq} \, K(q)
   + \frac{ \delta(q)}{q^{\nu+1}} \, \biggl(
      K(q_0)
      - \int_\epsilon^{q_0} d\tilde{q} \; \frac{d}{d\tilde{q}} \ms K(\tilde{q})
      \biggr)
   \Biggr]
   \notag \\
   & \quad
   - \int_0^\infty dz\, J_{\nu+1}(q z) \; \frac{d}{dq} \ms \frac{\ftil(z, q)}{z}
   \notag \\[0.3em]
   &=
   \lim_{\epsilon \to 0} \,
   \Biggl[
      \frac{\theta(q - \epsilon)}{q^{\nu+1}} \,
      \frac{d}{dq} \, K(q)
      + \frac{ \delta(q)}{q^{\nu+1}} \, K(\epsilon)
   \Biggr]
   - \int_0^\infty dz\, J_{\nu+1}(q z) \; \frac{d}{dq} \ms \frac{\ftil(z, q)}{z}
   \,.
\end{align}
This makes it explicit that the dependence on $q_0$ cancels, as it must be.
If $K(0) = 0$, one can take the $\epsilon = 0$ limit for each individual term on
the r.h.s.  The distributional part proportional to $\delta(q)$ then disappears,
and $I(q)$ is an ordinary function.

If $\ftil$ is independent of $q$, one readily finds
\begin{align}
   \int_0^q d\tilde{q}\; \tilde{q}^{\nu+1}\, I(\tilde{q})
   &=
   K(q)
   \,.
\end{align}
The representation in \eqref{BT-distr} and \eqref{K-def} thus generalises the
relation between the differential and cumulative $q$ spectra in \eqref{tmd-int}
and \eqref{tmd-cumul} to distributions and to $\nu \neq 0$.

As a simple illustration let us set $\ftil(z) = z$ and $\nu=0$.  One then has
$K(q) = 1$, and our definition \eqref{BT-distr} gives $I(q) = \delta(q) / q$ in
agreement with \eqref{delta-fct}.


\paragraph{Inverse Fourier-Bessel transform}
If $\ftil$ is independent of $q$, then \eqref{BT-distr} is an extension of the
ordinary Fourier-Bessel transform.  We now show that, in accordance with the
relations \eqref{FB-trf} and \eqref{FB-inverse} for ordinary functions, the
Fourier-Bessel transform of $I(q)$ gives back $\ftil(z) / z$, provided that
\begin{align}
   \label{K-conditions}
   q^2 \ms K(q)
   & \underset{q \to 0}{\rightarrow} 0
   \,,
   &
   q^{-1/2 - \nu} \, K(q)
   & \underset{q \to \infty}{\rightarrow} 0
   \,.
\end{align}
We do not investigate here for which functions $\ftil(z)$ these conditions are
satisfied, but we note that they do hold for the case $\ftil(z) = z \, \ln^k \!
z$ that we studied in \sect{\ref{sec:special-int}}.

Taking the Fourier-Bessel transform of \eqref{BT-distr} and using \eqref{plus-def} with $\varphi(q) = J_\nu(q y) / q^\nu$, we get
\begin{align}
   \tilde{I}(y)
   &=
   \int_0^\infty dq \, q \, J_\nu(q y) \, I(q)
   \notag \\[0.3em]
   &=
   \int_0^\infty dq \,
   \biggl[ \frac{J_\nu(q y)}{q^\nu} - \theta(q_0 - q) \ms J_\nu^{(0)} \biggr]
   \frac{d}{dq} \ms K(q)
   + J_\nu^{(0)} \, K(q_0)
\end{align}
with
\begin{align}
   J_\nu^{(0)}
   &=
   \lim_{q \to 0} \, \frac{J_\nu(q y)}{q^\nu}
   =
   \biggl( \frac{y}{2} \biggr)^\nu \frac{1}{\Gamma(1+\nu)}
   \,.
\end{align}
Integration by parts gives
\begin{align}
   \label{BT-inverse}
   \tilde{I}(y)
   &=
   - \int_0^\infty dq \;
   K(q)\, \frac{d}{dq} \ms \frac{J_\nu(q y)}{q^\nu}
   \,,
\end{align}
where the boundary terms have vanished by virtue of \eqref{K-conditions}. Using
\eqref{Bessel-diff-2} and \eqref{Bessel-closure}, we finally obtain the desired
result
\begin{align}
   \tilde{I}(y)
   &=
   y \int_0^\infty \! dq \, q \, J_{\nu+1}(q y)
     \int_0^\infty \! dz \, J_{\nu+1}(q z) \, \frac{\ftil(z)}{z}
   = \frac{\ftil(y)}{y}
   \,.
\end{align}


\phantomsection
\addcontentsline{toc}{section}{References}

\bibliographystyle{JHEP}
\bibliography{bessel.bib}

\end{document}